\documentclass[final,5p,times,twocolumn]{elsarticle}
\usepackage{graphicx}
\usepackage{dcolumn}
\usepackage{siunitx}

\usepackage{bm}
\usepackage{booktabs}
\usepackage{makecell}
\usepackage[flushleft]{threeparttable}
\usepackage{natbib}
\usepackage{calligra}
\usepackage{subcaption}
\usepackage{amsmath}
\usepackage[dvipsnames]{xcolor}
\usepackage{mathrsfs}
\usepackage{lmodern}
\usepackage{xr}
\usepackage{graphicx}
\usepackage{tikz}
\usepackage{alltt}
\usepackage{hyperref}
\usepackage{listings}
\usepackage{amsbsy}

\usepackage{color}
\definecolor{gray}{rgb}{0.4,0.4,0.4}
\definecolor{darkblue}{rgb}{0.0,0.0,0.6}
\definecolor{cyan}{rgb}{0.0,0.6,0.6}

\lstdefinelanguage{XML}
{
	morestring=[b]",
	morestring=[s]{>}{<},
	morecomment=[s]{<?}{?>},
	stringstyle=\color{black},
	identifierstyle=\color{darkblue},
	keywordstyle=\color{cyan},
	morekeywords={xmlns,version,type}
}

\makeatletter
\newcommand*{\textalltt}{}
\DeclareRobustCommand*{\textalltt}{%
	\begingroup
	\let\do\@makeother
	\dospecials
	\catcode`\\=\z@
	\catcode`\{=\@ne
	\catcode`\}=\tw@
	\verbatim@font\@noligs
	\@vobeyspaces
	\frenchspacing
	\@textalltt
}
\newcommand*{\@textalltt}[1]{%
	#1%
	\endgroup
}
\makeatother

\tikzstyle{rect} = [draw,rectangle,fill=white!20,text width=6em,text centered, minimum height=2em]
\tikzstyle{elli} = [draw,ellipse,fill=white!20,minimum height=2em]
\tikzstyle{circ} = [draw,circle,fill=white!20,minimum width=8pt, inner sep=10pt]
\tikzstyle{diam} = [draw,diamond,fill=white!20,text width=6em,text badly centered, inner sep=0pt]
\tikzstyle{line} = [draw,-latex]

\begin{document}

\begin{frontmatter}

\title{\texttt{BTE-Barna}: An extension of \texttt{almaBTE} for thermal simulation of devices based on 2D materials}

\author[a]{Mart\'i Raya-Moreno}
\author[a]{Xavier Cartoix\`a}
\author[b]{Jes\'us Carrete\corref{author}}

\cortext[author] {Corresponding author.\\\textit{E-mail address:} jesus.carrete.montana@tuwien.ac.at}
\address[a]{Departament d'Enginyeria Electr\`onica, Universitat Aut\`onoma de Barcelona, 08193 Bellaterra, Barcelona, Spain}
\address[b]{Institute of Materials Chemistry, TU Wien, A-1060 Vienna, Austria}


\begin{abstract}
	We present \texttt{BTE-Barna} (Boltzmann Transport Equation - Beyond the Rta for NAnosystems), a software package that extends the Monte Carlo (MC) module of the \texttt{almaBTE} solver of the Peierls-Boltzmann transport equation for phonons (PBTE) to work with nanosystems based on 2D materials with complex geometries. To properly capture how the phonon occupations evolve in momentum space as a result of scattering, we have supplemented the relaxation-time approximation with an implementation of the propagator for the full linearized version of the PBTE. The code can now find solutions for finite and extended devices under the effect of a thermal gradient, with isothermal reservoirs or with an arbitrary initial temperature distribution in space and time, writing out the temperature and heat flux distributions as well as their spectral decompositions. Besides the full deviational MC solver, a number of useful approximations for highly symmetric devices are also included.
\end{abstract}


\end{frontmatter}


{\bf PROGRAM SUMMARY}

\begin{small}
\noindent
{\em Program Title:} \texttt{BTE-Barna}                                         \\
{\em CPC Library link to program files:} (to be added by Technical Editor) \\
{\em Developer's repository link:} \url{https://github.com/sousaw/BTE-Barna} \\
{\em Code Ocean capsule:} (to be added by Technical Editor)\\
{\em Licensing provisions:}  Apache-2.0  \\
{\em Programming language:}   C++                            \\
{\em Nature of problem:} Calculation of temperature profiles, thermal flux and/or effective thermal conductivities for nanosystems based on 2D materials.\\
{\em Solution method:} \\
For highly symmetric systems confined along some direction(s), the linearized phonon Boltzmann transport equation is solved iteratively by partially suppressing the phonon lifetimes due to boundaries. For more complex geometries, an energy-based deviational Monte Carlo method including off-diagonal terms in the linearized scattering operator arising from the phonon-phonon interaction is used. \\
{\em Additional comments including restrictions and unusual features:} \\
Depends on the following external libraries: boost, Eigen, HDF5, MessagePack, MPI, oneTBB, spglib, and almaBTE.
\\
\end{small}

\section{\label{sec:Intro}Introduction}
The continuous shrinking of electronic components, following Moore's Law~\cite{Moores_law}, is pushing bulk semiconductor-based devices, such as silicon transistors, to their fundamental limits. In addition, this increase in the integration level leads to ever higher power densities and raises the Herculean challenge of dissipating the generated heat~\cite{MooreMatToday2014}.

In this context, two dimensional materials (2DMs), thanks to their atomic thickness, low surface roughness and density of dangling bonds~\cite{ChhowallaNatRevMat2016}, together with the possibility of stacking them to create heterostructures with tuned properties and their compatibility with CMOS technology, are quite promising candidates to replace III-V compounds and silicon in transistor channels~\cite{AkinwandeNatCom2014,JiangJEDS2019}. Understanding thermal transport in 2DMs is essential to optimize heat management in such devices. Since phonons are the main heat carriers in semiconductors, heat flux can be described through the Peierls-Boltzmann Transport Equation (PBTE)~\cite{ZimanEPH}. For highly symmetric structures (e.g.: bulk systems, nanowires, thin-films\ldots) the relaxation time approximation (RTA), where it is assumed that each phonon mode relaxes to equilibrium independently, has a long tradition of being used, but more recent advances enable an iterative solution beyond that crude approximation~\cite{OminiPRB1996,LiPRB2012}. Furthermore, the inclusion of phonon properties calculated from first-principles (frequencies, scattering rates\ldots) makes it possible to solve the PBTE even for novel materials where simpler models to describe those properties are lacking~\cite{WardPRB2009,LindsayPRB2013}. These iterative first-principles-based PBTE solvers are available to the community in software packages such as \texttt{ShengBTE}~\cite{ShengBTE}, \texttt{almaBTE}~\cite{almaBTE} or \texttt{Phono3py}~\cite{Phono3py}.

Despite those advances in solving the PBTE for simple systems, using direct or iterative methods for its solution becomes impractical for more intricate configurations, such as the ones required by micro or nanosized devices. A usual approach to overcome such limitations is to use a Monte Carlo method to integrate the PBTE. Notwithstanding their success in other applications, in classical Monte Carlo methods to solve the PBTE the accuracy is hindered by several factors, most notably the inability of the scattering algorithm to conserve energy. Indeed, additional algorithms are required to keep the system energy constant~\cite{ChenJHT2005,MeiJAP2014,JeanJAP2014}, which can bias the distribution in unknown ways~\cite{PeraudARHT2014}. Moreover, classical methods suffer from high statistical noise plus the fact that most of the computational time is wasted simulating the equilibrium part of the distribution~\cite{almaBTE,PeraudPRB2011,PeraudARHT2014}. In that context, RTA-based deviational energy Monte Carlo methods have proven themselves as good alternative ways to overcome the limitations~\cite{PeraudPRB2011,PeraudAPL2012,PeraudARHT2014} of more traditional methods, as they naturally conserve the energy by operating with bundles of energy instead of phonons and reduce the required number of computational particles by simulating only the deviation from a reference equilibrium distribution. 

However, the validity of the RTA approach for 2D materials is questionable, and it has been shown to yield a very poor description of thermal properties for several of them~\cite{CepellottiNatCom2015,LindsayJAP2019}. For those cases one might need to use an energy deviational Monte Carlo method based on the full collision operator~\cite{LandonJAP2014}.

In this work we present the \texttt{BTE-Barna} software package, an extension of \texttt{almaBTE} to tackle 2D systems both within and beyond the RTA, so that now it can address finite and/or periodic 2D materials and their heterojunctions under the effect of thermal gradients and isothermal reservoirs. We analyze a selection of test cases and discuss the validity of the RTA. Additionally, the iterative solver in \texttt{almaBTE} is extended to provide the effective thermal conductivity for nanoribbons and nanowires.

The paper is structured as follows: after displaying the general structure of \texttt{BTE-Barna} in Sec.~\ref{sec:CodeStructure} and discussing the theoretical background in Sec.~\ref{sec:Methods}, we provide test cases of the implementation in Sec.~\ref{sec:validation} and present illustrative example applications of our package simulators in Sec.~\ref{sec:Result}. Our summary and conclusions are given in Sec.~\ref{sec:Conclusions}. The appendices provide a discussion about the solution of the PBTE in nanowires, additional examples and detailed documentation. 

\section{\label{sec:CodeStructure}\texttt{BTE-Barna} structure}
\begin{figure*}
	\centering
	\includegraphics[width=0.85\textwidth]{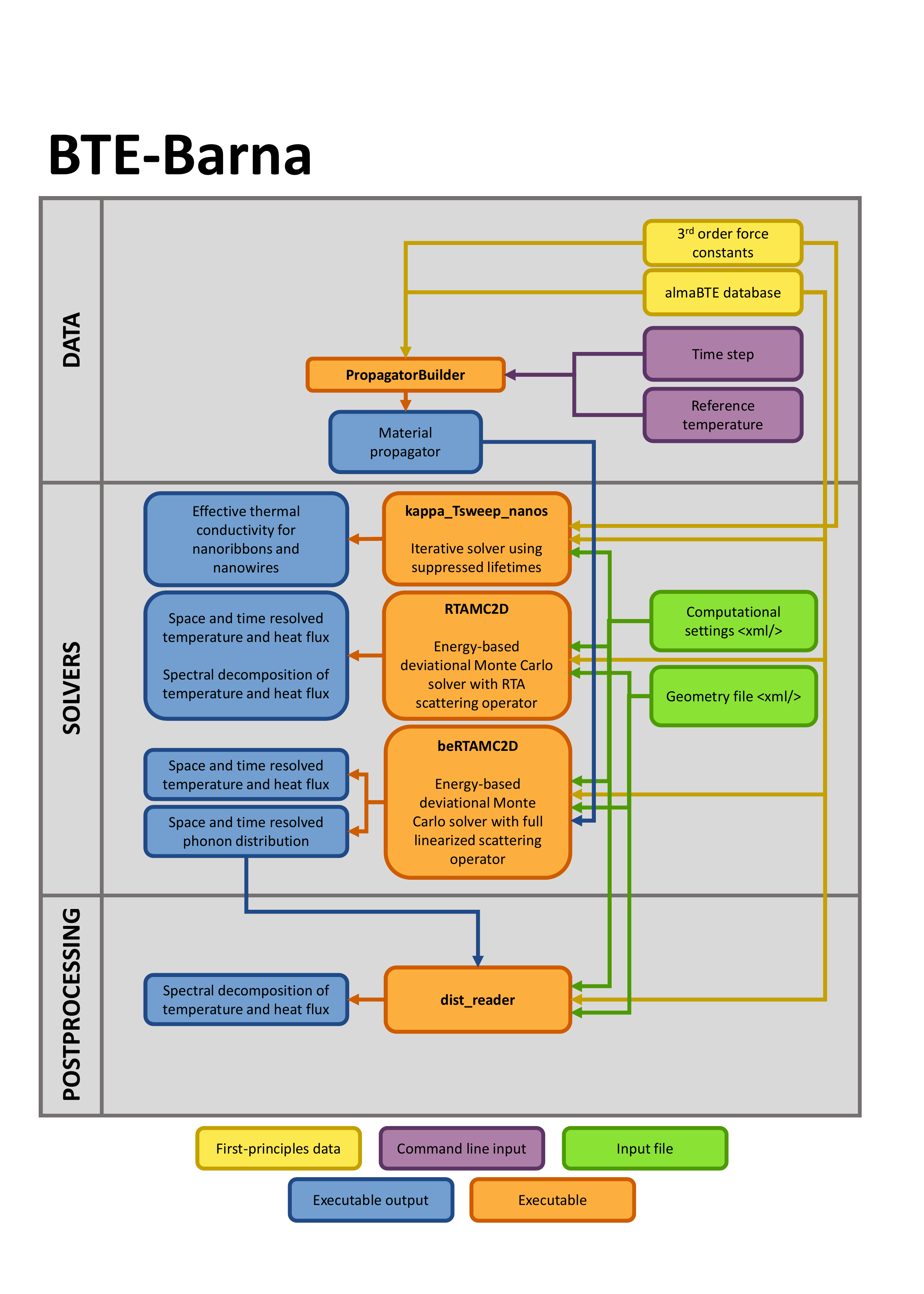}
	\caption{General structure of \texttt{BTE-Barna}  package.}
	\label{BCNBARNA}
\end{figure*}

Fig.~\ref{BCNBARNA} shows the different pieces of \texttt{BTE-Barna} package and how they relate to each other. The whole package heavily relies on \verb|almaBTE| library routines, which have been extended to allow for the solution of PBTE in finite devices based on 2D materials. Moreover, all the executables use the mode-resolved phonon properties as inputs, which are read from \verb|almaBTE|-generated HDF5 files.
A more detailed explanation of all the executables, their inputs, and outputs can be found in the ~\ref{inputNANOS} for the iterative solver, and in ~\ref{inputfiles} for the Monte Carlo solvers, their input generators, and post-processing tools.

\section{\label{sec:Methods}Methodology}

\subsection{\label{sec:Methods_nanoribbons}Effective thermal conductivity for nanoribbons}
Despite the fact that an exact solution of the PBTE for highly symmetric systems like nanoribbons or nanowires would require a discretization in space, it is possible to obtain an approximate solution by using averages under the assumption of fully dispersive boundaries~\cite{LiPRB2012}. In such a way, it becomes possible to obtain an effective thermal conductivity ($\kappa_\text{nano}$) by simply  introducing suppression factors in the lifetimes ($\tau_{\lambda}^\text{nano} = \tau^0_\lambda S_\lambda^\text{nano}$) and then solving the PBTE like in bulk under homogeneous gradients~\cite{ShengBTE}. 

For nanoribbons contained in the XY plane, the suppression factors ($S^\text{nr}_{\lambda}$) can be calculated by evaluating the integrals in Eq.~(8) of Ref.~\citenum{LiPRB2012} (see \ref{APPnanowires} for nanowires), obtaining:
\begin{gather}
	S^\text{nr}_{\lambda} = 1 + \left[ \frac{M^\text{nr}_{\lambda}}{L} \left(e^{-\frac{L}{M^\text{nr}_{\lambda}}}-1\right)\right] \\
	M^\text{nr}_{\lambda} = \left\lvert\left[\begin{pmatrix}
		u_y  & u_x  \\
		-u_x & u_y 
	\end{pmatrix}^{-1}{v_\lambda}\right]\cdot {e_1}\right\rvert \tau_{\lambda},
\end{gather}
where $u$ is a normalized vector pointing along the unbounded direction of the system, ${v_\lambda}$ and $\tau_{\lambda}$ are the velocity and bulk lifetime of $\lambda$-th mode, $L$ is the nanoribbon width and ${e_1}$ is the first column of identity matrix. Details on implementation, executable, inputs and outputs can be found in~\ref{inputNANOS}.

\subsection{\label{sec:Methods_RTA}RTA Monte Carlo}

The implementation for 2D materials of the RTA Monte Carlo is based on code already in \texttt{almaBTE}~\cite{almaBTE}, whose formulation was proposed by P\'eraud~\textit{et al.}~\cite{PeraudAPL2012}. The algorithm simulates the space and time evolution of deviational power (emitted by sources and absorbed by sinks) by splitting its distribution into discrete packets---the deviational particles---and tracking their trajectories in a linearized regime. The validity of the existing implementation rests on the assumption that differences in temperature are small enough that a single reference temperature can be defined for the whole system. We now take a look at the main improvements of our code upon that baseline; see~\ref{rtaalgo} for a detailed explanation of the whole algorithm.

The original implementation, \texttt{steady\_montecarlo1d}, was designed to investigate one-dimensional steady-state situations in materials/heterostructures embedded between two isothermal reservoirs having the same cross section, which owing to finite thickness might not be true for 2D-material-based systems/devices. Consequently, we extended the geometric algorithms to deal with 2D systems using the boost::geometry library~\cite{boostgeometry}. In this implementation the system is composed of different computational boxes, which are defined by the user as convex hulls of points, therefore enabling the creation of complex geometries. Additionally, as the finiteness of real 2D devices requires dealing with boundary scattering, we implemented a full diffusive condition in which the out-state is randomly selected from a Lambert cosine law distribution~\cite{SabattiJHFT2016} defined by transition probabilities from a state $i$ to $f$:
\begin{equation}
	\label{lambert}
	P_{i \rightarrow f} = \frac{{v}_f \cdot \hat{{e}}_{\perp}\delta(\omega_f-\omega_i)}{\sum_j {v}_j \cdot \hat{{e}}_{\perp}\delta(\omega_j-\omega_i)}
\end{equation}
where ${v}_i$ and $\omega_i$ are the group velocity and frequency of the $i-\mathrm{th}$ phonon mode, $\hat{{e}}_{\perp}$ is the wall normal vector pointing inwards the material and $\delta(\omega_j-\omega_i)$ is regularized using adaptive smearing~\cite{ShengBTE}. Besides the steady-state, the code allows for the exploration of time evolution determined by boundary conditions. This is done by sampling the trajectories on a time grid on top of the spatial grid, in such a way that the contribution from a trajectory path inside a computational box to deviational energy ($e^d$) and heat flux (${j}$) grid point is:  

\begin{gather}
	\label{edrta}
	e^d(k,i) = \frac{1}{V_{k}}\sigma\varepsilon_d (t_f-t_0)\left[\Theta(t-t_i)-\Theta(t-t_{i+1})\right]
	\\
	\label{jrta}
	{j}(k,i) = \frac{1}{V_{k}}{v}\sigma\varepsilon_d (t_f-t_0)\left[\Theta(t-t_i)-\Theta(t-t_{i+1})\right]
\end{gather}
where $k$ is the computational box id, $V_{k}$ is the volume of the $k$-th box, $\sigma$ is the particle sign, $\varepsilon_d$ is the deviational power carried per particle, ${v}$ is the particle velocity and $(t_f-t_0)\left[\Theta(t-t_i)-\Theta(t-t_{i+1})\right]$ represents the interval between $t_0$ and $t_f$ that belongs to the $i$-th point of the time grid, with $\Theta$ representing the Heaviside function. Such contribution is computed each time a particle changes its state or its spatial bin. Finally, by integrating Eqs.~(\ref{edrta})-(\ref{jrta}) one can obtain the evolution of temperature $T(k,t_i) $ and heat flux ${J}(k,t_i)$:
\begin{gather}
	T(k,t_i) = T_\text{ref} + \frac{1}{C_v(T_\text{ref})}\sum_{l=0}^{i}e^d(k,l) \\
	{J}(k,t_i) = \sum_{l=0}^{i}{j}(k,i)
\end{gather}
where $T_\text{ref}$ is the reference temperature and $C_v(T_\text{ref})$ is the volumetric heat capacity at $T_\text{ref}$.

\subsubsection{A note on extended systems with applied gradients\label{RandrianalisoaAlgo}}
For extended systems where no isothermal boundaries are present, e.g.\ an infinitely long nanoribbon with an applied thermal gradient, particles cannot exit the structure. It is therefore necessary to introduce an alternative way to collect them.
For this purpose, and taking into account that transient to steady-state is normally not of interest for those cases, we implemented a modified version of the algorithm originally proposed by Randrianalisoa \textit{et al.}~\cite{RandrianalisoaJAP2008,RandrianalisoaJHT2008} and adapted to such cases by P\'eraud \textit{et al.}~\cite{PeraudARHT2014}. We now list our modifications to the algorithm, with respect to the finite systems described in~\ref{rtaalgo}, for dealing with extended systems under applied gradients:
\begin{enumerate}
	\item\label{Randrianalisoa-1} Generate particles from a gradient (source) generator, evolve them until they scatter (intrinsically or at borders or interfaces) and compute their contribution to steady-state.
	\item\label{Randrianalisoa-2} Calculate the net number of particles that intrinsically scatter $N_j = \sum_{i}^{N_{intrinsic}}\sigma[r\in j]$ at each computational box and delete those particles.
	\item\label{Randrianalisoa-3} Generate $|N_j|$ particles with sign $\sigma=\mathrm{sgn}(N_j)$ from the postscattering distribution $\frac{C_{k'}(j)/\tau_{k'}(j)}{\sum_i C_i(j)/\tau_k(j)}$.
	\item\label{Randrianalisoa-4} If the number of particles is 0, no new particles are introduced or the properties have converged, end the simulation; otherwise repeat step \ref{Randrianalisoa-1}.
\end{enumerate}
It is noteworthy that the cancellation scheme in step~\ref{Randrianalisoa-2} introduces a non-negligible error of second order with respect to the space mesh~\cite{PeraudARHT2014}. 

\subsubsection{Interface model for stacked layered systems: localized diffuse mismatch model\label{sandwichDMM}}
\begin{figure}
	\centering
	\includegraphics[width=\linewidth]{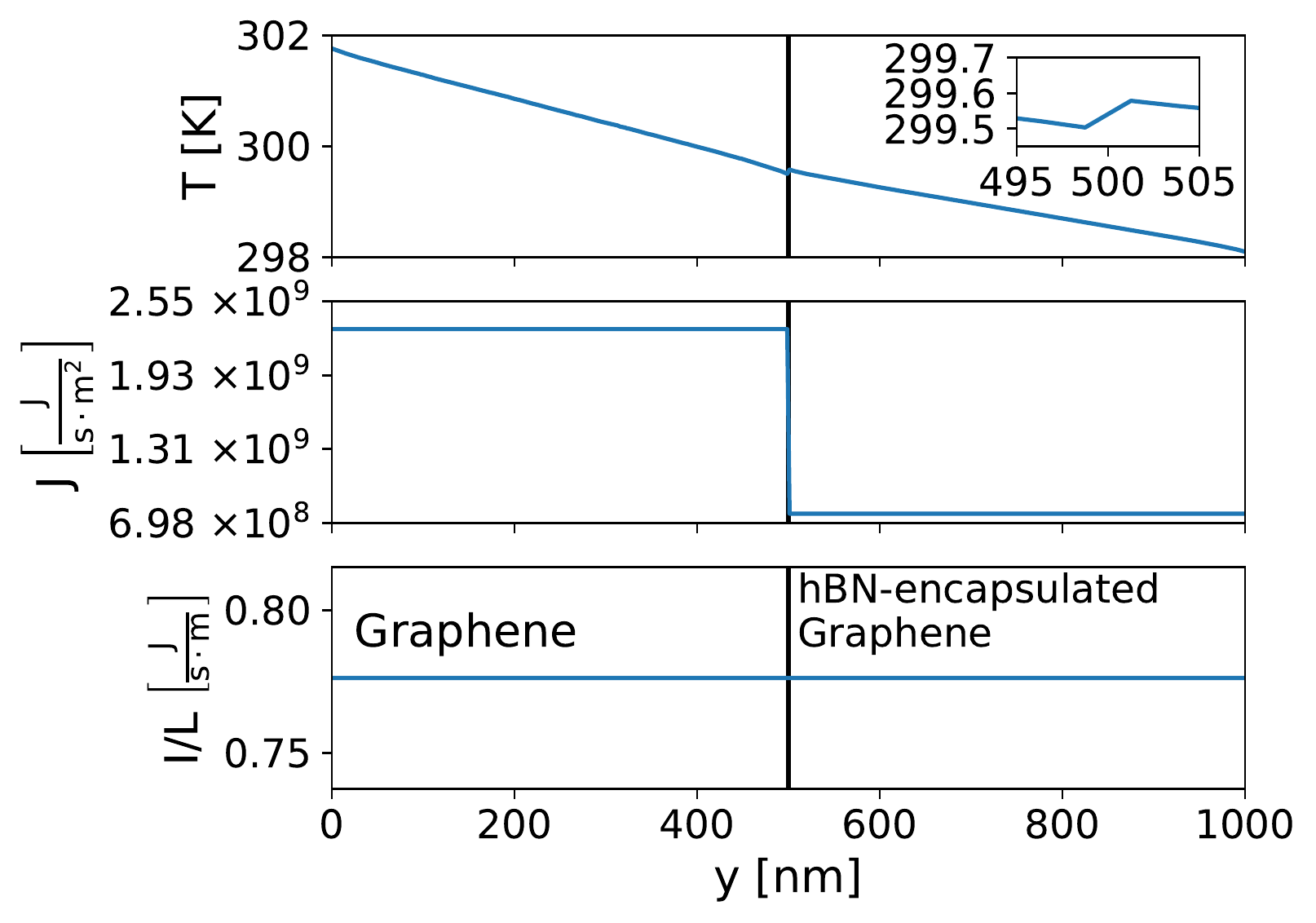}
	\caption{Thermal profile, flux and heat intensity per unit of length in graphene/hBN-encapsulated graphene obtained using the RTA and the traditional DMM to model interface scattering. Inset: Zoom of thermal profile at the interface.}
	\label{dmm}
\end{figure}

\begin{figure}
	\centering
	\includegraphics[width=\linewidth]{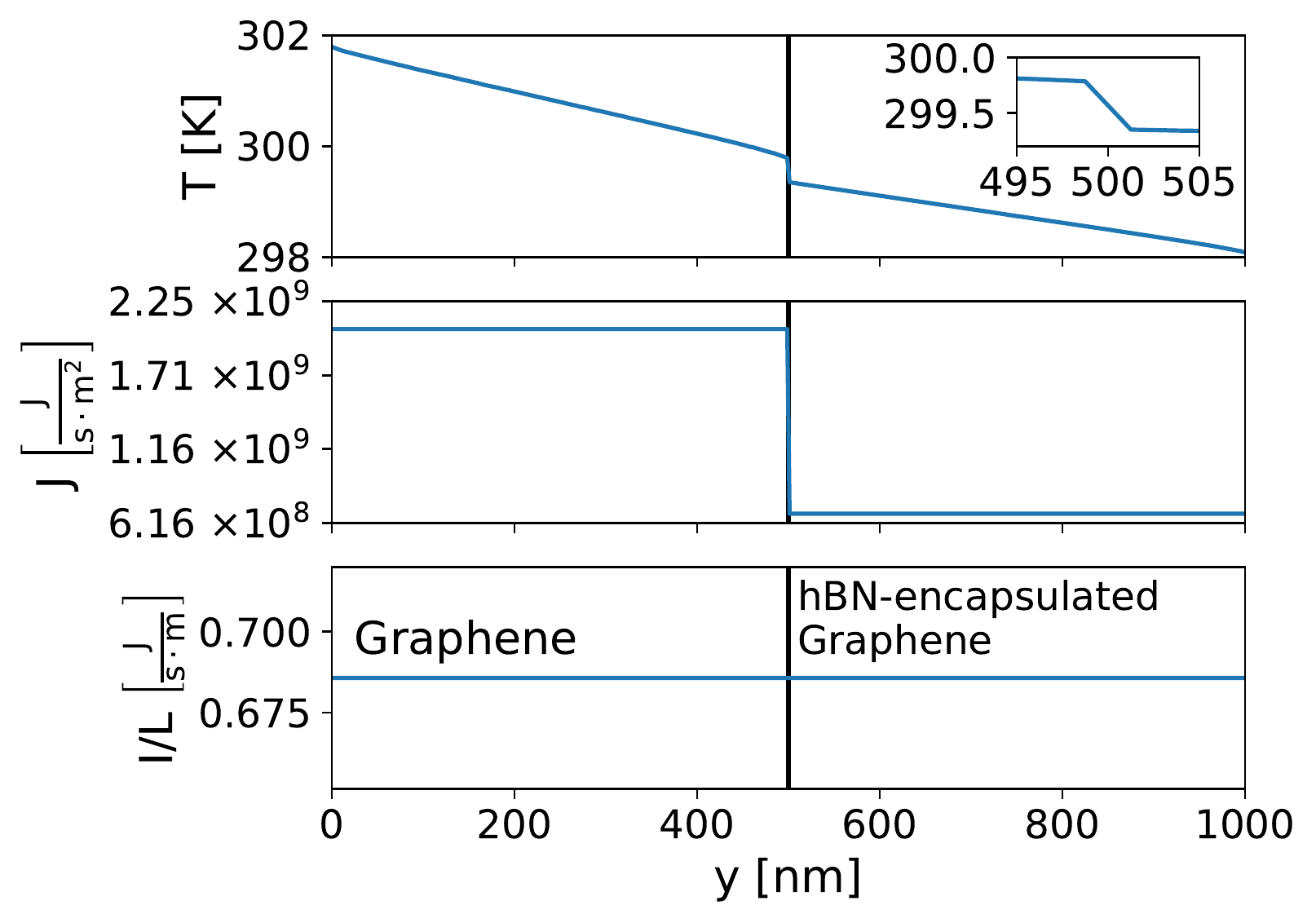}
	\caption{Thermal profile, flux and heat intensity per unit of length in graphene/hBN-encapsulated graphene obtained using the RTA and the LDMM to model interface scattering. Inset: Zoom of thermal profile at the interface}
	\label{ldmm}
\end{figure}

The traditional diffuse mismatch model (DMM) implemented for the treatment of interface scattering in \texttt{steady\_montecarlo1d} is a purely elastic model, allowing the coupling between modes at each side of the interface with energy conservation as the sole requirement. Despite this crude approach, the DMM has been proven to qualitatively describe the interface thermal resistance (ITR) for several interfaces of bulk 3D materials such as Si/Ge~\cite{almaBTE}. However, for systems comprised of stacked layers, such as the interface between graphene and encapsulated graphene~\cite{WenJAP2014}, energy matching as the single condition for transmission is no longer a valid assumption. Under such conditions, well localized modes in unconnected layers---e.g.: encasing hBN layers vs. the bare monolayer at the graphene/encapsulated-graphene interface---would be predicted to be coupled, which is quite unrealistic and would lead to too low (or even negligible or negative) ITR values (see Fig.~\ref{dmm})~\cite{WenJAP2014}. Consequently, we developed the localized DMM (LDMM) to account for mode localization at layers.
To that end, we define the localization vector at the $I$-th layer for a given mode ($\mathscr{L}_{I,\lambda}^{A}$) as:
\begin{equation}
	\mathscr{L}_{I,\lambda}^{A} = \frac{\sum_{j}^{N}\sum_{\alpha}^{x,y}|{\zeta}_{\lambda,\alpha,j}|^2~[j \in I]}{\sum_{j}^{N}\sum_{\alpha}^{x,y}|{\zeta}_{\lambda,\alpha,j}|^2}
\end{equation}
where $I$ is the layer index of $A$ side, $\lambda$ is the phonon mode index, $j$ is the atomic index, $\alpha$ is the cartesian axis and ${\zeta}_{\lambda,\alpha,j}$ is the eigenvector. Therefore, when calculating the coupling strength we multiply the classical DMM expression by the coupling factor $\mathscr{C}_{\lambda,\lambda'}^{A,B}$:
\begin{equation}
	\mathscr{C}_{\lambda,\lambda'}^{A,B} = 1 - \mathrm{JSD}({\mathscr{L}}_{\lambda}^{A}||{\mathscr{L}}_{\lambda'}^{B})
\end{equation}
where $\mathrm{JSD}({\mathscr{L}}_{\lambda}^{A}||{\mathscr{L}}_{\lambda'}^{B})\in\left[0,1\right]$ is the Jensen-Shannon divergence between localization functions. Therefore, $\mathscr{C}_{\lambda,\lambda'}^{A,B}$ is one for perfectly matching localization and zero for modes fully localized at different layers. As the Jensen-Shannon divergence requires both vectors to be of same length, for different materials the localization vectors are modified in such a way that connected layers remain untouched and paired while the remaining unconnected layers have their values summed up and added at the back of the localization vector; we provide an example for reference:
\begin{equation}
	{\mathscr{L}}_{\lambda}^C =  \left(
	\begin{matrix}
			\mathscr{L}_{\lambda,i}^C         \\
			\mathscr{L}_{\lambda,j}^C         \\
			\sum_{k}\mathscr{L}_{\lambda,k}^C \\
		\end{matrix}
	\right),~~~
	{\mathscr{L}}_{\lambda'}^D =  \left(
	\begin{matrix}
			\mathscr{L}_{\lambda',\alpha}^D      \\
			\mathscr{L}_{\lambda',\beta}^D       \\
			\sum_{\mu}\mathscr{L}_{\lambda',k}^D \\
		\end{matrix}
	\right)
\end{equation}
In this example, layers $i$ and $j$ of $C$ are connected to the $\alpha$ and $\beta$ layers on the $D$ side, and $k$ and $\mu$ denote all unconnected layers on each side or a zero term if no additional layers exist. In Fig~\ref{ldmm} we show results for the same case studied in Fig.~\ref{dmm} but using the LDMM, enabling us to obtain a non-negligible ITR as expected in this kind of system~\cite{WenJAP2014}. It should be noted that, despite the improvement in the results, the model is still incapable of describing the thermal rectification predicted in such systems~\cite{WenJAP2014,ChenACSI2020} because of the intrinsic symmetry of the DMM and the Jensen-Shannon divergence used to model the interface scattering.

\subsection{\label{sec:Methods_beyondRTA}Beyond RTA Monte Carlo}

Given the RTA's failure to describe the thermal properties of 2DMs~\cite{CepellottiNatCom2015,LindsayJAP2019}, we implemented in \verb|beRTAMC2D| (see ~\ref{app:progbRTA}) a linearized ab initio phonon-low variance deviational simulation Monte Carlo (LAIP-LVDSMC)~\cite{LandonJAP2014} simulator to overcome such limitation. This LAIP-LVDSMC algorithm, henceforth referred to as {\it beyond RTA} (bRTA), solves the deviational energy linearized PBTE:
\begin{equation}
	\label{PBTE}
	\frac{\partial n^d_i}{\partial t} + {v_i} \cdot {\nabla_r} n^d_i + \frac{\partial n^0_i}{\partial T} {v_i} \cdot {\nabla_r} T = \sum_{j} B_{ij} n^d_j
\end{equation}
where $n^d_i$ is the deviational energy distribution, $n^0_i$ is $\hbar\omega_if^0_i = \hbar\omega_i/(\mathrm{exp}(\hbar\omega_i/k_BT) - 1)$ and $B_{ij}$ is the linearized scattering operator comprising three-phonon and isotopic (mass disorder) scattering terms:
\begin{multline}
	\label{bmat}
	B_{ij} = \sum_{sml}P^{3ph}_{s+m\rightarrow l}\left[\frac{\omega_s}{\omega_m}(f^0_l-f^0_s)\delta_{is}\delta_{jm}\right. + \\ \left. \frac{\omega_s}{\omega_l}(f^0_s + f^0_m + 1)\delta_{is}\delta_{jl}+(f^0_l-f^0_m)\delta_{is}\delta_{ji}\right] \\
	+ \frac{1}{2} \sum_{sml} P^{3ph}_{s\rightarrow n+l} \left[ \frac{\omega_s}{\omega_m}(f^0_l - f^0_s)\delta_{is}\delta_{jm} + \frac{\omega_s}{\omega_l}(f^0_m - f^0_s)\delta_{is}\delta_{jl} \right. \\ \left. - (f^0_l + f^0_m + 1)\delta_{is}\delta_{ji}\right] + \\ \sum_{sm} P^{iso}_{s\rightarrow m}\left[\frac{\omega_s}{\omega_m}\delta_{is}\delta_{jm} - \delta_{is}\delta_{ji}\right]
\end{multline}
where $P^{3ph}_{s+m\rightarrow l}$ and $P^{3ph}_{s\rightarrow m + l}$ are the intrinsic transition probabilities for three-phonon absorption and emission processes, derived from perturbation theory~\cite{ZimanEPH,ShengBTE} and $P^{iso}_{s\rightarrow m}$ is the intrinsic isotopic transition calculated using Tamura's model~\cite{TamuraPRB1983}. The latter was not included in Landon's original algorithm. The PBTE is then solved using a Monte Carlo scheme with each step split in two parts: advection, in which we introduce, evolve and scatter particles at boundaries (see \ref{app:bRTA} for more details) and scattering within the material, in which the distribution evolves according to
\begin{equation}
	\label{propagator}
	{n}(t+\Delta t) = P(\Delta t) {n}(t) = e^{B\Delta t} {n}(t)
\end{equation}
where the propagator $P(\Delta t)$ is a non-Markovian transition matrix, as it has negative or higher than unity elements~\cite{LandonThesis}, which makes its direct implementation difficult. A more implementation-friendly form can be obtained by using the power series:
\begin{equation}
	\label{powerseries}
	n_i(t+\Delta t) = \sum_j \frac{P_{ij}(\Delta t)}{\mathscr{P}_j}\left(\sum_{n=0}^{\infty}\left(2\frac{\mathscr{P}_j^-}{\mathscr{P}_j}\right)^{n}\right)n^d_j(t)
\end{equation}
where $\mathscr{P}_j = \sum_{k}^{N_\text{states}}|P_{kj}|$ and $\mathscr{P}_j^- = \sum_{k}^{N_\text{states}}|P_{kj}|[P_{kj}<0]$. The recursive Eq.~(\ref{powerseries}) can be implemented stochastically for a particle in state $j$ with sign $\sigma$ through the following strategy:
\begin{enumerate}
	\item Sample a random number $R$ in $[0,1)$, and find the lower bound $f$ of $\sum_k |P_{kj}(\Delta t)|/\mathscr{P}_j$ for $R$.
	\item Set the new sign to $\sigma' = \mathrm{sgn}(P_{fj}\sigma)$.
	\item If $\sigma'\neq \sigma$, generate two particles with state $j$ and time $t$.
\end{enumerate}
Finally, the deviational energy density $\rho^d$ and the heat flux ${J}$ are obtained from the distribution in the $i$-th computational box as:
\begin{align}
	\rho^d(i) = \frac{\sum_{j}^{N_\text{particles}} \varepsilon_d \sigma_j [{r_j}\in \text{$i$-th box}]}{V_{i,box}} \\
	{J}(i) = \frac{\sum_{j}^{N_\text{particles}} {v_j}\varepsilon_d \sigma_j [{r_j}\in \text{$i$-th box}]}{V_{i,box}},
\end{align}
where $\varepsilon_d$ is the deviational energy per particle and ${r_j}$ is the position of the $j$-th particle. 

\subsubsection{\label{Bcalculation}$B_{ij}$ calculation and enforcement of conservation laws}
As noted by Landon \textit{et al.}, it is important for $B_{ij}$ to respect crystal symmetries and microscopic reversibility~\cite{LandonJAP2014}. The original \texttt{almaBTE} routines for calculating scattering amplitudes lead to vialoations of those constraints because the smearing method implemented there does not enforce the symmetry between emission and absorption processes~\cite{ShengBTE,broadening}.

To enforce symmetry, matrix elements are built from a single representative of each equivalence class in the quotient group of $q$ points using a new symmetric adaptive smearing scheme for energy conservation
\begin{gather}
	\sigma_{ijk} = a \sqrt{\sigma_i^2 + \sigma_j^2  + \sigma_k^2} \\
	\sigma_i = \frac{1}{\sqrt{12}}\lVert\{G_{\mu\alpha}^T\cdot  N_{\mu\mu}^{-1}\}^T \cdot ({v}_i)_{\alpha}\rVert
\end{gather}
where $i$, $j$ and $k$ are the phonon modes taking part in the three-phonon process, $\mu$ indicates a reciprocal-space lattice vector, $\alpha$ indicates a Cartesian axis, $G_{\mu\alpha}$ is the reciprocal lattice basis matrix, $N_{\mu\mu}$ is a diagonal matrix whose elements are the size of the $q$-point grid, and $a$ is a broadening factor. The theoretically optimal value of $a$ is $1$, but it can often be decreased with significant gains in performance and little degradation in accuracy.

Next, those matrix elements are expanded using crystal symmetry and microscopic time reversibility and averaged to eliminate possible asymmetries.  This lookup, together with matrix building, is parallelized via MPI and OpenTBB, with stable summations following Neumaier's algorithm for matrix collapse and gathering~\cite{NeumaierZAMM1974}.

Finally, it must be noted that $B_{ij}$ requires a rather strict conservation of energy, which is violated by the broadening scheme. Therefore, we add a correction extracted from a Lagrange-multiplier approach~\cite{LandonJAP2014} to our matrix to enforce it. On top of that, Landon \textit{et al.} also discussed the necessity of including a momentum correction to make normal processes conserve the momentum. However, we found that including such correction was unnecessary and, in fact, results in spurious effects such as nonnegligible fluxes in directions perpendicular to the thermal gradient for homogeneous bulk systems.

\subsubsection{\label{Pcalculation}Efficient propagator calculation}

The scattering algorithm requires the explicit calculation of the propagator matrix $P(\Delta t) = e^{B\Delta t}$ [see Eq.~(\ref{powerseries})]. Although the matrix exponential is a well defined mathematical operation given by:
\begin{equation}
	e^{ B\Delta t} = \sum_{n=0}^{\infty}\frac{\Delta t^n}{n!}B^n,
\end{equation}
its practical computation is cumbersome and still a topic under active research, with lots of methods available~\cite{ExponentialMatrixMethods}. One of the most common approaches to computing $e^A$ is the scaling-and-squaring method~\cite{NicholasSIAMRev2009,eigenEXP}, also chosen by Landon in his original work~\cite{LandonThesis}. The method is based on squaring the matrix to reduce its norm, then computing the exponential using a Pad\'e approximant and undoing the squaring, with an overall computational cost of at best $20N^3$ operations for dense matrices of size $N$~\cite{eigenEXP}. For reference, the $B$-matrix of the prototypical 2DM, graphene, contains approximately $1.5\times10^{9}$ elements when a $80\times80\times1$ grid is used, so the scaling and squaring method is not suited for our problem. 

In contrast, Krylov subspace methods are especially suited for big matrices, where the action of the exponential matrix ($e^A$) on vector (${b}$) can be approximated using much more smaller matrices. The Krylov subspace ($\mathscr{K}_n(A,{b})$) of order $n$ is a vector subspace spanned by $\{{b},A{b},A^2{b},\ldots,A^{n-1}{b}\}$, an orthonormal basis ($S_n$) of which can be build via Arnoldi iteration~\cite{ArnoldiQAM1951}. The problem can be then recast in terms of $\mathscr{K}_n(A,{b})$ as~\cite{ExponentialMatrixMethods,WasshuberThesis}:
\begin{equation}
	\label{krylovEXP}
	e^A{b} \approx \lVert {b} \rVert S_n e^{H_n}{e}_1
\end{equation}
where $H_n$ is the projection of $A$ on the basis $S_n$ (of size $n\times n$) and ${e}_1$ is the first column of the identity matrix. 
The Krylov subspace size and therefore the dimensions of $H_n$ may be truncated down to a desired precision via the error bound $\lVert e^A{b} - \lVert {b} \rVert S_n e^{H_n} {e_1} \rVert \leq 2\lVert {b} \rVert\frac{\lVert A \rVert^n e^{\lVert A \rVert}}{n!}$~\cite{SaadSIAMJNA}. In fact, small values of $n$ tend to give good approximations and enable a calculation of the small $n\times n$ sized $e^{H_n}$-matrix efficiently through the scaling and squaring method.
Despite its efficiency and suitability for our case, the Krylov subspace method is limited to the calculation of arbitrary matrix-vector products $e^A{b}$, not of $e^A$ itself. Nevertheless, one can easily recover each column of $e^A$ by using canonical basis vectors as ${b}$-vectors. This way of calculating $e^A$ has the added advantage of being straightforward to parallelize, as each column can be calculated independently.

\subsubsection{Linear interpolation of the propagator for systems with multiple reference temperatures}

From Eq.~(\ref{bmat}) it becomes clear that different reference temperatures would require different propagators. This is not problematic per se, but the fact that each propagator occupies a big amount of RAM can be a problem for simulations with variable reference temperatures~\cite{PeraudARHT2014}. To relieve the memory burden for such simulations we use on-the-fly linear interpolation of $P(\Delta t)$ between pairs of temperatures, thus requiring memory storage only for a few reference propagators. Linear interpolation was chosen because it ensures energy conservation at the interpolated temperatures. We tested the performance of these linear interpolants against the corresponding exact propagators by calculating the error per element between the phosphorene propagator at \SI{305}{\kelvin} and the interpolated result using \SI{300}{\kelvin} and \SI{310}{\kelvin} as knots. We also did the same for \SI{320}{\kelvin} using \SI{300}{\kelvin} and \SI{340}{\kelvin} as knots. In both cases, we obtained an error per element in the order of $10^{-6}$.

\section{\label{sec:validation}Code Validation}

As previously mentioned, 2DMs are being extensively studied as possible substitutes of silicon in MOSFET~\cite{Wang2012NatNano,Schwierz2010NatNano,Chen2019NanoLet}. Amid all candidates to succeed silicon, the monolayer, also known as phosphorene, and few-layer black phosphorous (bP) have attracted lots of attention due to its electronic properties, such as its high mobility when compared to other candidates like transition metal dichalcogenides~\cite{HaratipourACSNano2016}. Indeed, is it possible to find several examples of fully functional MOSFETS based on few-layer bP~\cite{LuoEDL2014,Li2014NatNano,HaratipourACSNano2016}. The work of Wu \textit{et al.} is of interest, presenting high-performance MOSFETs with reconfigurable polarities \cite{Wu2021NatElec}. Furthermore, phosphorene has been proposed to be an important actor in the survival of Moore's law down to atomic sizes~\cite{IlatikhamenehSR2016} thus increasing the importance of controlling heat transport for phosphorene at the device level.

Therefore, in this section we present phosphorene-based test cases. To that end, we have used first-principles data---i.e.: atomic positions and interatomic force constants of second and third order---of Ref.~\citenum{SmithAM2017} to obtain phonon properties (frequencies, eigenvectors, lifetimes, group velocities, etc.) alongside with the propagator. Second-order interatomic force constants were renormalized to enforce crystal symmetry, translational invariance and rotational invariance necessary for a proper description of quadratic acoustic bands~\cite{CarreteMRL2016}, the broadening parameter was fixed to $1$ for energy conservation and the layer thickness was set to \SI{0.533}{\nano\meter}~\cite{CastellanosGomez2DMat2014}. The phonon properties and the propagator were calculated on a $\Gamma$-centered $q$-mesh of $50\times50\times1$ points, for which thermal bulk conductivity is found to be converged---with less than a 5\% change with respect to a higher quality mesh of $100\times100\times1$ points---at \SI{300}{\kelvin}. The propagator for the bRTA calculations was calculated using a time step of \SI{0.25}{\pico\second}.

\subsection{RTA code validation}
\begin{figure}
	\centering
	\includegraphics[width=\linewidth]{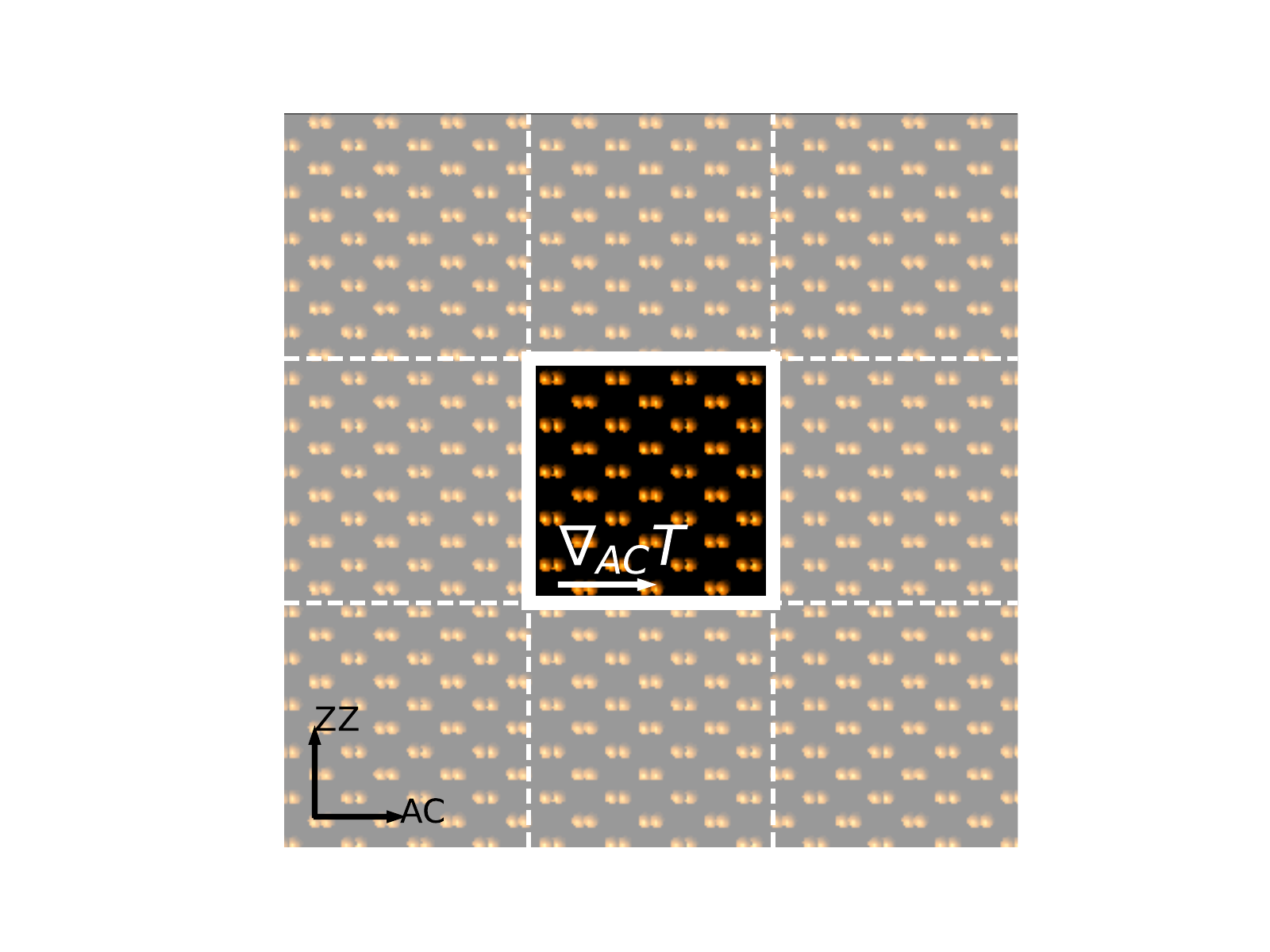}
	\caption{Sketch of the simulation setup for perfect phosphorene with a  thermal gradient applied in the AC direction. Replicas illustrating the periodic boundary conditions are depicted as off-color boxes.}
	\label{Homogeneous}
\end{figure}

To validate the RTA code, we simulated an infinitely large piece of phosphorene with an applied thermal gradient represented as a source generator (see Sec.~\ref{RandrianalisoaAlgo} and \ref{rtaalgo}) as depicted in Fig.~\ref{Homogeneous}.

\begin{table}
	\centering
	\caption{\label{table-kappasRTA} Calculated $\kappa$ for phosphorene at \SI{300}{\kelvin} along the armchair (AC) and zigzag (ZZ) directions using the RTA.}
	\begin{threeparttable}
		\begin{tabular}{@{}llll@{}}
			\toprule
			\hline \hline                                                                                                  \\
			                                                 & \makecell[tc]{AC                                            \\ $\left[\mathrm{\frac{W}{m\cdot K}}\right]$} & \makecell[tc]{ZZ \\ $\left[\mathrm{\frac{W}{m\cdot K}}\right]$} \\ \\ \midrule
			\hline                                                                                                         \\
			\makecell[tl]{$\kappa_\mathrm{almaBTE}$}         & \makecell[tc]{20.7}         & \makecell[tc]{57.8}           \\
			\\
			\makecell[tl]{$\kappa^\mathrm{MC}_\mathrm{RTA}$} & \makecell[tc]{20.3$\pm$0.1} & \makecell[tc]{56.2$\pm$0.2}   \\
			\\
			                                                 &                             &                             & \\ \bottomrule
			\hline \hline
		\end{tabular}
	\end{threeparttable}
\end{table}

$\kappa^\mathrm{MC}_\mathrm{AC,RTA}$ and $\kappa^\mathrm{MC}_\mathrm{ZZ,RTA}$ were calculated via Fourier's law from fluxes ($J_h = -\kappa_l\nabla_rT$). The results are quite close to the ones obtained using \texttt{almaBTE}'s bulk thermal conductivity calculator, \texttt{kappa\_Tsweep} (see Table~\ref{table-kappasRTA}); with the differences being less than the typical experimental error of $5\%$ for thermal conductivities~\cite{ZhaoJEP2016}.

\begin{figure}
	\centering
	\includegraphics[width=\linewidth]{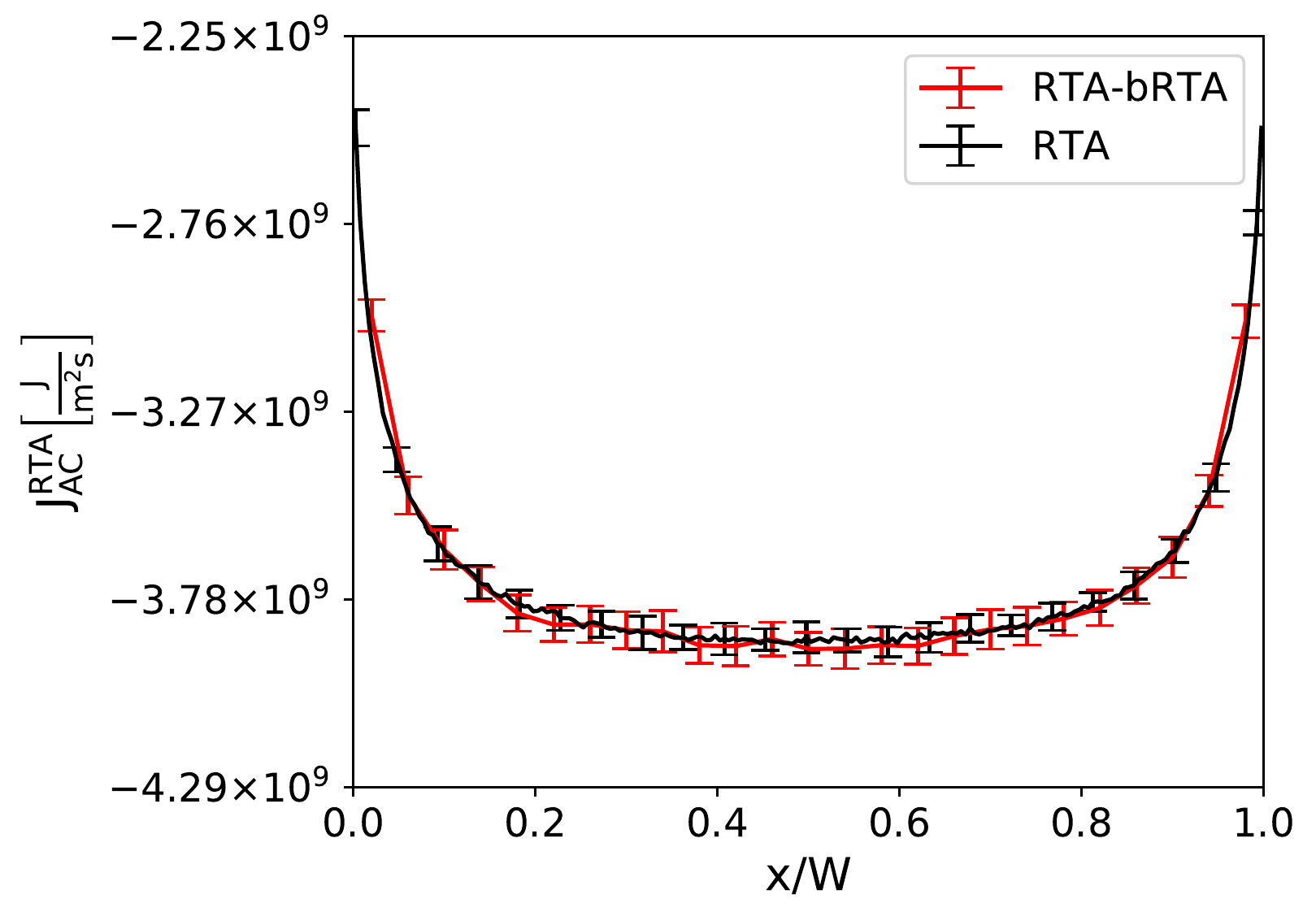}
	\caption{Comparison of RTA (black) and RTA-bRTA (red) heat flux in AC direction as function of normalized position for a phosphorene nanoribbon of \SI{400}{\nano\meter} of width with $\nabla_\mathrm{AC}T = \SI{0.2}{\kelvin\per\nano\meter}$.}
	\label{RTAtest2}
\end{figure}

We conducted an additional test of this RTA algorithm by comparing it to an RTA version of bRTA (see \ref{RTALVDSMC}). To do so we simulated an infinite nanoribbon (NR) in the AC direction with an applied gradient of \SI{0.2}{\kelvin\per\nano\meter} along the NR. The heat flux profiles from both methods, plotted in Fig.~\ref{RTAtest2}, are in excellent agreement. 

\subsection{Beyond RTA: $B$-matrix validation}
To validate our $B_{ij}$ construction algorithm we used the resulting matrix to obtain the lattice thermal conductivity ($\kappa_l$) by iteratively solving the linear system:
\begin{equation}
	\left(\frac{\partial n^0_i}{\partial T}{v_i} \cdot {\nabla_r} T\right)_i = \sum_{j} B_{ij} n^d_j
\end{equation}
using the RTA solution ($n^{d,RTA}_j = \frac{1}{B_{jj}}\frac{\partial n^0_j}{\partial T}{v}_j \cdot {\nabla_r} T$) as an initial guess, and we then compared the results against \texttt{almaBTE}'s \texttt{kappa\_Tsweep} for the case of phosphorene (see Table~\ref{table-kappas}). The agreement between both methods and the fact that they are in line with other theoretical calculations provide support to our methodology.

\begin{table}
	\centering
	\caption{\label{table-kappas} Calculated $\kappa$ for phosphorene at 300 K for armchair (AC) and zigzag (ZZ) directions. Other theoretical results are provided for comparison.}
	\begin{threeparttable}
		\begin{tabular}{@{}llll@{}}
			\toprule
			\hline \hline                                                                                        \\
			                                                   & \makecell[tc]{AC                                \\ $\left[\mathrm{\frac{W}{m\cdot K}}\right]$} & \makecell[tc]{ZZ \\ $\left[\mathrm{\frac{W}{m\cdot K}}\right]$} \\ \\ \midrule
			\hline                                                                                               \\
			\makecell[tl]{$\kappa_\mathrm{almaBTE}$}           & \makecell[tc]{27.501} & \makecell[tc]{82.878}   \\
			\\
			\makecell[tl]{$\kappa_\mathrm{B}$}                 & \makecell[tc]{27.499} & \makecell[tc]{82.860}   \\
			\\
			\makecell[tl]{Ref.~\citenum{ZhuPRB2014}$^\dagger$} & \makecell[tc]{23.9}   & \makecell[tc]{82.1}     \\
			\\
			\makecell[tl]{Ref.~\citenum{JainSR2015}$^\dagger$} & \makecell[tc]{35.5}   & \makecell[tc]{108.3}    \\
			\\
			\makecell[tl]{Ref.~\citenum{SmithAM2017}}          & \makecell[tc]{22.0}   & \makecell[tc]{63.2}     \\
			\\
			                                                   &                       &                       & \\ \bottomrule
			\hline \hline
		\end{tabular}
		\begin{tablenotes}
			\small
			\item $^\dagger$ These results are rescaled to take into account differences in assumed thickness. 
		\end{tablenotes}
	\end{threeparttable}
\end{table}

\subsection{Beyond RTA: Propagator and bRTA validation}
To validate $P(\Delta t)$ together with the rest of the bRTA implementation, we simulated an infinitely large piece of phosphorene with an applied thermal gradient represented as a source generator [see Eq.~(\ref{gradGen})], as shown in Fig.~\ref{Homogeneous}.

\begin{figure}
	\begin{subfigure}[b]{0.45\textwidth}
		\centering
		\includegraphics[width=\textwidth]{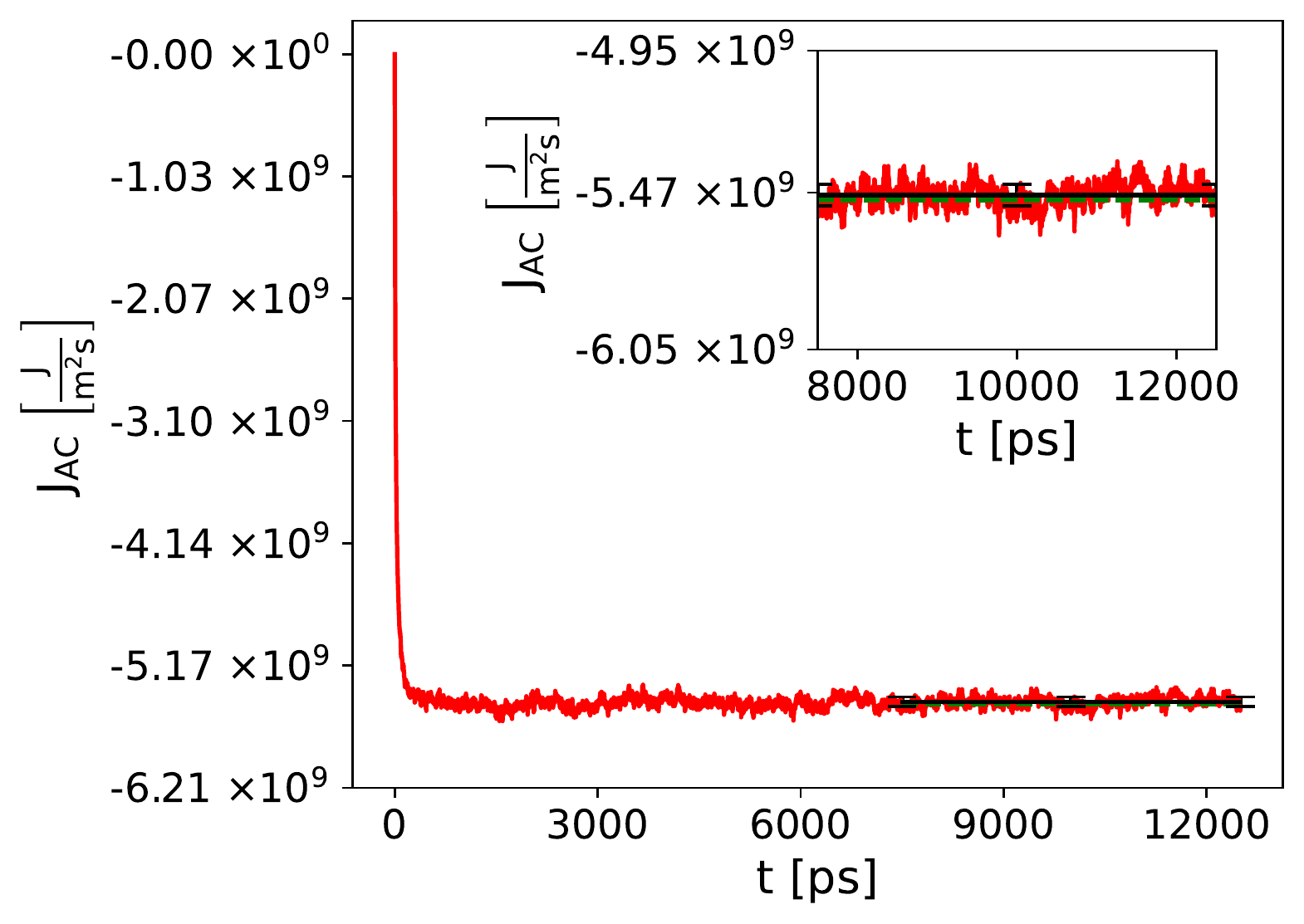}
		\label{homoX}
	\end{subfigure}
	\begin{subfigure}[b]{0.45\textwidth}
		\centering
		\includegraphics[width=\textwidth]{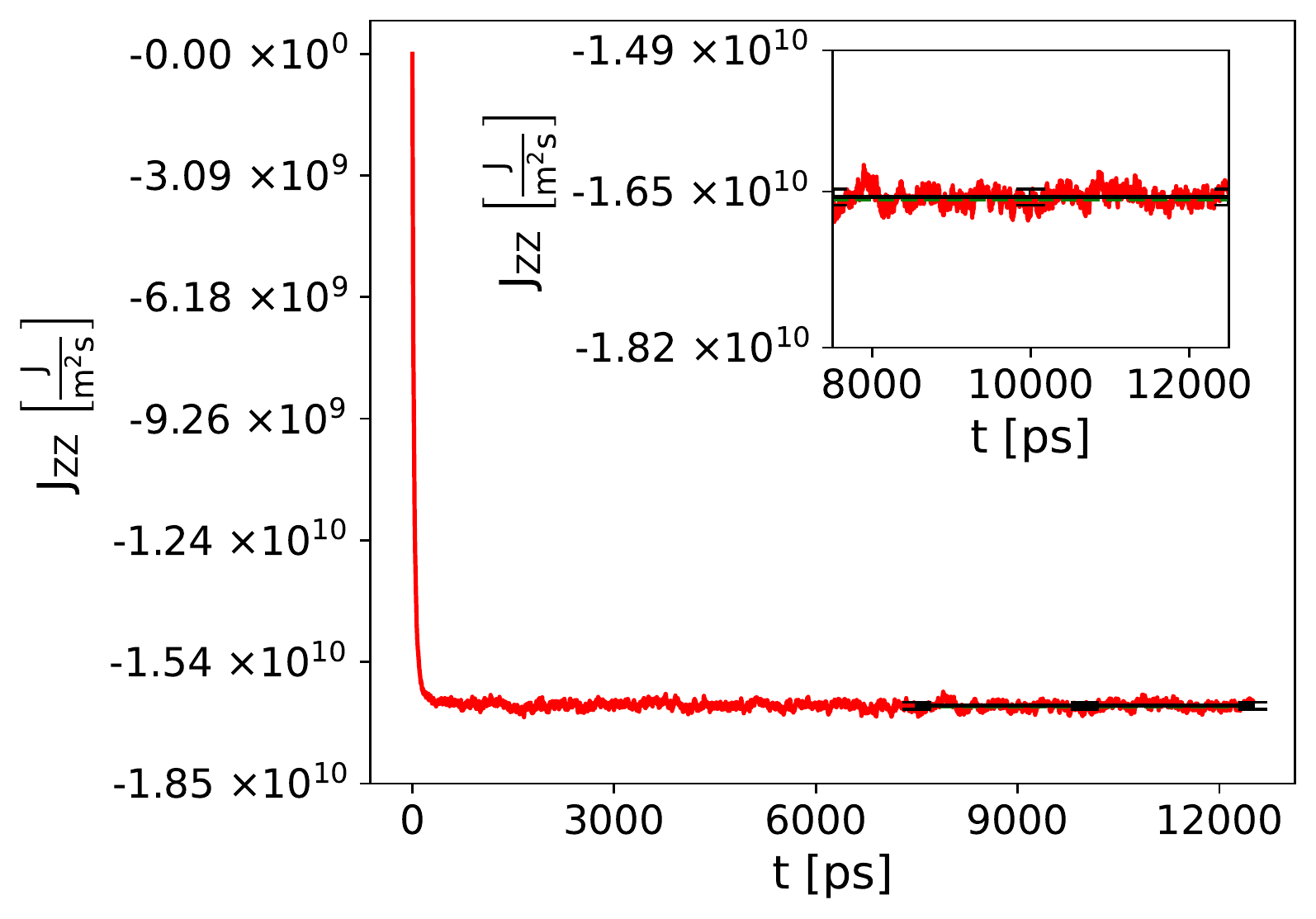}
		\label{homoY}
	\end{subfigure}
	\caption{bRTA heat flux as function of simulation time (red) and average steady-state flux (black) in AC (top) and ZZ (bottom) for infinitely large phosphorene at \SI{300}{\kelvin} under an applied thermal gradient of \SI{0.2}{\kelvin\per\nano\meter} in the transport direction. The iterative result calculated via Fourier's law with $\kappa^\text{almaBTE}$ is given in both cases for comparison (dashed green). Inset: zoomed view of the steady-state region used to compute the mean flux.}
	\label{Heat-fluxes-homo}
\end{figure}
The MC heat fluxes for the infinite phosphorene under thermal gradients along the ZZ and AC directions are plotted in Fig.~\ref{Heat-fluxes-homo}. $\kappa^\mathrm{MC}_\mathrm{AC}$ and $\kappa^\mathrm{MC}_\mathrm{ZZ}$ are calculated via Fourier's law ($J_h = -\kappa_l\nabla_rT$) to be $27.4\pm0.2 ~ \mathrm{W/(m\cdot K)}$ and $82.8\pm0.5 ~ \mathrm{W/(m\cdot K)}$ respectively. Those results show an excellent agreement between iterative and MC solutions, thus validating our bRTA implementation.

\section{\label{sec:Result}Results}

\subsection{Phosphorene devices}

In this section we present thermal transport results for different phosphorene-based configurations/devices using the simulators developed in previous sections.

\subsubsection{\label{Sec:nanoribbons}Nanoribbons}
\begin{figure}
	\begin{subfigure}[b]{0.45\textwidth}
		\centering
		\includegraphics[width=\textwidth]{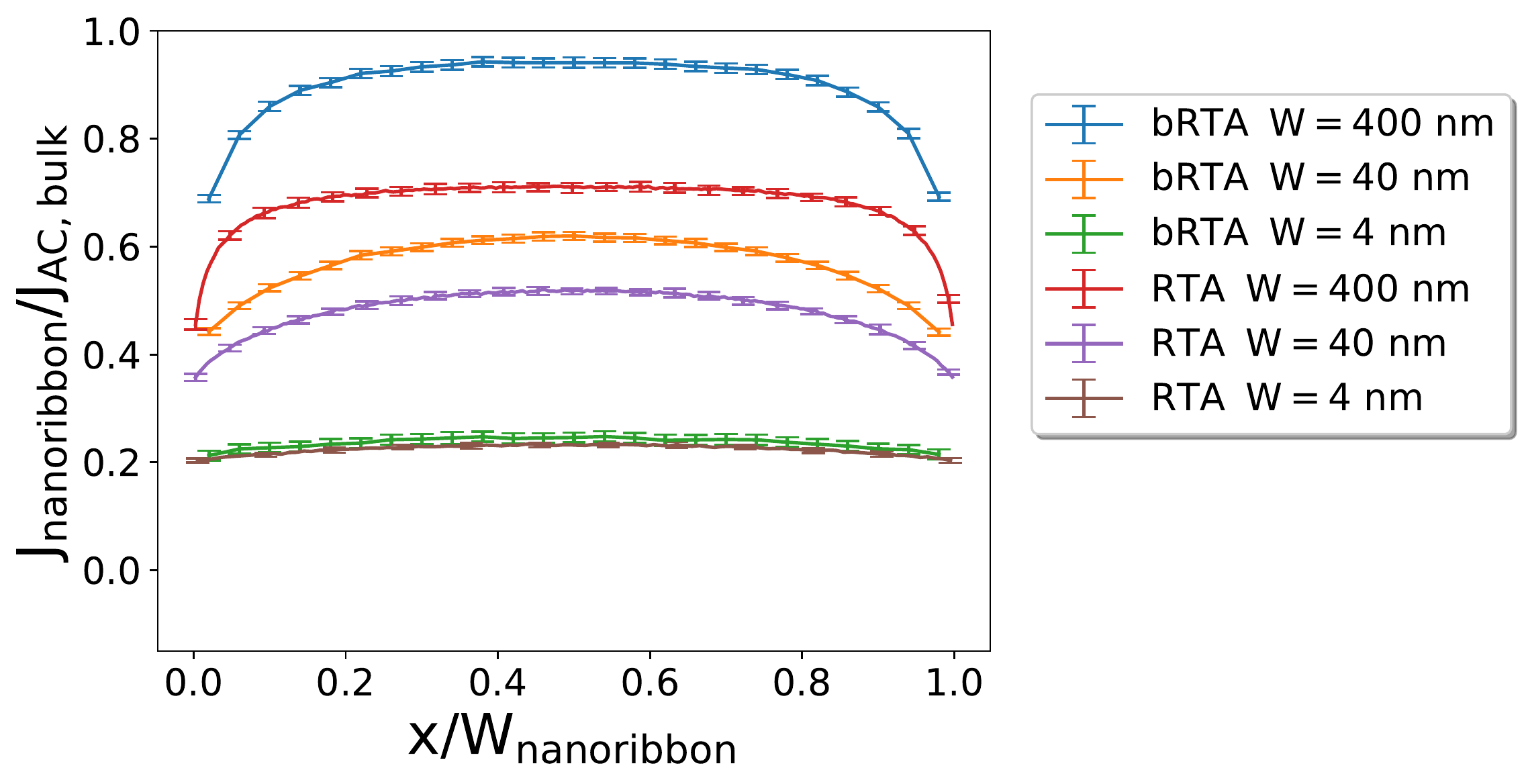}
		\caption{~}
		\label{nanoribbons-a}
	\end{subfigure}
	\begin{subfigure}[b]{0.45\textwidth}
		\centering
		\includegraphics[width=\textwidth]{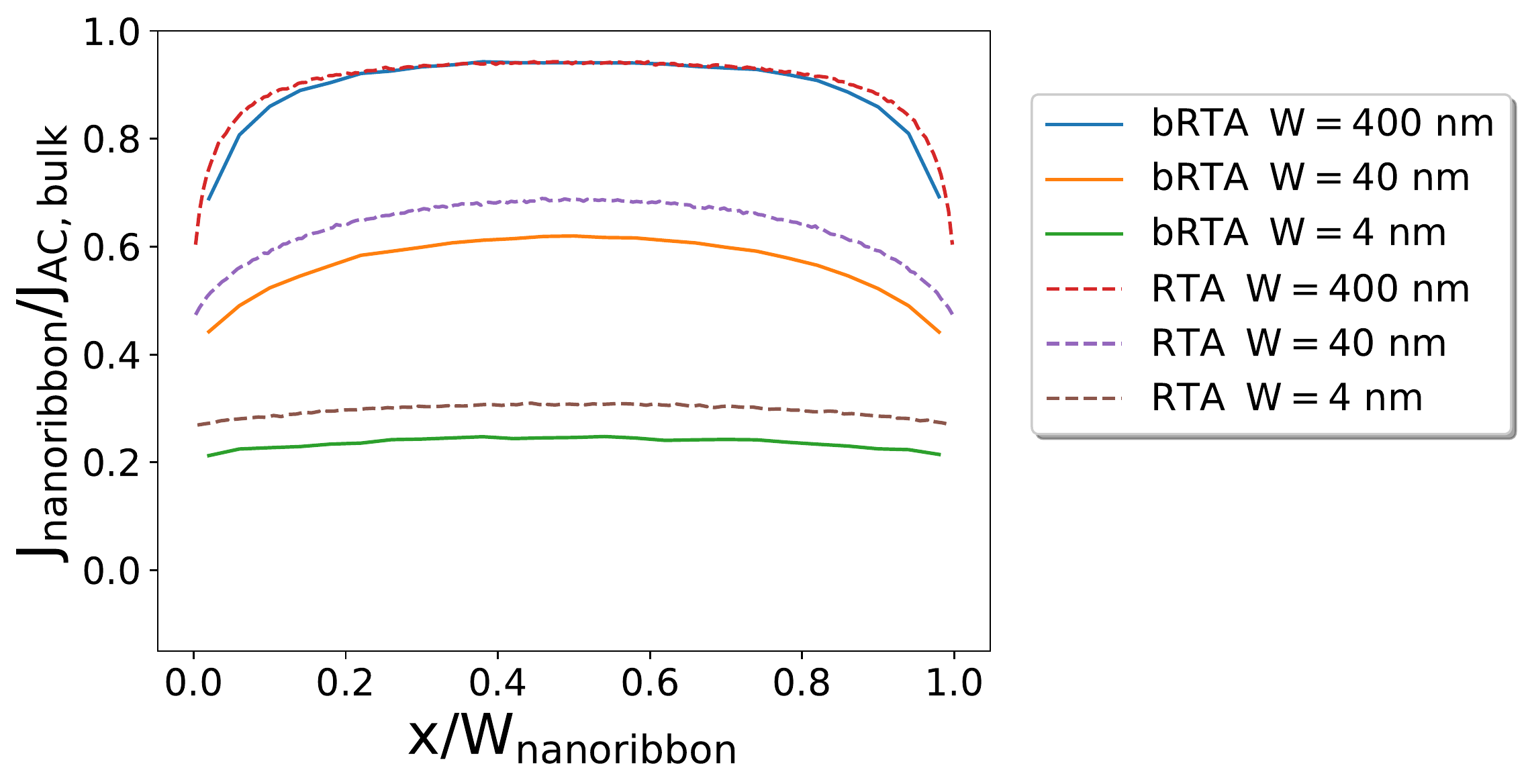}
		\caption{~}
		\label{nanoribbons-b}
	\end{subfigure}
	\caption{Top: RTA and bRTA bulk-normalized heat flux in the AC direction as a function of the normalized position for a phosphorene nanoribbon with $\nabla_\mathrm{AC}T = \SI{0.2}{\kelvin\per\nano\meter}$. Bottom: Comparison of bulk-normalized bRTA and  $\kappa_\mathrm{almaBTE}/\kappa_\mathrm{RTA}$-rescaled RTA heat fluxes.}
	\label{nanoribbons}
\end{figure}

Among the simplest 2D-based systems used in devices are nanoribbons~\cite{Schwierz2010NatNano,Chen2019NanoLet}. We have calculated the heat transport in infinite phosphorene AC nanoribbons under the effect of a thermal gradient of \SI{0.2}{\kelvin\per\nano\meter}. The results of the normalized heat flux relative to the bulk value, together with an RTA rescaled version for three different widths are shown in Fig.~\ref{nanoribbons}.

As expected, boundary scattering increases and becomes dominant over other mechanisms in thinner nanoribbons. This is clearly seen in the reduction of heat flux with decreasing width and the fact that the difference between the RTA and beyond-RTA methods vanishes for smaller ribbons (see the \SI{4}{\nano\meter} case in Fig.~\ref{nanoribbons-a}),since boundary scattering is not dependent on the approach used to describe intrinsic anharmonic and isotopic scattering. Consequently, for wide nanoribbons in which anharmonic and isotopic scattering are dominant it should be possible to obtain a good approximation to bRTA results by simply using a $\kappa_\mathrm{almaBTE}/\kappa_\mathrm{RTA}$-rescaled RTA (see Fig.~\ref{nanoribbons-b}). Indeed, the main differences between this estimate and bRTA for the \SI{400}{\nano\meter}-ribbon are in the regions near the boundaries, in which the  bRTA flux is lower than the rescaled version. Although also visible in other cases, this is more pronounced in the wider nanoribbon.

In view of the above, the RTA clearly overestimates the momentum destruction due to intrinsic scattering leading to more diffusive flux profiles when compared to bRTA results. The latter yields more Poiseuille-like profiles, with stronger hydrodynamic features, by properly capturing the coupling between phonon modes~\cite{CepellottiNatCom2015,Lee2015NatCom}.

To further explore those hydrodynamic signatures, we fitted our nanoribbon results to a mesoscopic equation based on Sellitto \textit{et al.}'s work~\cite{SellittoPRSMPES2015}: 
\begin{multline}
	\label{Seillitto}
	J(x) = -\kappa \left\{ 1 - \left[\frac{1}{1 + C \tanh\left(\frac{W}{2\ell}\right)}\right]\right.\\\left.\frac{\cosh\left(\frac{x}{\ell}\right)}{\cosh\left(\frac{W}{2\ell} \right)}  \right\}\nabla T 
\end{multline}
where $W$ is the nanoribbon width, $x$ is the distance from the center of the nanoribbon, $\ell$ is the non-local length~\cite{SendraPRB2021} and $C$ is related to wall properties, taking a value of $2$ in our case because we have assumed completely diffusive walls.

\begin{figure}
	\centering
	\includegraphics[width=\linewidth]{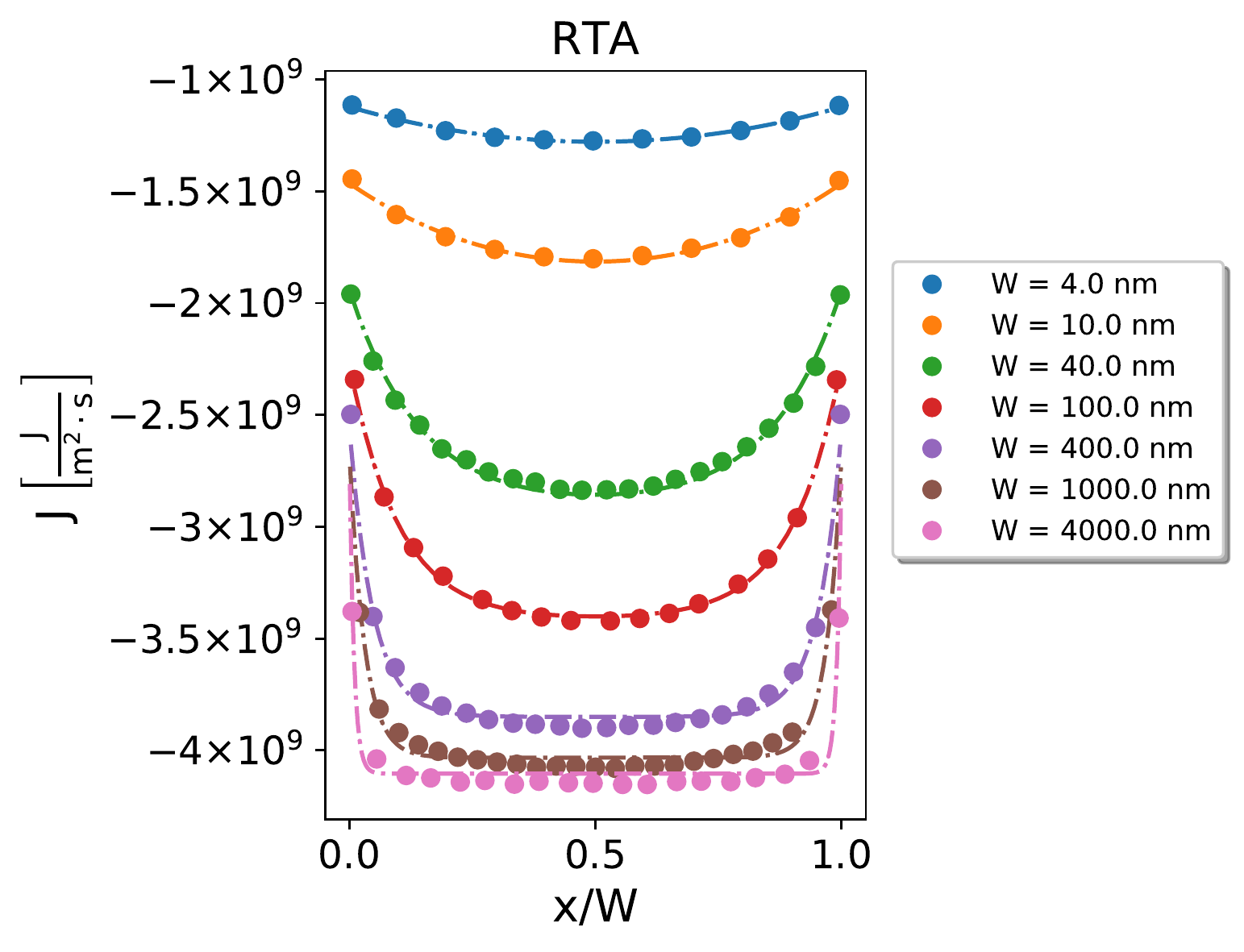}
	\caption{Fitting to to Eq.~\ref{Seillitto} (lines) of RTA-MC calculated heat flux (points) as a function of normalized position for phosphorene AC nanoribbons of different widths under the effect of $\nabla_\mathrm{AC}T = \SI{0.2}{\kelvin\per\nano\meter}$.}
	\label{nrhydroRTA}
\end{figure}

\begin{figure}
	\centering
	\includegraphics[width=\linewidth]{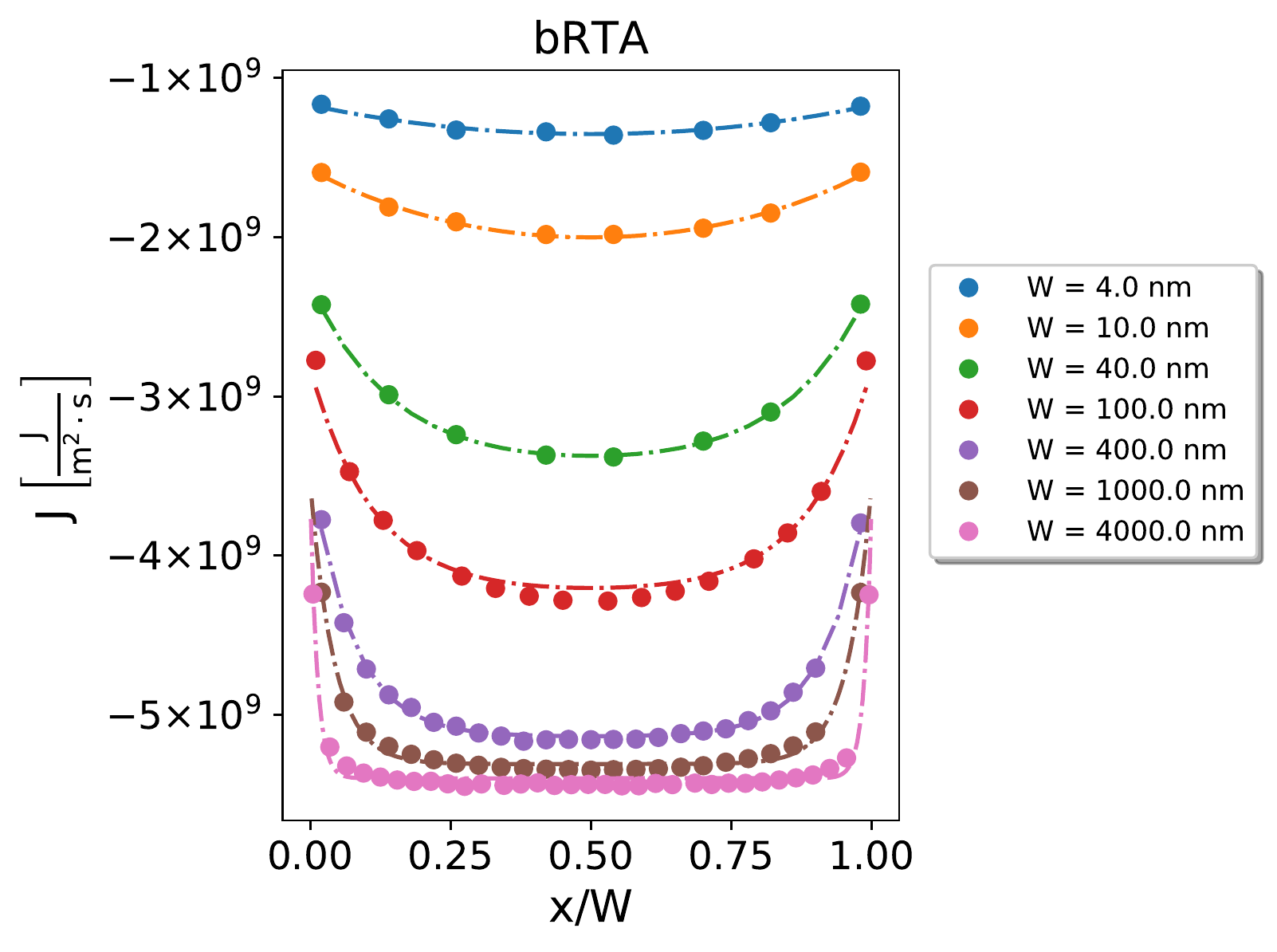}
	\caption{Fitting to Eq.~\ref{Seillitto} formula (lines) of bRTA calculated heat flux (points) as a function of normalized position for phosphorene AC nanoribbons of different widths under the effect of $\nabla_\mathrm{AC}T = \SI{0.2}{\kelvin\per\nano\meter}$.}
	\label{nrhydroBERTA}
\end{figure}

In Figs.~\ref{nrhydroRTA} and ~\ref{nrhydroBERTA} the fits of our RTA and bRTA simulator results to hydrodynamic mesoscopic equation for nanoribbons are shown; in both we obtain a set of parameters that accurately match the simulation results. The agreement afforded by the beyond-RTA method is, however, slightly better.

\begin{figure}
	\centering
	\includegraphics[width=\linewidth]{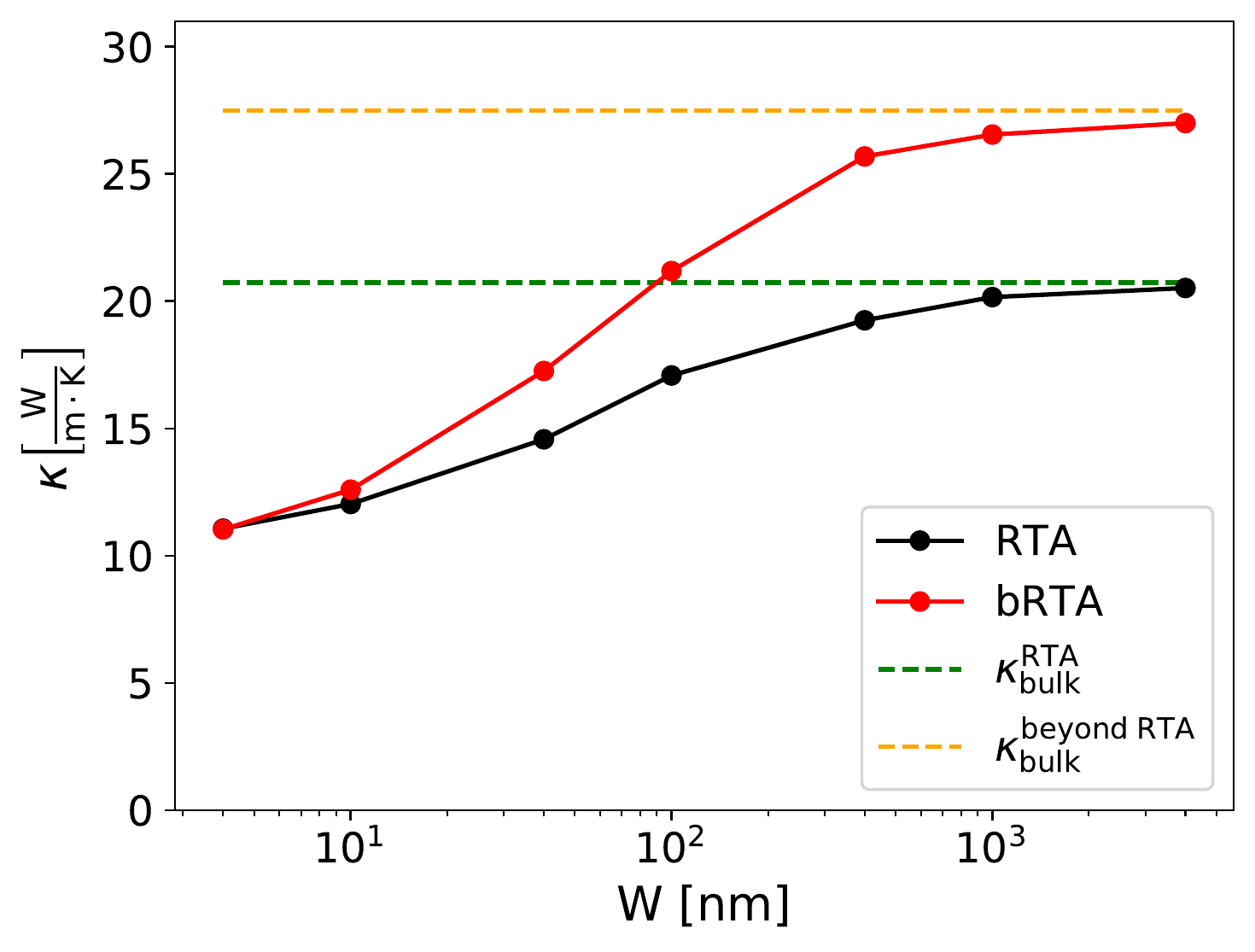}
	\caption{Fitted thermal conductivity as a function of AC nanoribbon width for RTA (black) and bRTA (red) MC calculations. RTA (green) and beyond RTA (orange) bulk values are given for reference.}
	\label{kapparibbons}
\end{figure}

\begin{figure}
	\centering
	\includegraphics[width=\linewidth]{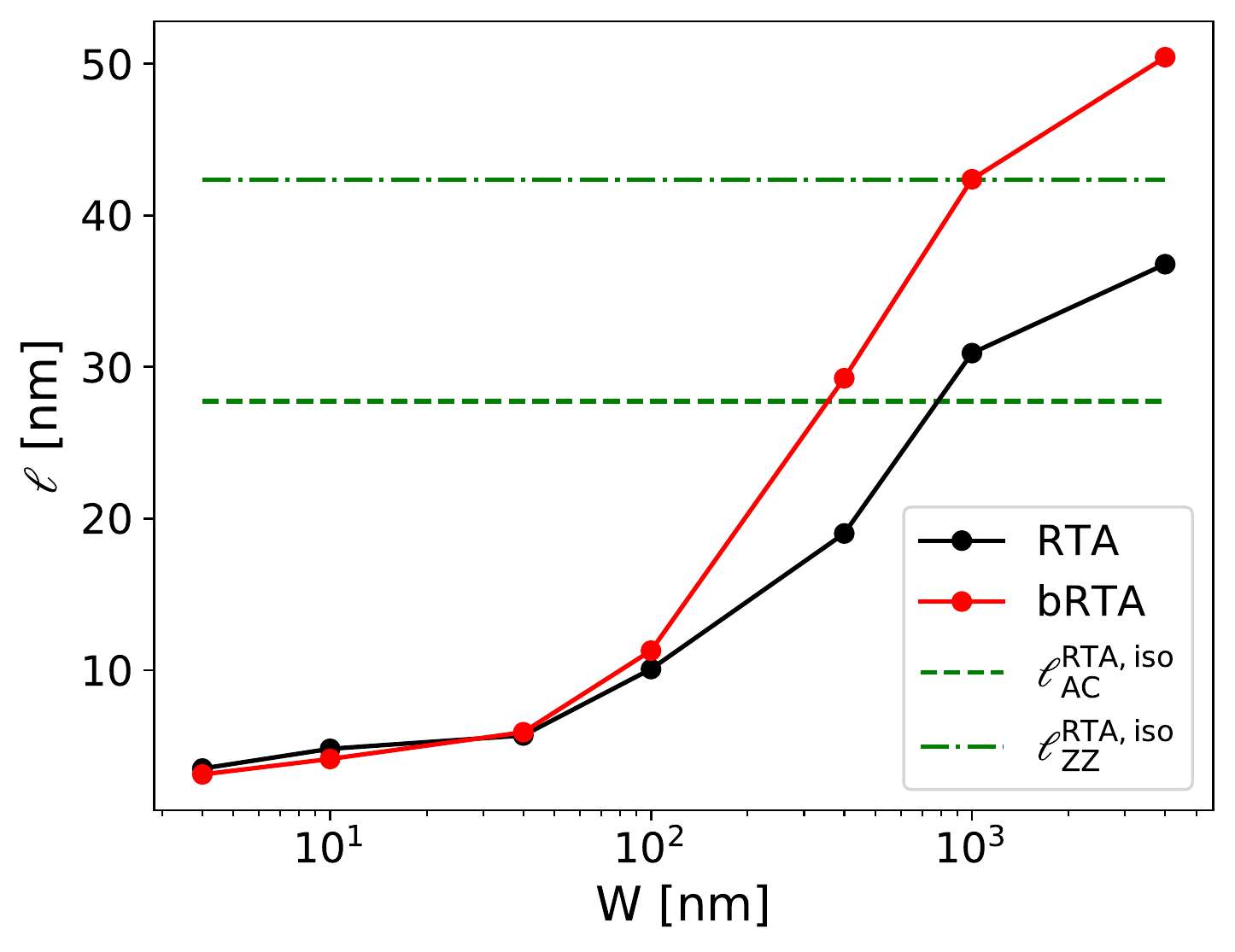}
	\caption{Fitted non-local distance $\ell$ as a function of AC nanoribbon width for RTA (black) and bRTA (red) MC calculations. RTA bulk values of $\ell$ calculated using Sendra et. al.'s formula~\cite{SendraPRB2021} are given for reference (green).}
	\label{lgkribbons}
\end{figure}

As for the fitted parameters, we can observe a relative fast convergence of thermal conductivity towards bulk values as the nanoribbon gets wider and boundary scattering effects become negligible in the middle of the strip (see Fig.~\ref{kapparibbons}). The value of the non-local length ($\ell$) rises towards a converged value as the ribbon becomes wider (see Fig.~\ref{lgkribbons}), in agreement with what is expected from a microscopic description of the value~\cite{SendraPRB2021}. The observed differences between RTA and the higher beyond-RTA $\ell$ values, especially for larger widths, can be easily interpreted by keeping in mind that in RTA all scattering processes are deemed as resistive and introduce artifactual modifications of the heat flux. It should be noted that the theoretical formulas used to obtain the $\ell^{\mathrm{RTA,iso}}$ values are derived under the assumption of isotropy and are therefore are expected to be useful only as approximations in our case. For reference, the results for ZZ-nanoribbons are also given in \ref{nanoribbons-ZZ}.


\begin{figure}
	\centering
	\includegraphics[width=\linewidth]{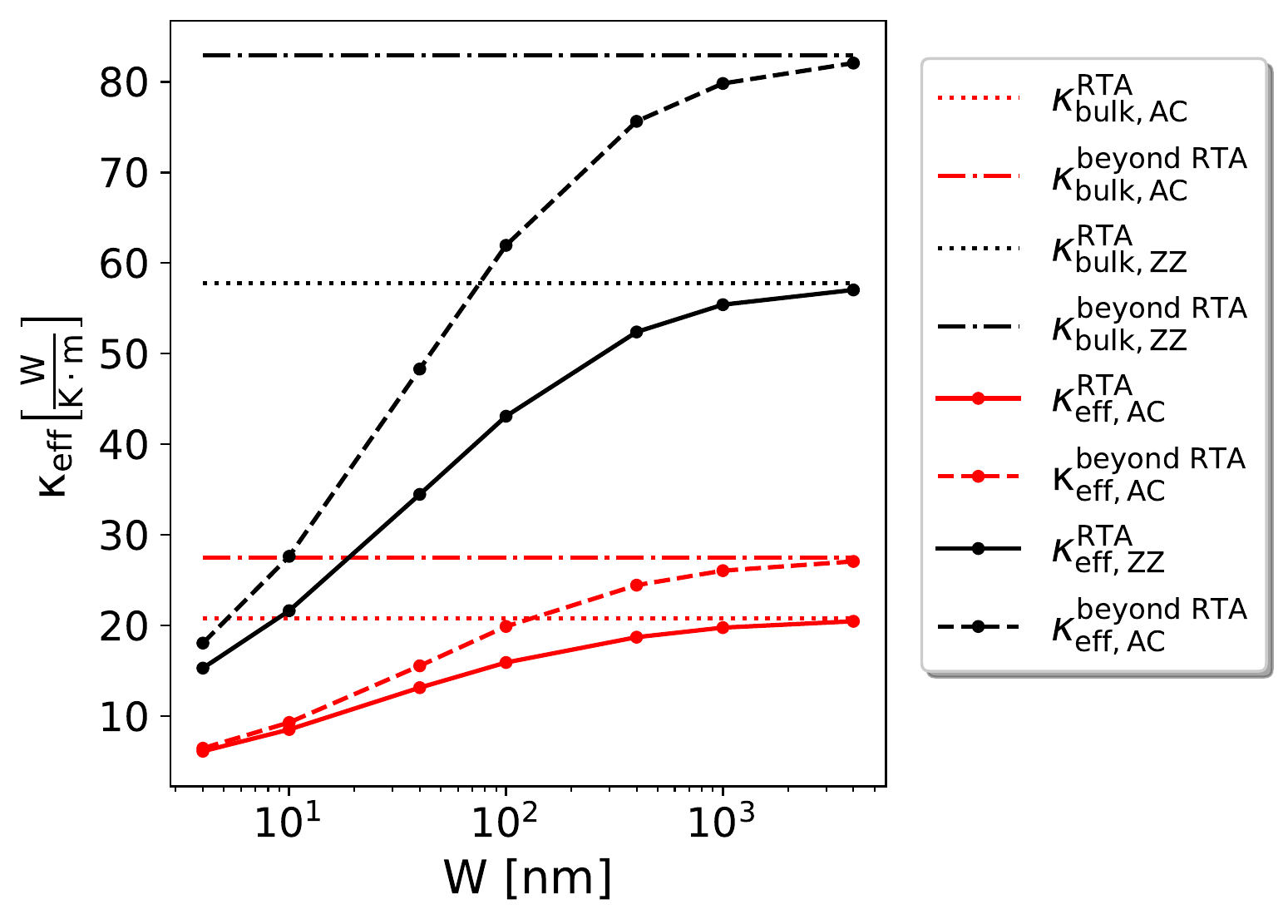}
	\caption{Effective thermal conductivity for AC and ZZ nanoribbons of different widths at \SI{300}{\kelvin} obtained through the direct (RTA) and iterative (beyond the RTA) solution of the linerized-PBTE. Bulk values are provided as reference.}
	\label{keff_nanoribbons}
\end{figure}

\begin{figure*}
	\begin{subfigure}[b]{0.45\textwidth}
		\centering
		\includegraphics[width=\textwidth]{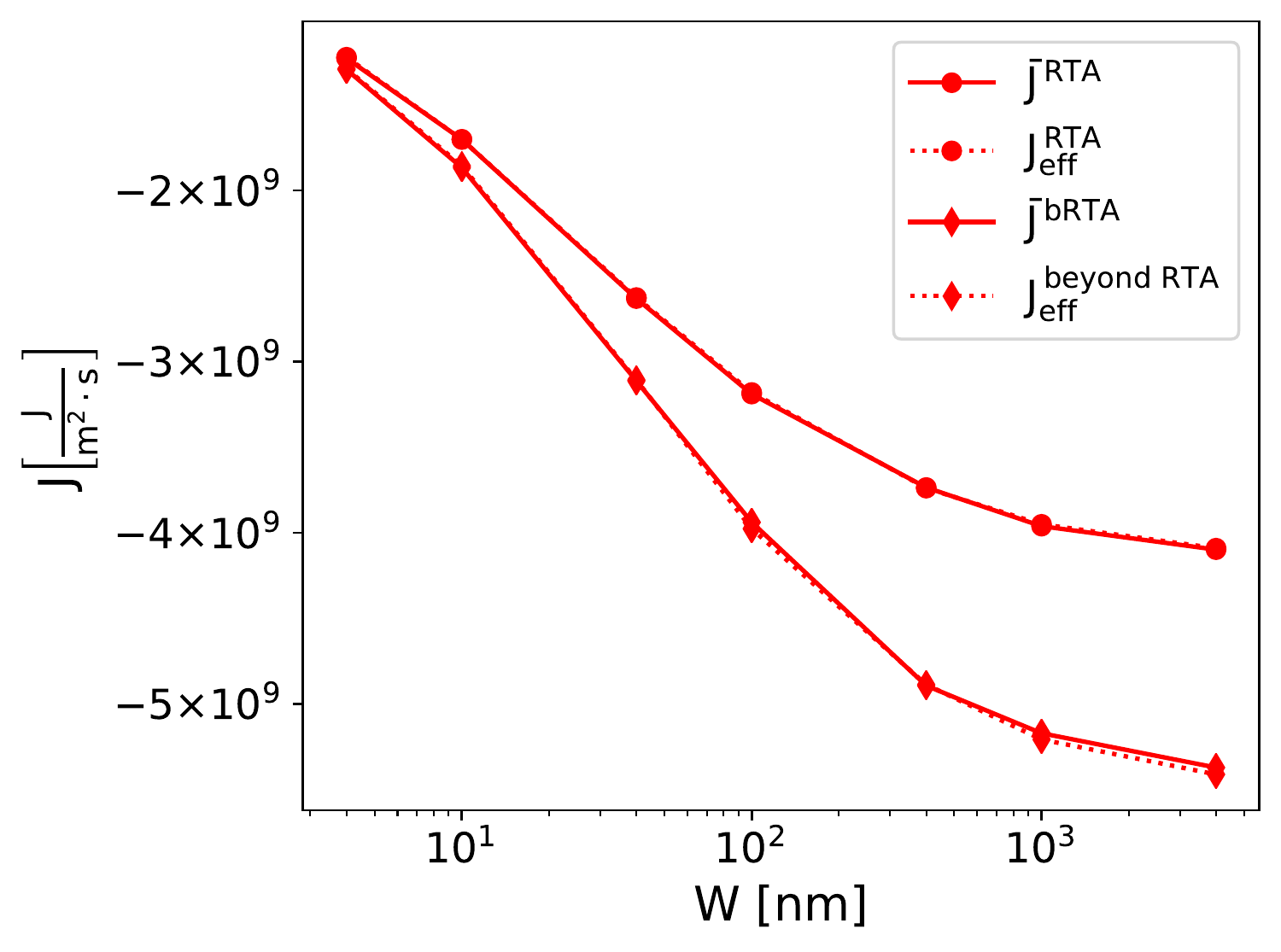}
	\end{subfigure}
	\begin{subfigure}[b]{0.45\textwidth}
		\includegraphics[width=\textwidth]{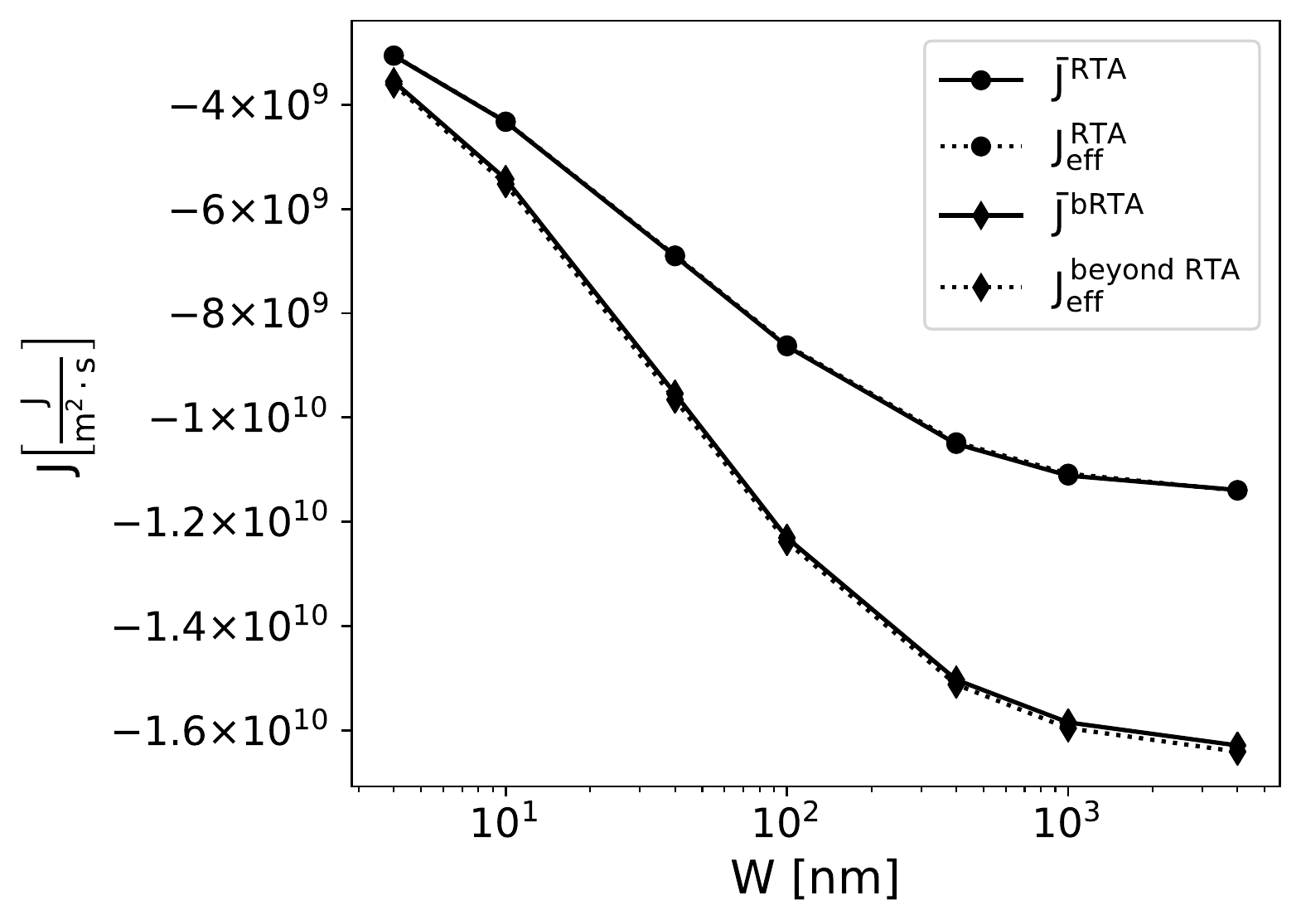}
	\end{subfigure}
	\caption{Comparison between effective RTA and beyond RTA heat fluxes for AC (left) and ZZ (right) nanoribbons and the respective Monte Carlo obtained fluxes for a \SI{0.2}{\kelvin\per\nano\meter} gradient in the unbound direction.}
	\label{jeff_nanoribbons}
\end{figure*}

Finally, we have also obtained $\kappa_\text{nano}$ for nanoribbons of several widths with both types of edges using the methodology for solving the PBTE in systems with edges described in Sec.~\ref{sec:Methods_nanoribbons} (see Fig. \ref{keff_nanoribbons}). For consistency check, we compared the effective flux obtained via Fourier using $\kappa_\text{nano}$ with the flux average over width obtained from Monte Carlo simulators, obtaining, as can be seen in Fig.\ref{jeff_nanoribbons}, an excellent match between both methods. 

\subsubsection{RTA, bRTA and Fourier heat equation comparison}

\begin{figure}
	\centering
	\includegraphics[width=\linewidth]{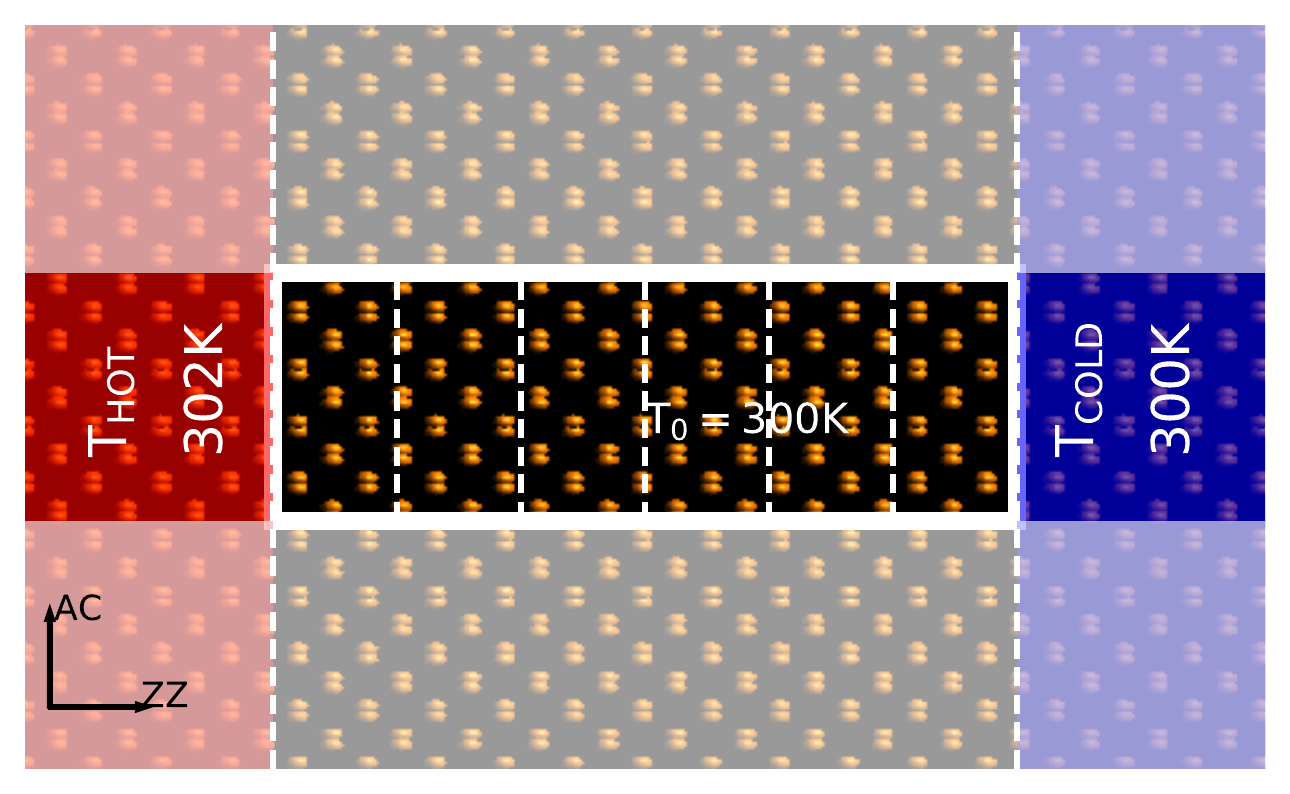}
	\caption{Sketch of phosphorene heating simulation in AC direction. Periodic boundary conditions are depicted with off-color boxes.}
	\label{bar}
\end{figure}

Taking into account that operational frequencies of microprocessors are limited to the GHz by cooling constraints~\cite{IlatikhamenehSR2016,BernsteinPIEEE2010}, it is of interest to be able to study heating dynamics at short times.
As an example of capabilities to simulate short heating dynamics, we have studied the temperature time evolution for a piece of phosphorene initially at \SI{300}{\kelvin} with periodic boundary conditions in the AC direction and sandwiched between two isothermal reservoirs, at \SI{300}{\kelvin} and \SI{302}{\kelvin}, in the ZZ direction (see Fig.~\ref{bar}). 

\begin{figure}
	\centering
	\includegraphics[width=\linewidth]{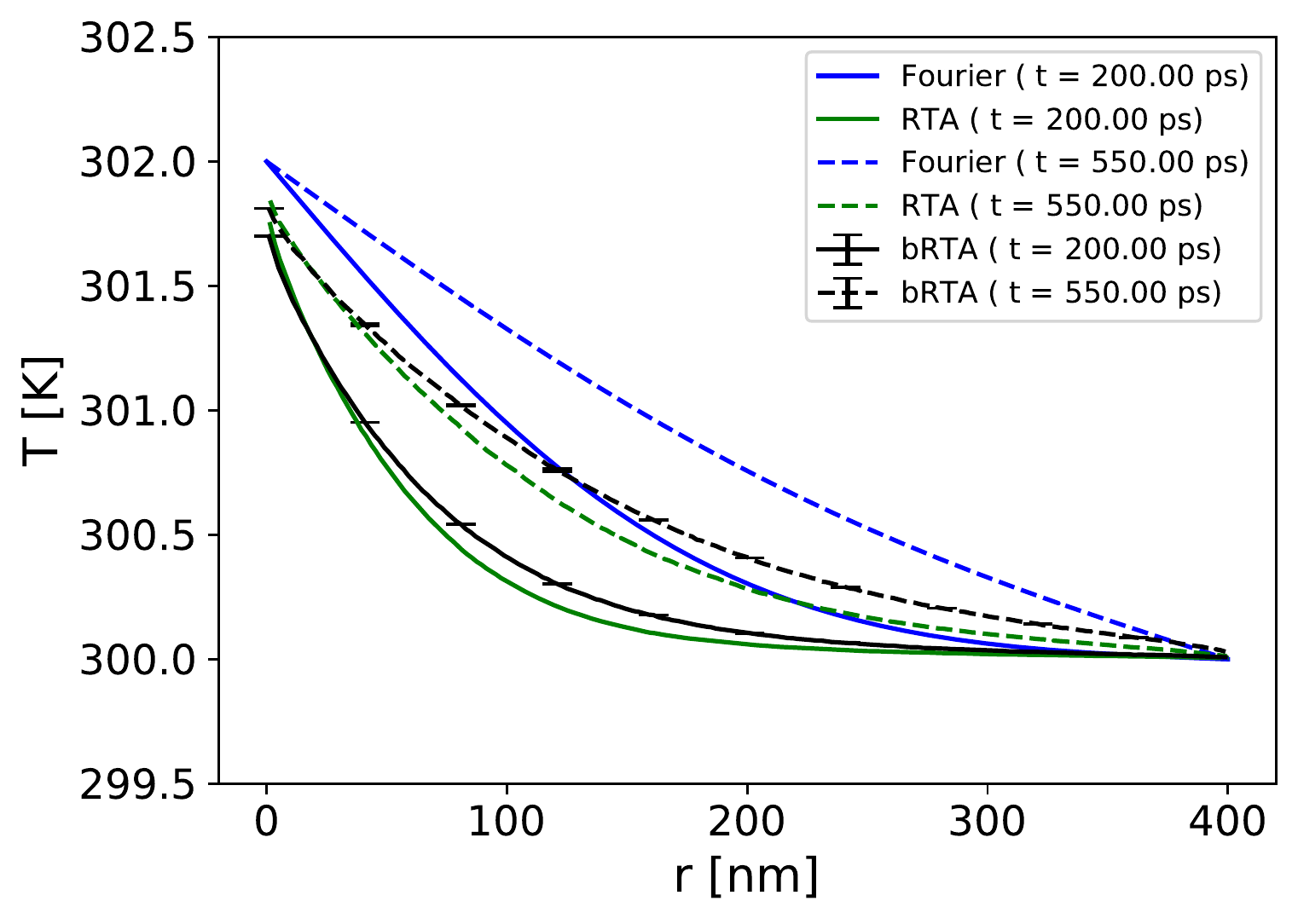}
	\caption{Temperature profiles obtained with the RTA (green), bRTA (black) and Fourier (blue) approaches as functions of position for an AC-phosphorene bar at \SI{200}{\pico\second} (solid) and \SI{550}{\pico\second} (dashed).}
	\label{barheating}
\end{figure}

Fig.~\ref{barheating} shows the RTA, bRTA and Fourier ($\frac{\partial T}{\partial t} = \alpha \nabla^2 T$, where $\alpha$ is the thermal diffusivity: $\alpha = \kappa/C_v$) heat profiles at two different times. The difference between the Fourier heat equation and BTE results at short times is a well known shortcoming of the former~\cite{MinnichPRB2011,VermeerschPRB2015}. Regarding PBTE solutions, bRTA and RTA results clearly differ at this instance. The bRTA shows a faster heating, which is not surprising since RTA scattering completely randomizes momentum, thus dampening the fluxes and leading to a lower thermal conductivity. 

\begin{figure}
	\centering
	\includegraphics[width=\linewidth]{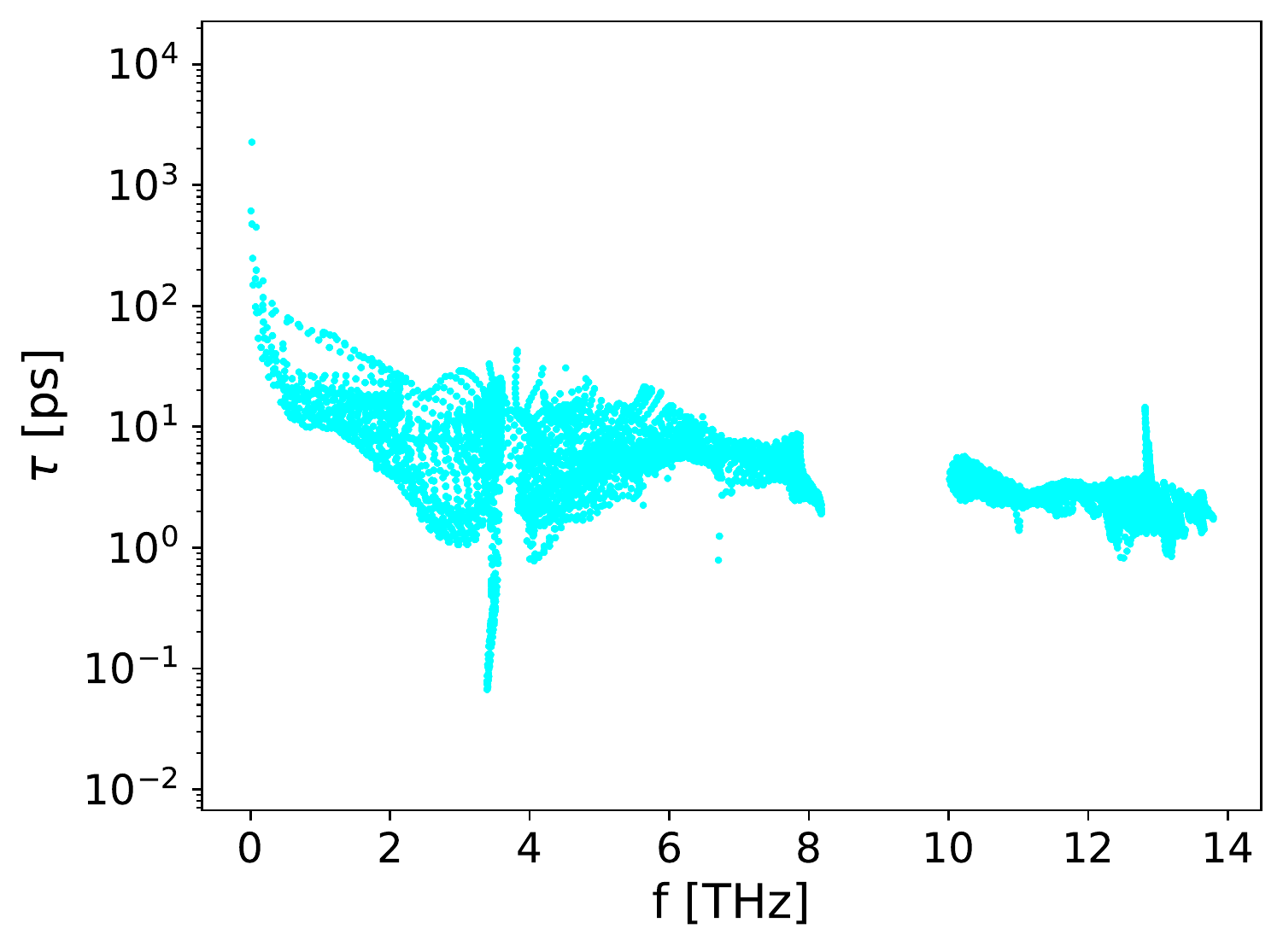}
	\caption{Bulk-phosphorene lifetimes as function of frequency at \SI{300}{\kelvin}.}
	\label{tau300}
\end{figure}

\begin{figure*}
	\centering
	\begin{subfigure}[b]{0.45\textwidth}
		\includegraphics[width=\textwidth]{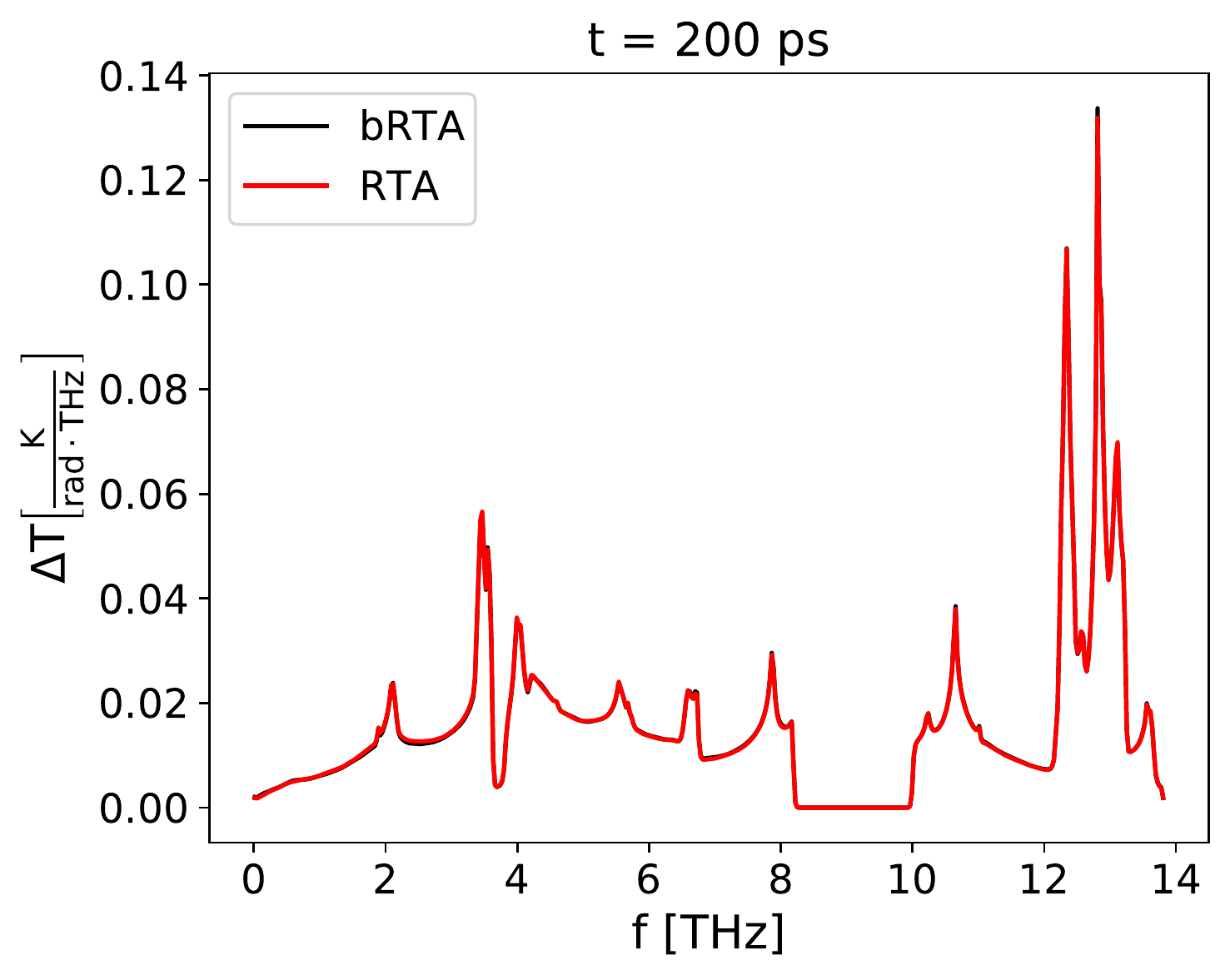}
		\caption{~}
		\label{heatingpops-a}
	\end{subfigure}
	\begin{subfigure}[b]{0.45\textwidth}
		\includegraphics[width=\textwidth]{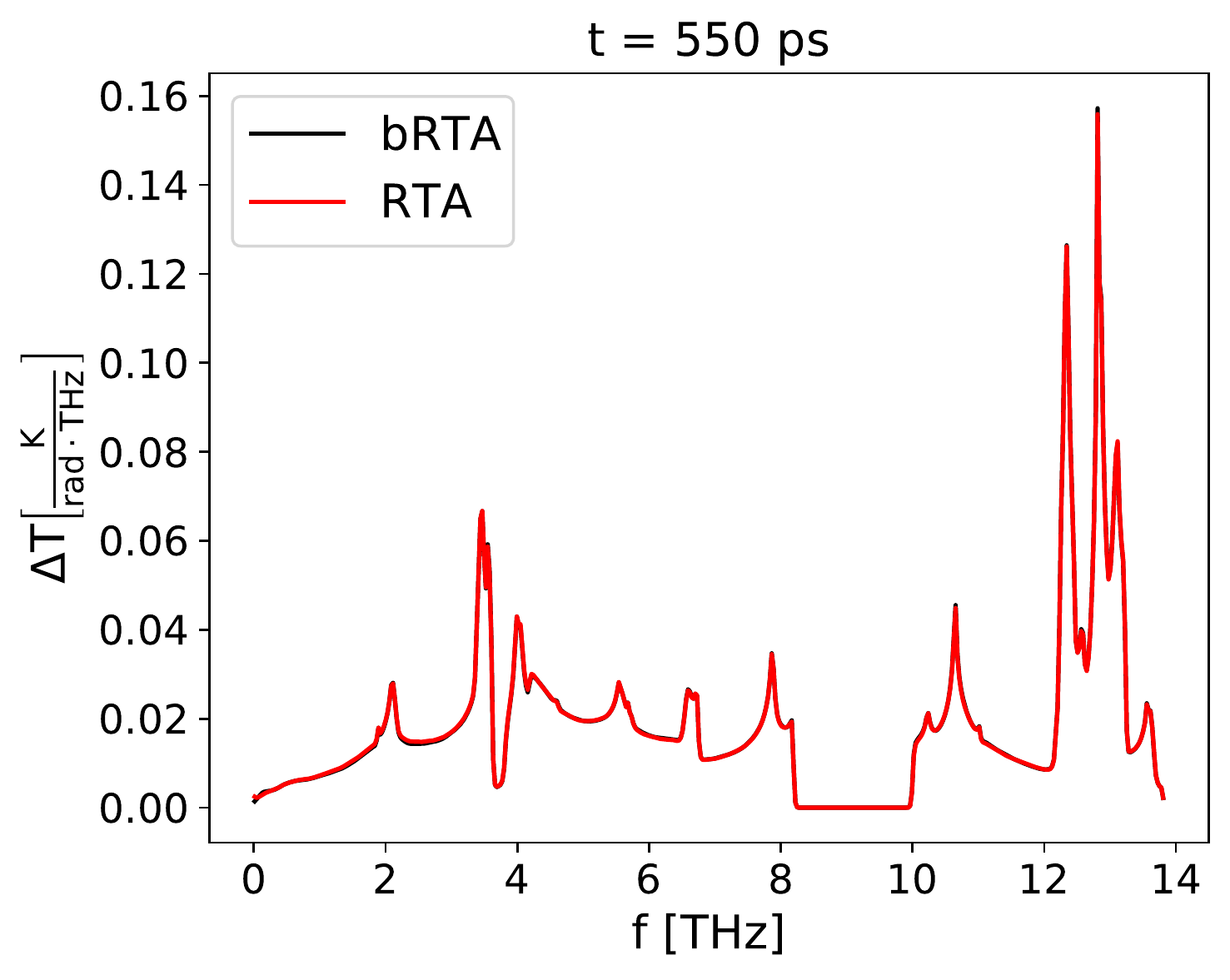}
		\caption{~}
		\label{heatingpops-b}
	\end{subfigure}
	\begin{subfigure}[b]{0.45\textwidth}
		\includegraphics[width=\textwidth]{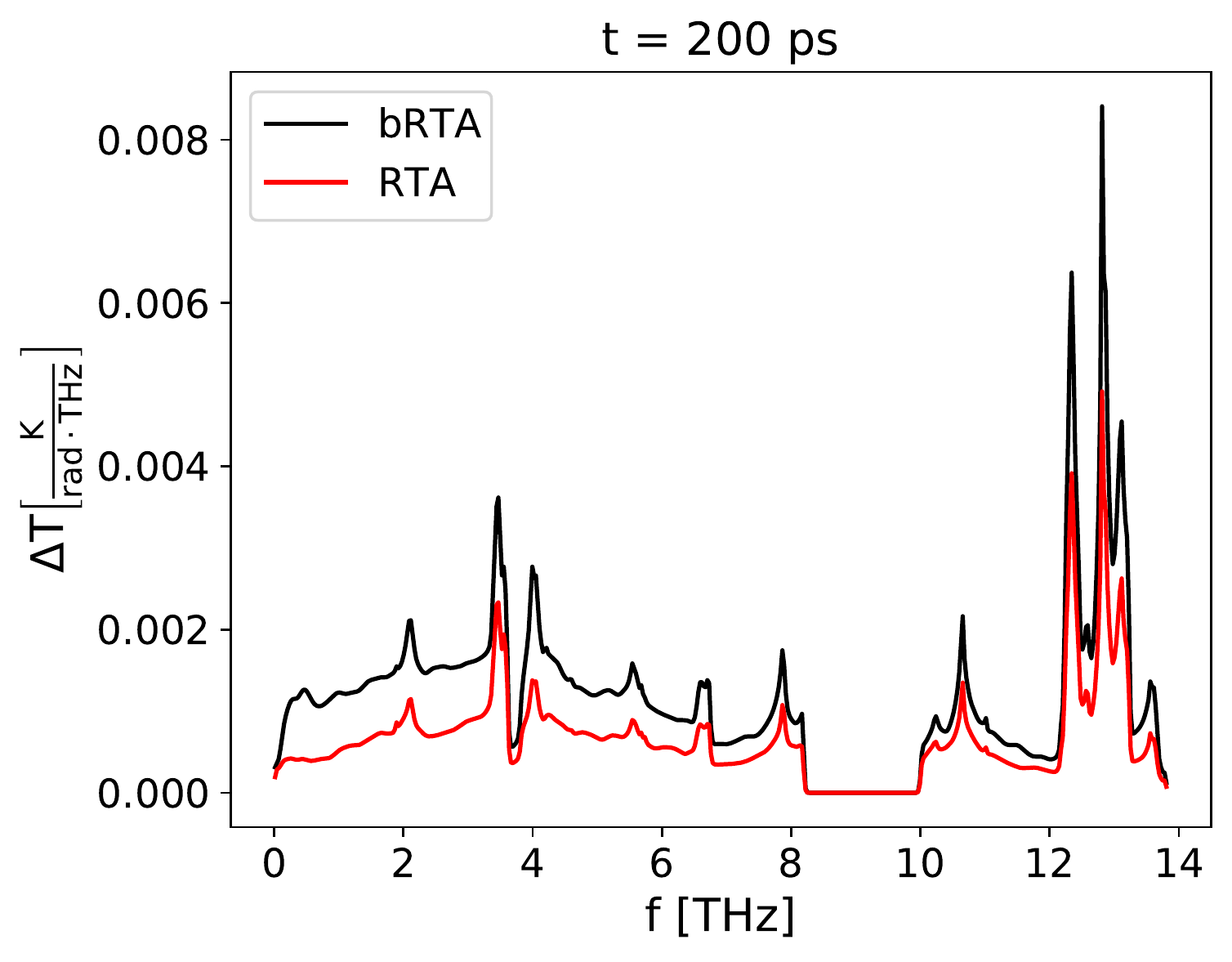}
		\caption{~}
		\label{heatingpops-c}
	\end{subfigure}
	\begin{subfigure}[b]{0.45\textwidth}
		\includegraphics[width=\textwidth]{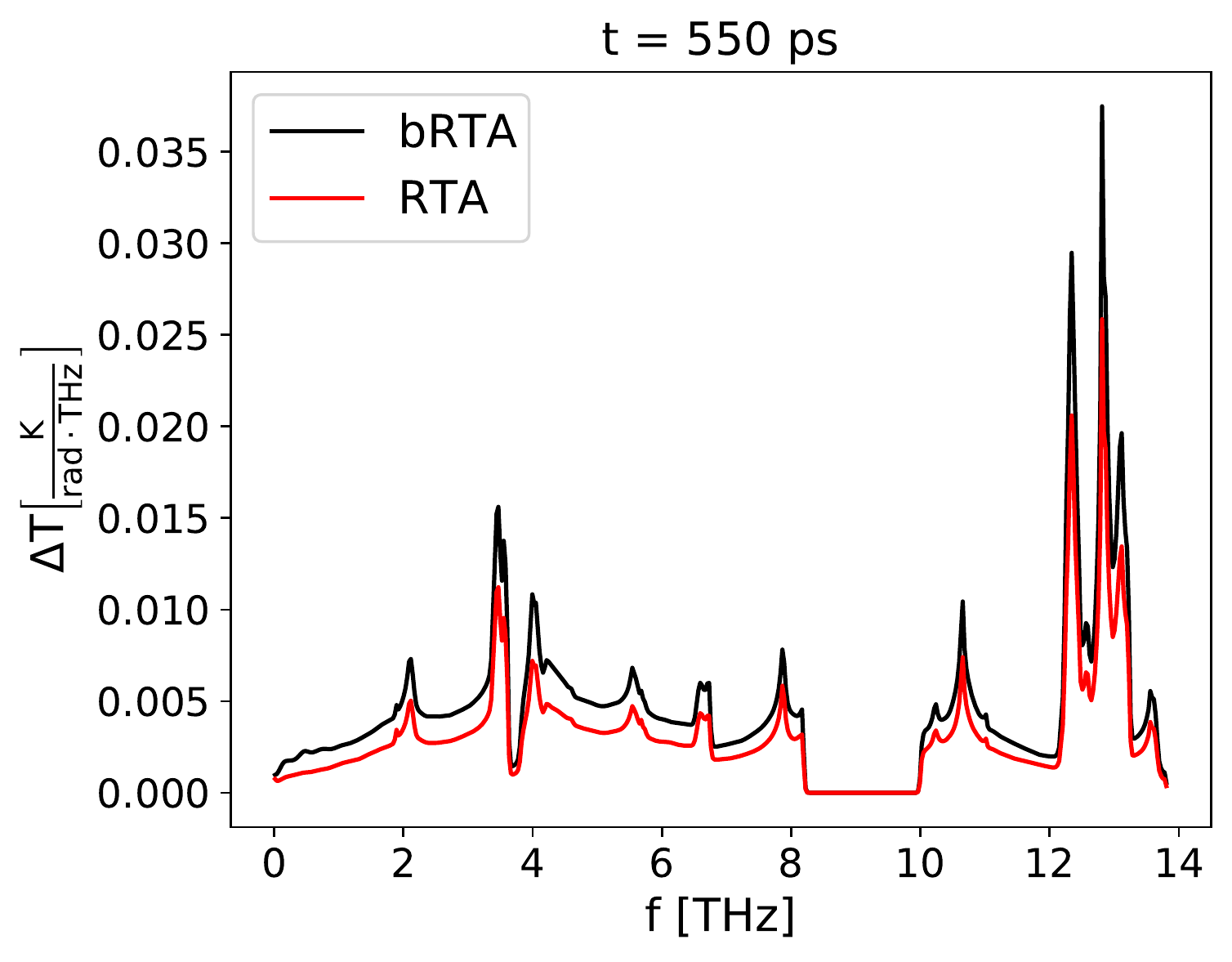}
		\caption{~}
		\label{heatingpops-d}
	\end{subfigure}
	\caption{RTA (red) and bRTA (black) spectral decomposed temperature deviations for \SI{400}{\nano\meter} ZZ-phosphorene bar at \SI{15}{\nano\meter} (a and b) and \SI{203}{\nano\meter} (c and d) from the hot edge, at times \SI{200}{\pico\second} and \SI{550}{\pico\second}.}
	\label{barheatingpops}
\end{figure*}

In Fig.~\ref{barheatingpops} we plot the spectral decomposition of the contributions to the deviational temperature at the middle and near the hot edge of the beam. In keeping with the fact that differences between the RTA and bRTA solutions are the largest in the middle of the bar (see Fig.~\ref{barheating}), spectral decompositions of the deviational temperature deviate the most at the middle as opposed to the edges. They also vanish with time as both tend to the same temperature profile (see Figs. \ref{heatingpops-a}-\ref{heatingpops-d}). Moreover, from the spectral decomposition it can also be seen that differences are more prominent at low frequencies, corresponding to phonons with longer intrinsic lifetimes (see Fig.~\ref{tau300}), which indicates that the decay of such modes is clearly much more overestimated than for high-frequency ones. This explains the large disparity between the RTA and bRTA conductivities.  
It is therefore advisable to resort to the bRTA method for modeling fast/short heat dynamics, for example when studying heat dissipation in state-of-the-art electronic devices, as less sophisticated approximations fail to describe it accurately. 

\subsubsection{\label{res:finite_device} Finite device examples}

\begin{figure*}
	\begin{subfigure}[b]{0.45\textwidth}
		\centering
		\includegraphics[width=\textwidth]{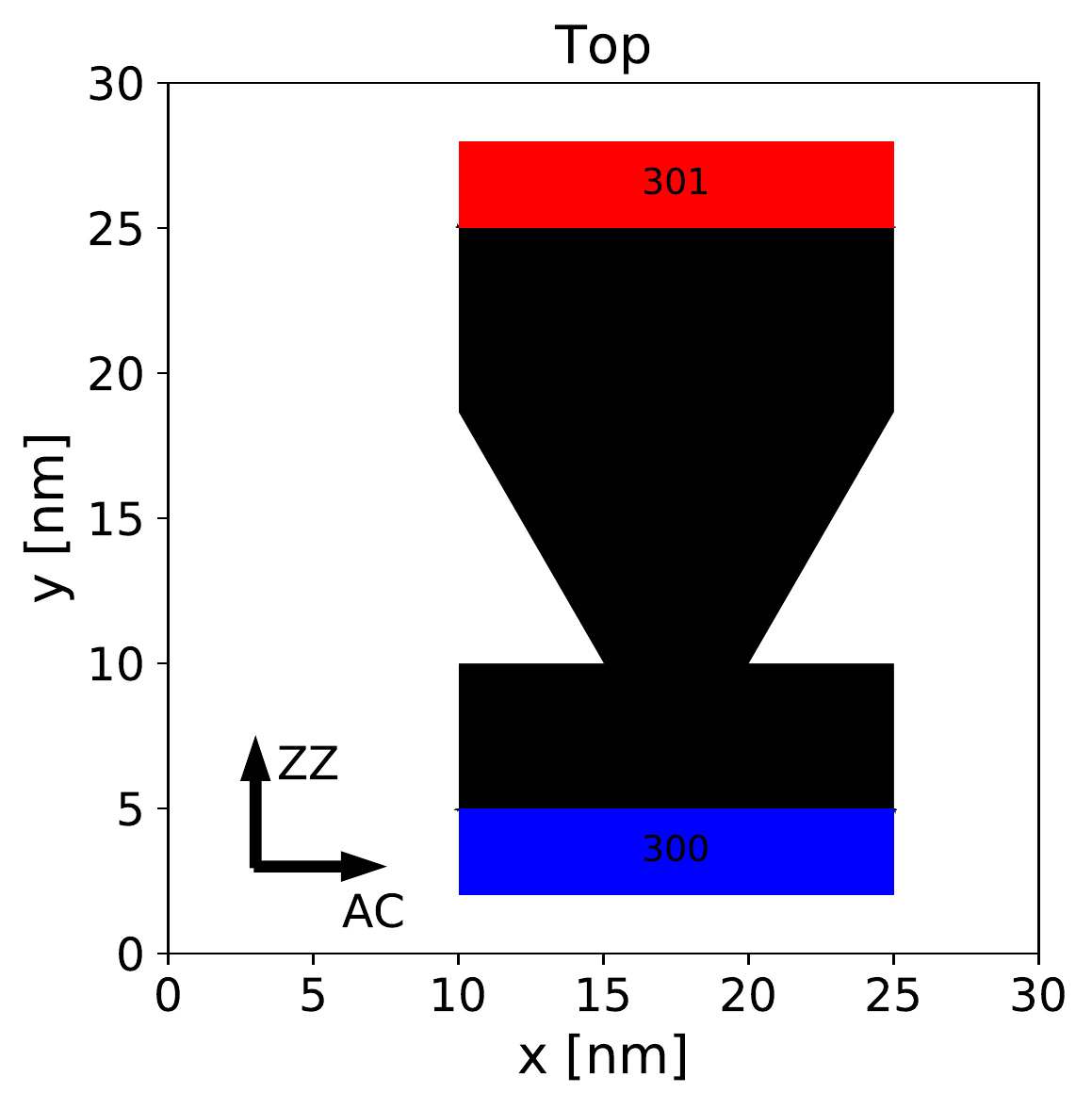}
		\label{CUNYAtop}
	\end{subfigure}
	\begin{subfigure}[b]{0.45\textwidth}
		\centering
		\includegraphics[width=\textwidth]{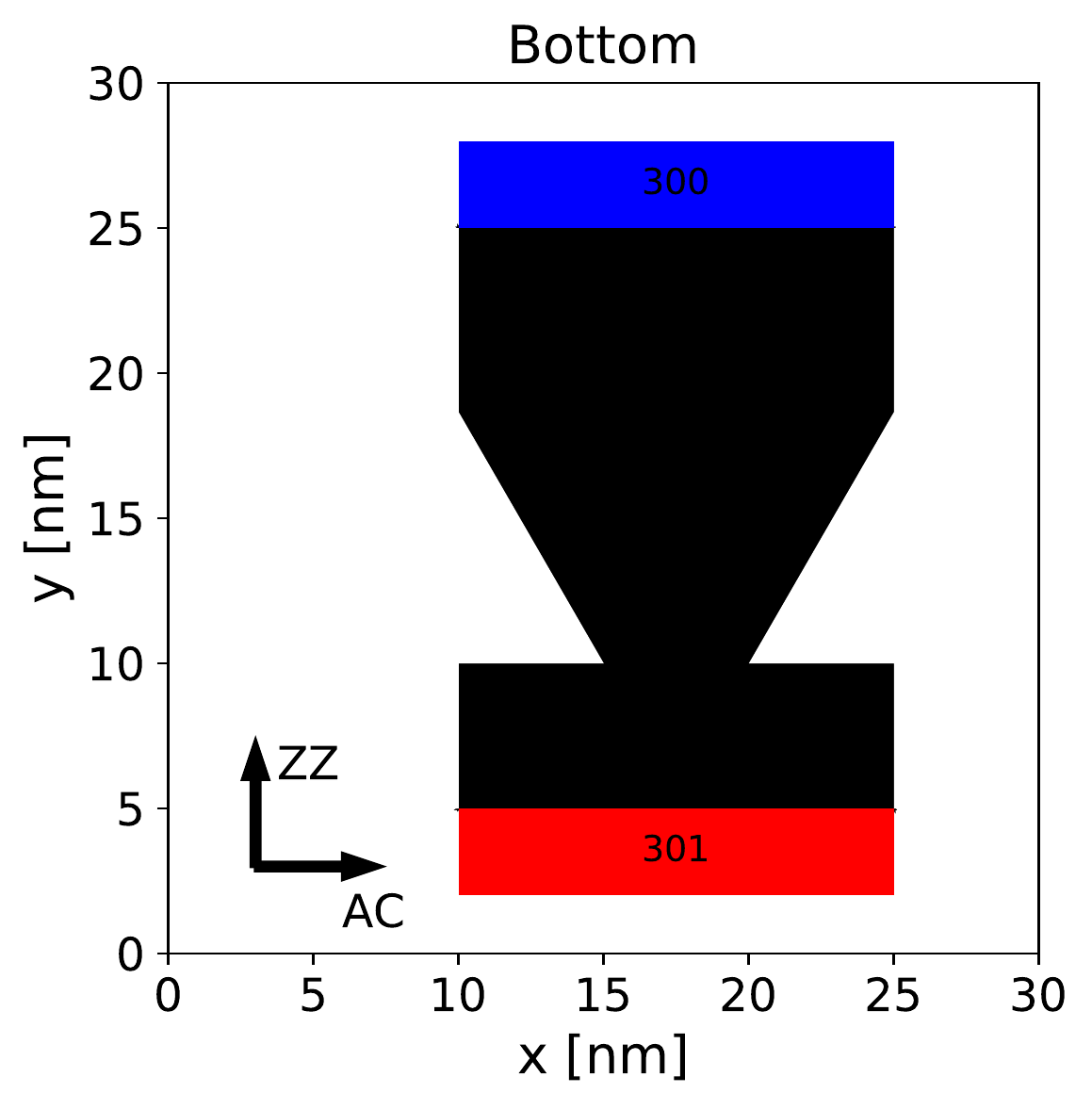}
		\label{CUNYAbot}
	\end{subfigure}
	\caption{Phosphorene wedge-like geometry with hot reservoir at the top or at the bottom.}
	\label{CUNYAS}
\end{figure*}

As previously mentioned, being able to predict thermal transport in complex devices and geometries is of key importance. To this end, we show examples for more complex systems, either because they have geometrical elements which are difficult to model computationally, such a wedge geometry (see Fig.~\ref{CUNYAS}), or because they present interesting elements from a simulator capability point of view such as more than two terminals, which is a common experimental setup.

\begin{figure*}
	\centering
	\begin{subfigure}[b]{0.45\textwidth}
		\includegraphics[width=\textwidth]{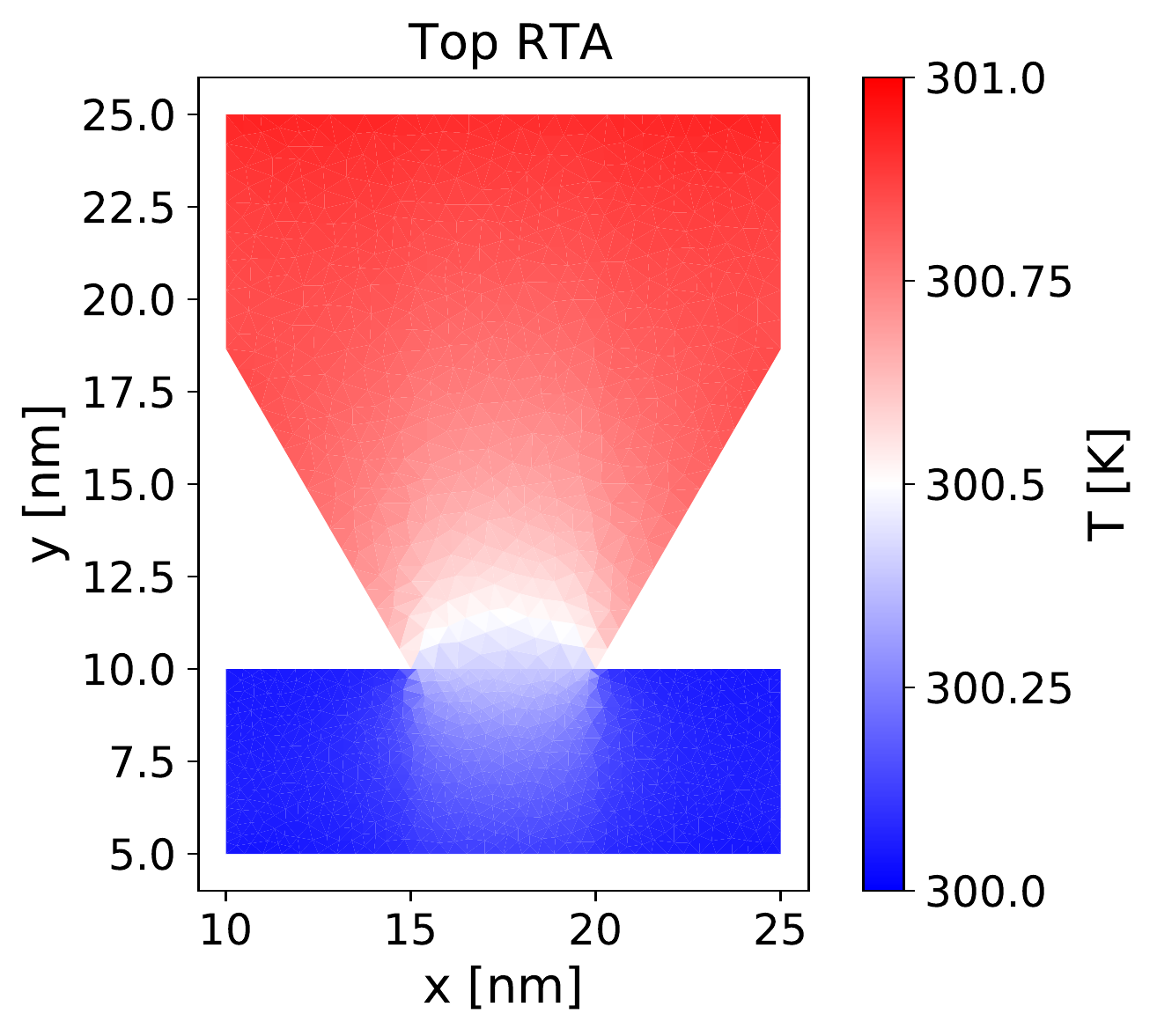}
		\label{CUNYAtop-T}
	\end{subfigure}
	\begin{subfigure}[b]{0.45\textwidth}
		\includegraphics[width=\textwidth]{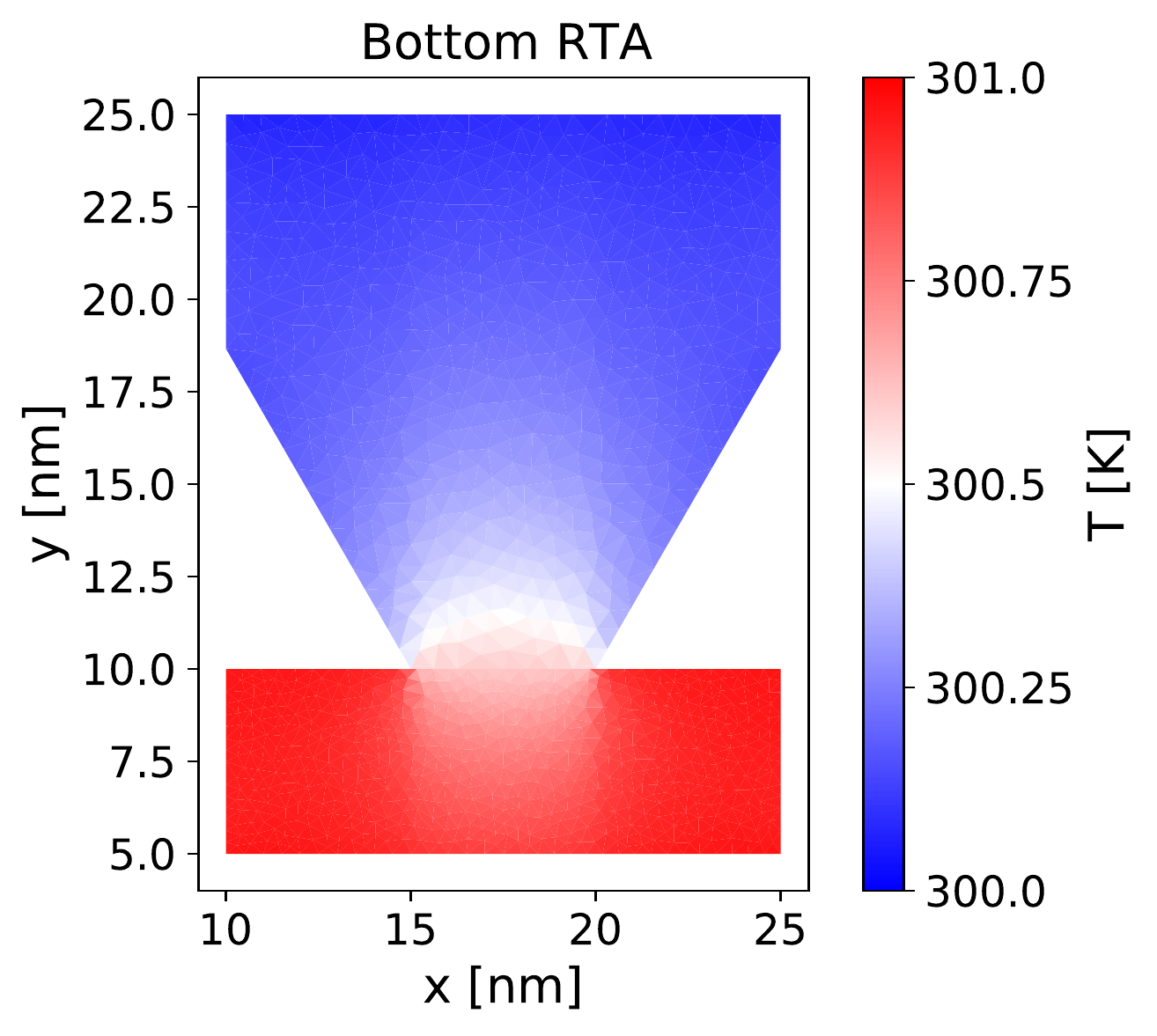}
		\label{CUNYAbot-T}
	\end{subfigure}
	\begin{subfigure}[b]{0.45\textwidth}
		\includegraphics[width=\textwidth]{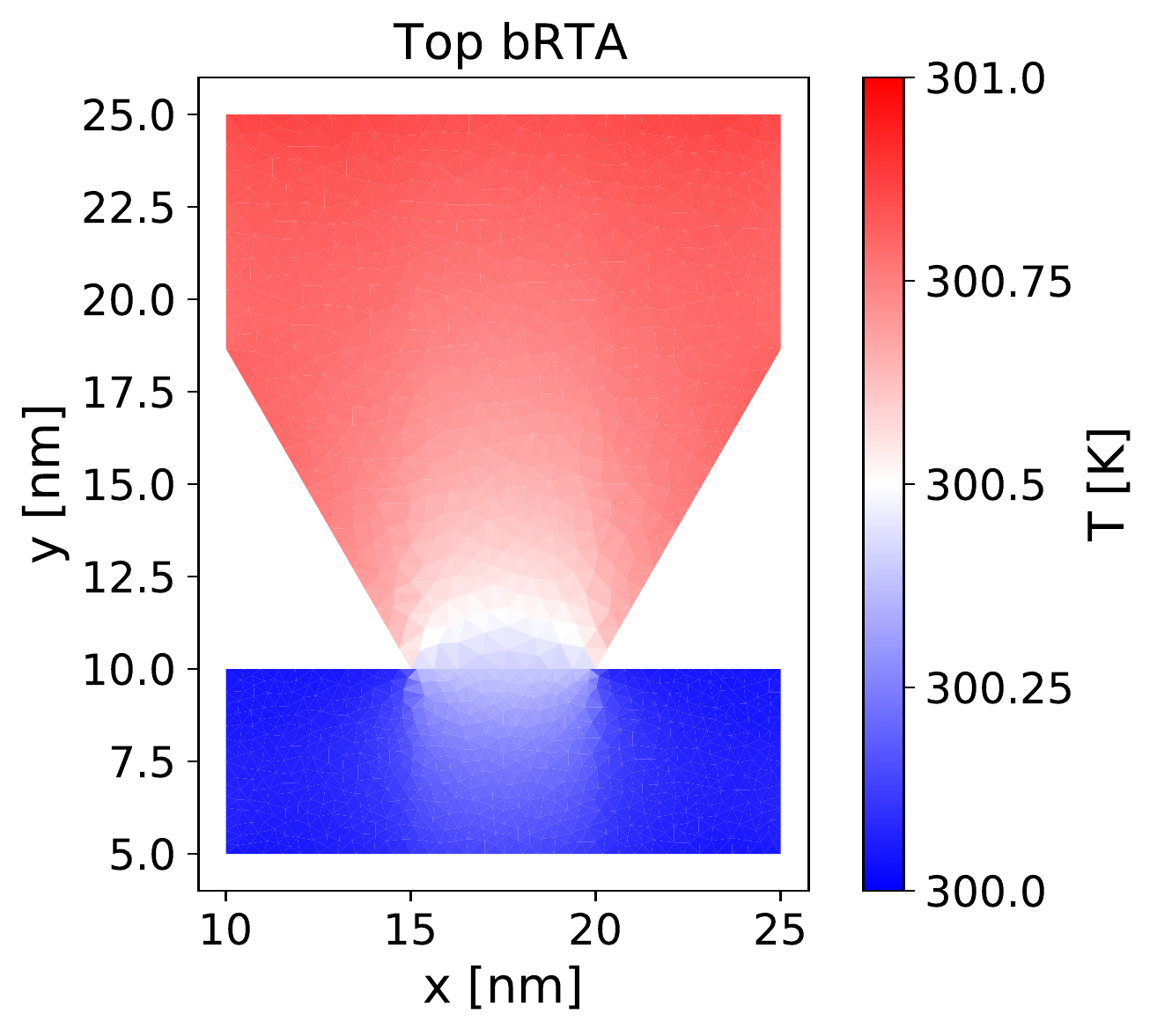}
		\label{CUNYAtop-T-b}
	\end{subfigure}
	\begin{subfigure}[b]{0.45\textwidth}
		\includegraphics[width=\textwidth]{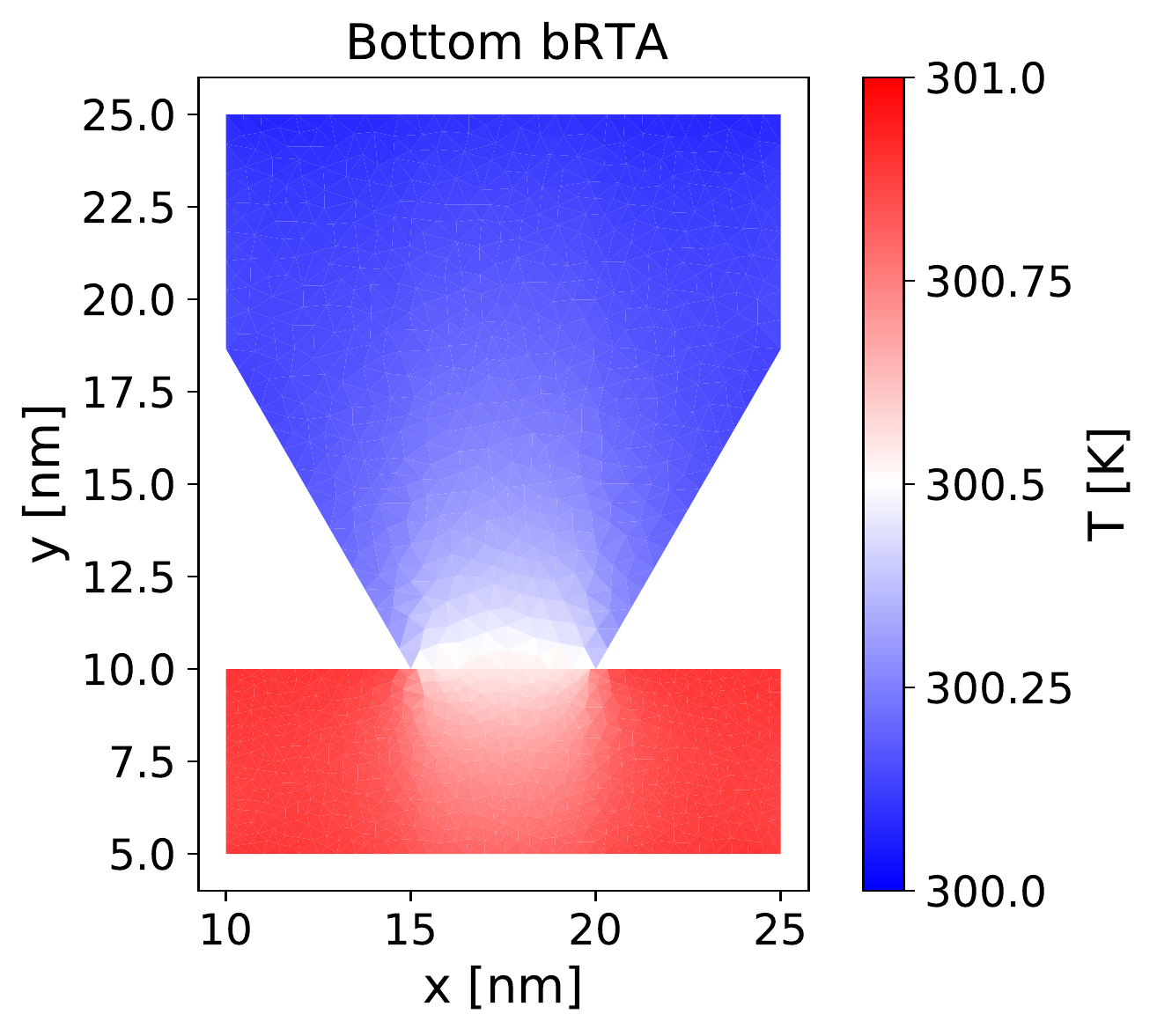}
		\label{CUNYAbot-T-b}
	\end{subfigure}
	\caption{RTA and beyond RTA temperature profiles for Fig.~\ref{CUNYAS} configurations at steady-state.}
	\label{CUNYAS-T}
\end{figure*}

\begin{figure*}
	\centering
	\begin{subfigure}[b]{0.45\textwidth}
		\includegraphics[width=\textwidth]{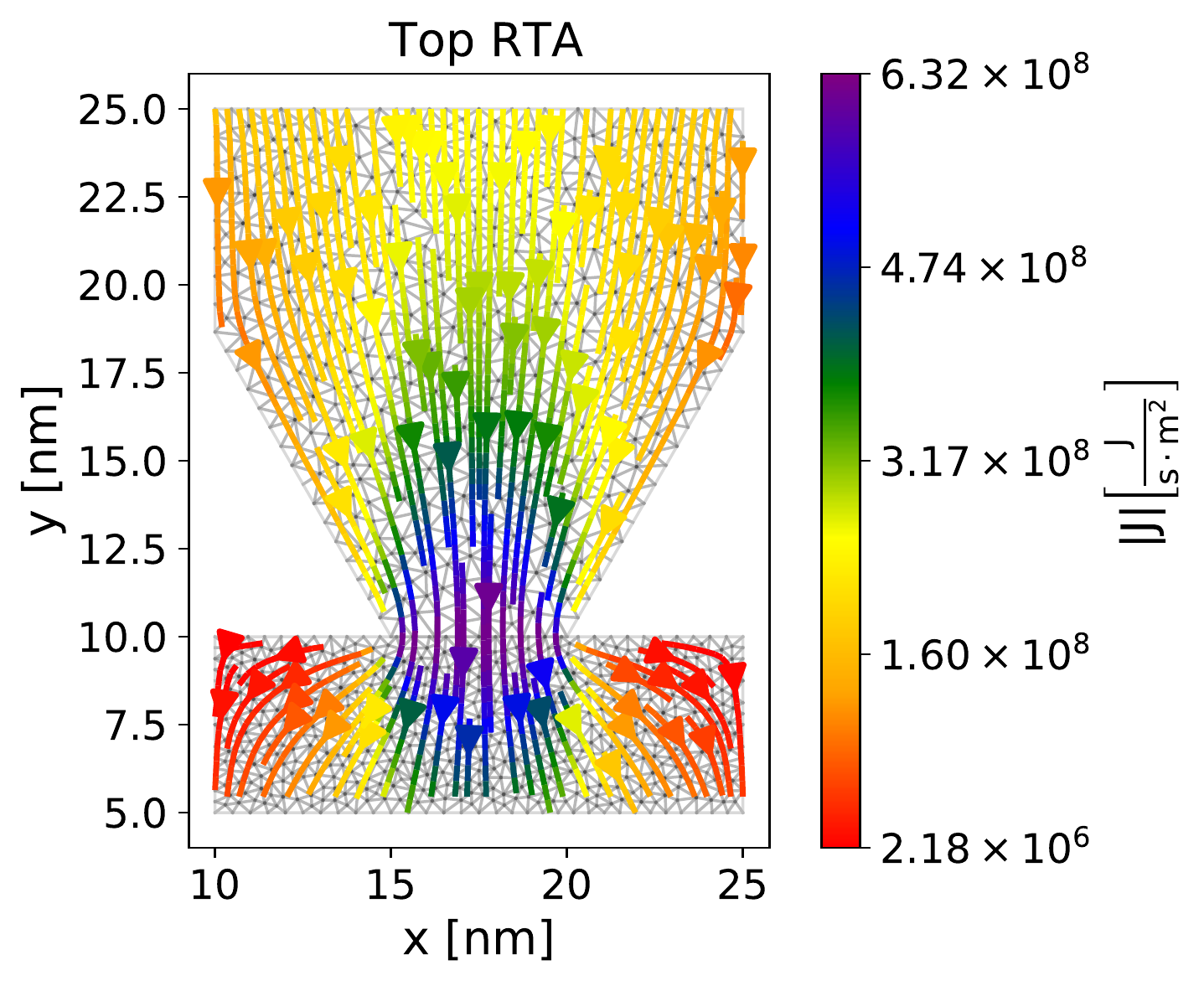}
		\label{CUNYAtop-J}
	\end{subfigure}
	\begin{subfigure}[b]{0.45\textwidth}
		\includegraphics[width=\textwidth]{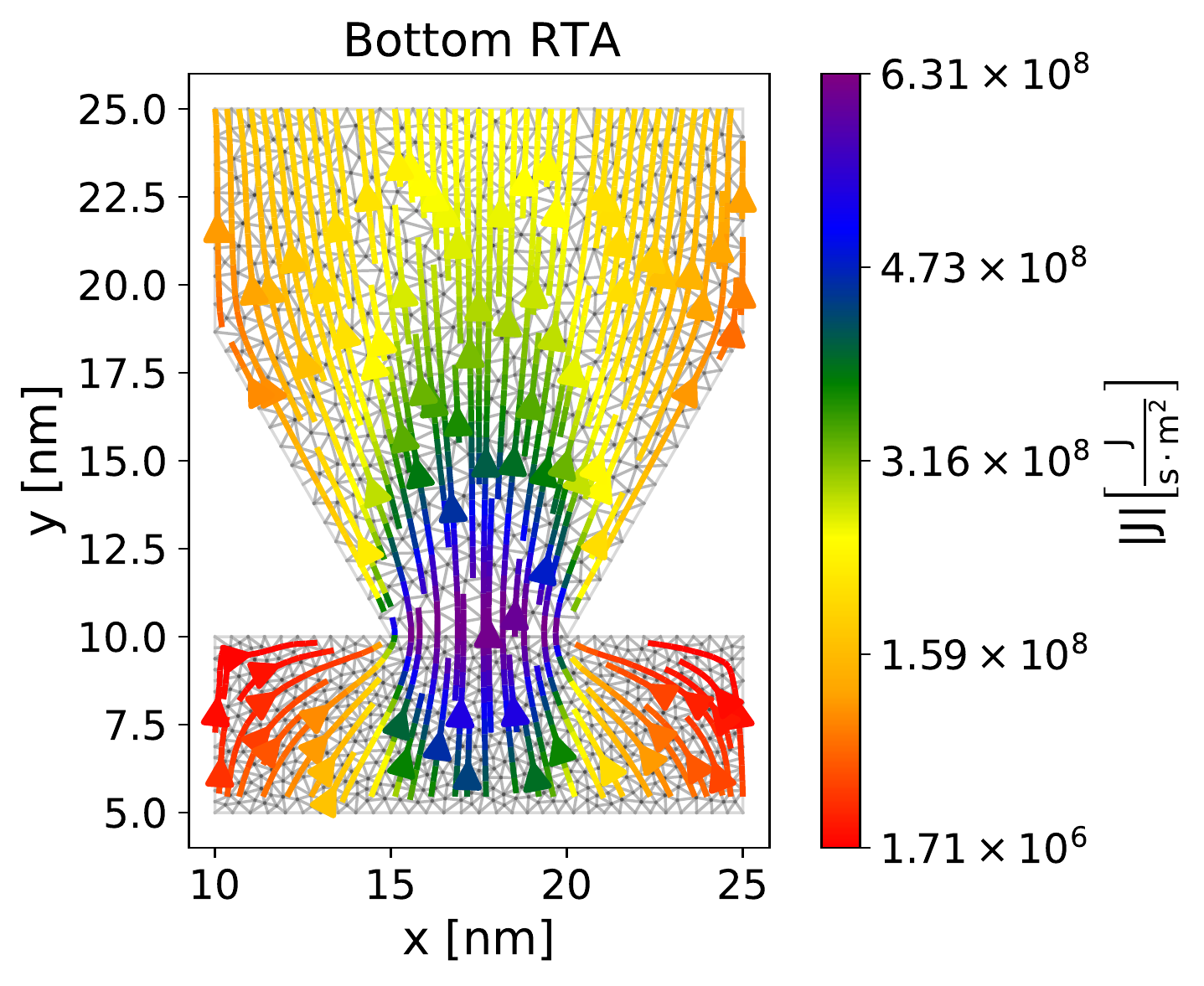}
		\label{CUNYAbot-J}
	\end{subfigure}
	\begin{subfigure}[b]{0.45\textwidth}
		\includegraphics[width=\textwidth]{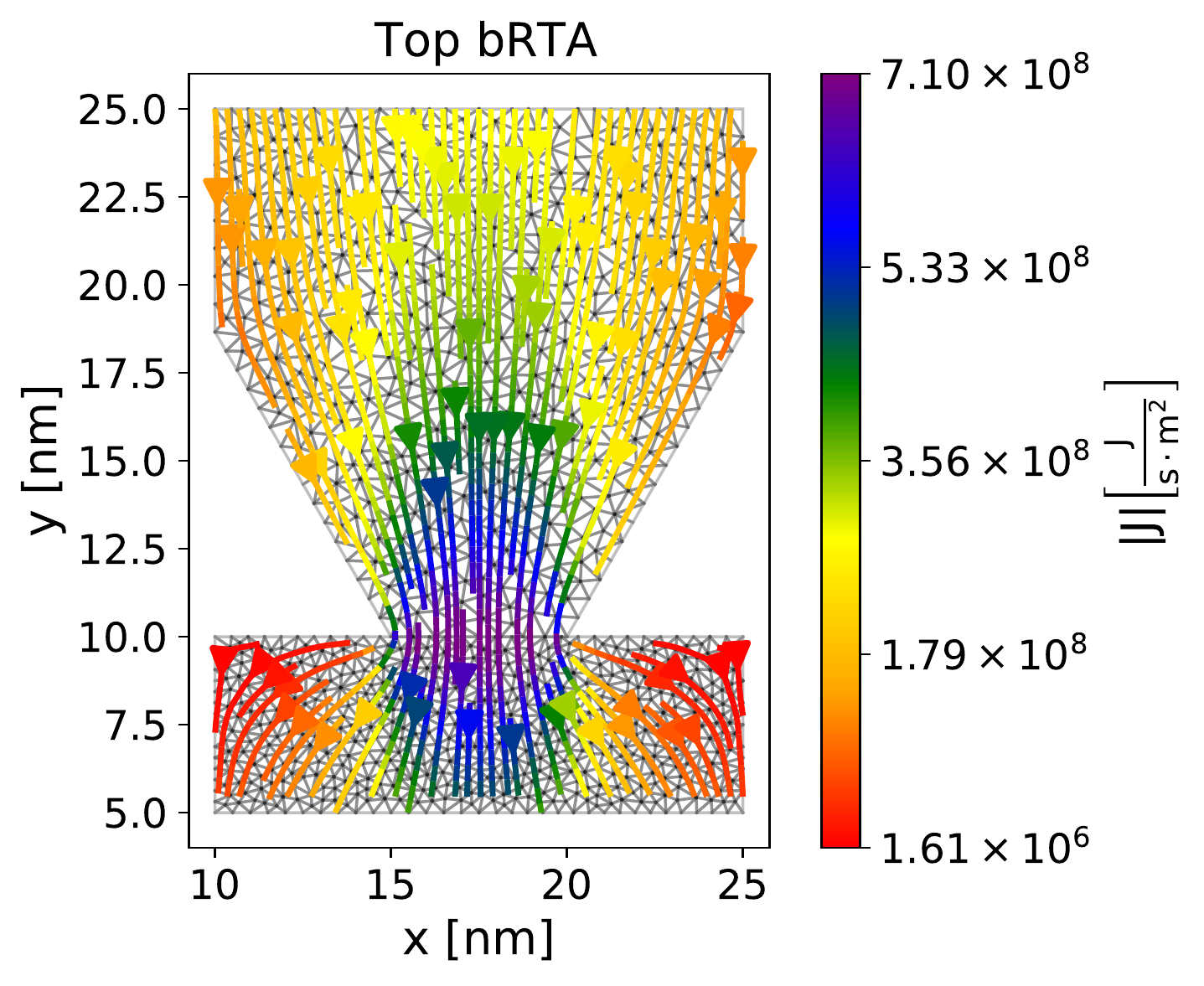}
		\label{CUNYAtop-J-b}
	\end{subfigure}
	\begin{subfigure}[b]{0.45\textwidth}
		\includegraphics[width=\textwidth]{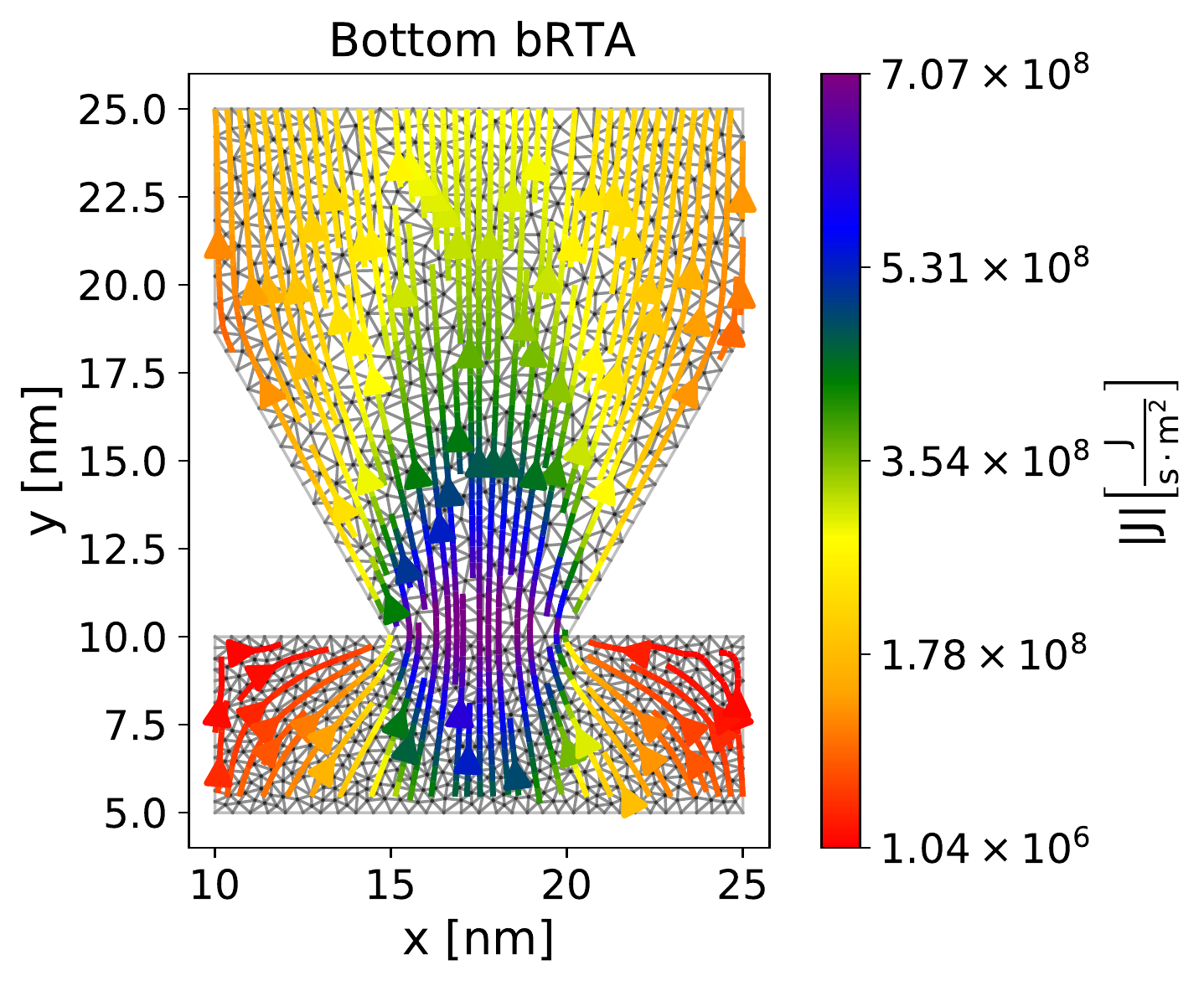}
		\label{CUNYAbot-J-b}
	\end{subfigure}
	\caption{Steady-state RTA and beyond-RTA heat fluxes for the configurations depicted in Fig.~\ref{CUNYAS}.}
	\label{CUNYAS-J}
\end{figure*}

Steady-state temperature profiles and heat fluxes for the wedge-like geometry within and beyond the RTA for two different configurations, depending on which terminal is put at \SI{301}{\kelvin} (the one at the top or the one at the bottom) while the other is kept at the reference temperature of \SI{300}{\kelvin}, can be seen at Figs.~\ref{CUNYAS-T} and~\ref{CUNYAS-J} respectively. Similar asymmetric devices are used as thermal rectifiers, but we cannot expect to detect rectification here because this model is based on the bulk spectrum and therefore does not account for phenomena such as device-reservoir interactions or size-dependent vibrational spectra~\cite{WangNL2014, LeeNL2012}. Moreover, it should be noted that thermal differences used here are too small for any sign of thermal rectification to be significant over statistical noise~\cite{WangNL2014}, or to activate the rectification mechanism based on different temperature-dependent behavior~\cite{WangNL2014,DamesJHT2009,CartoixaNL2015}. Higher thermal differences are however unattainable with the current implementation as the error introduced by the linearization of the collision operator would be too high.

\begin{figure}
	\centering
	\includegraphics[width=\linewidth]{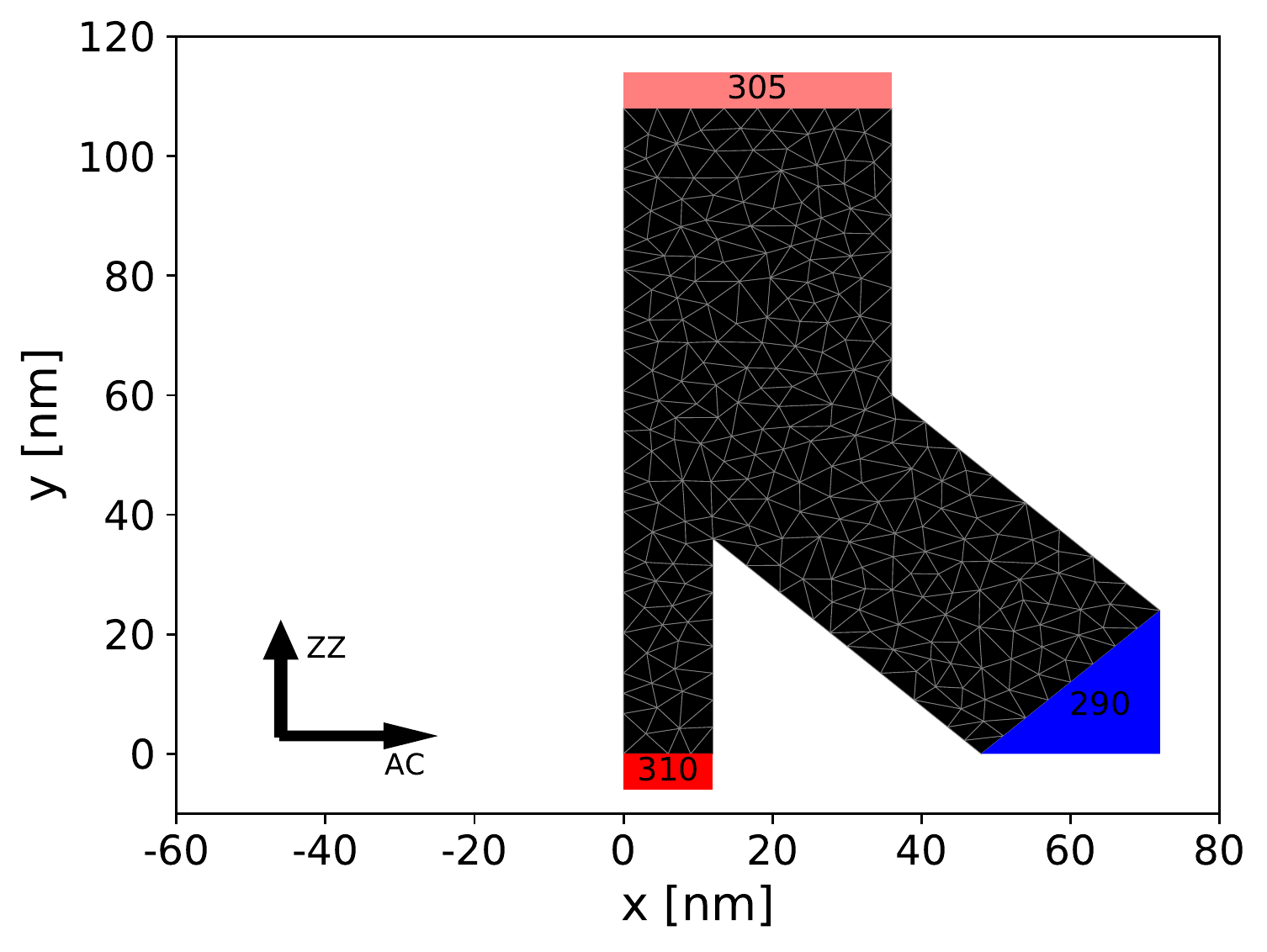}
	\caption{Example phosphorene structure with multiple terminals (isothermal reservoirs) at \SI{310}{K}, \SI{305}{K} and \SI{290}{K}.}
	\label{pipe}
\end{figure}

\begin{figure*}
	\begin{subfigure}[b]{0.45\textwidth}
		\centering
		\includegraphics[width=\textwidth]{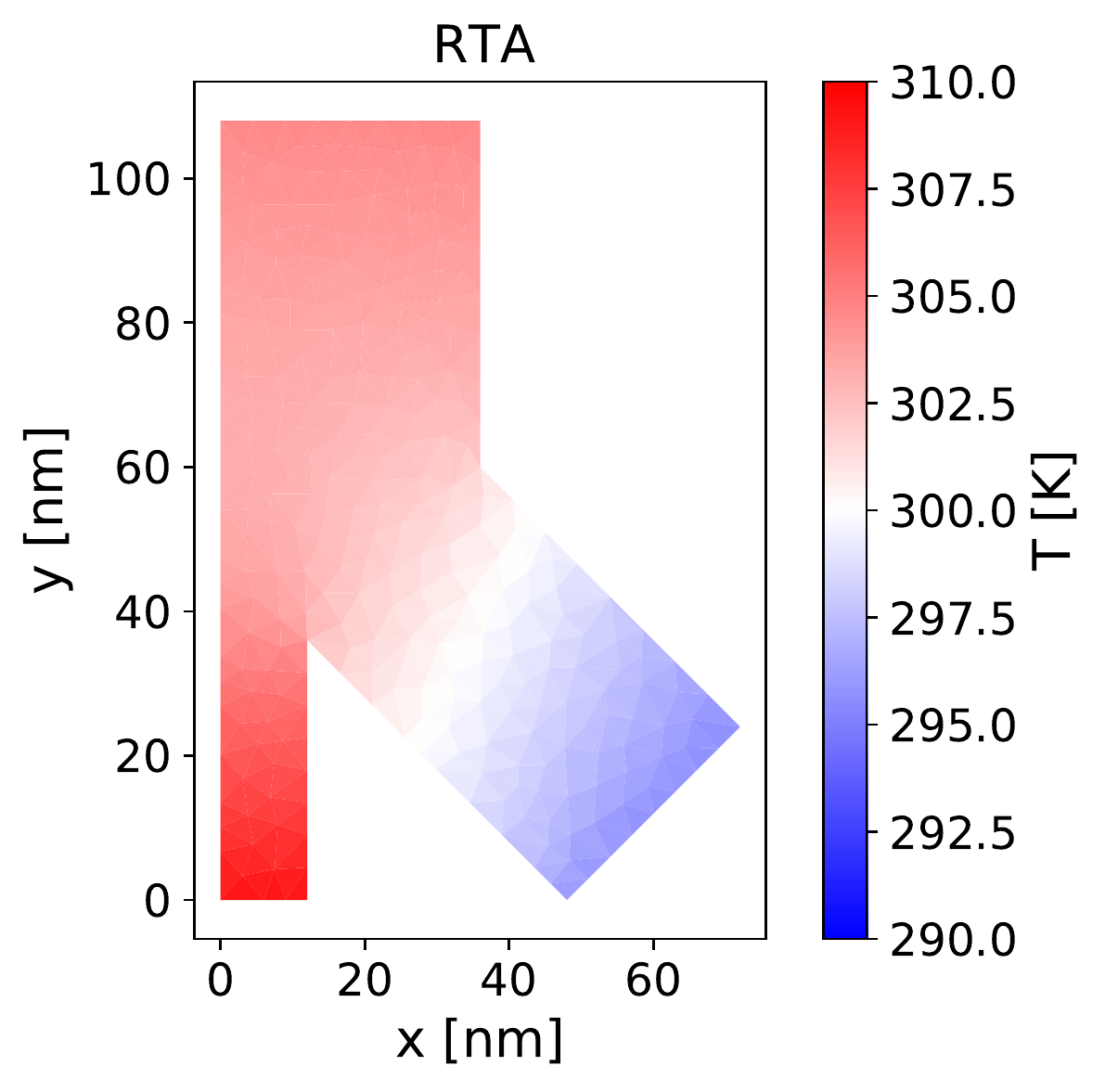}
		\label{pipe-T}
	\end{subfigure}
	\begin{subfigure}[b]{0.45\textwidth}
		\centering
		\includegraphics[width=\textwidth]{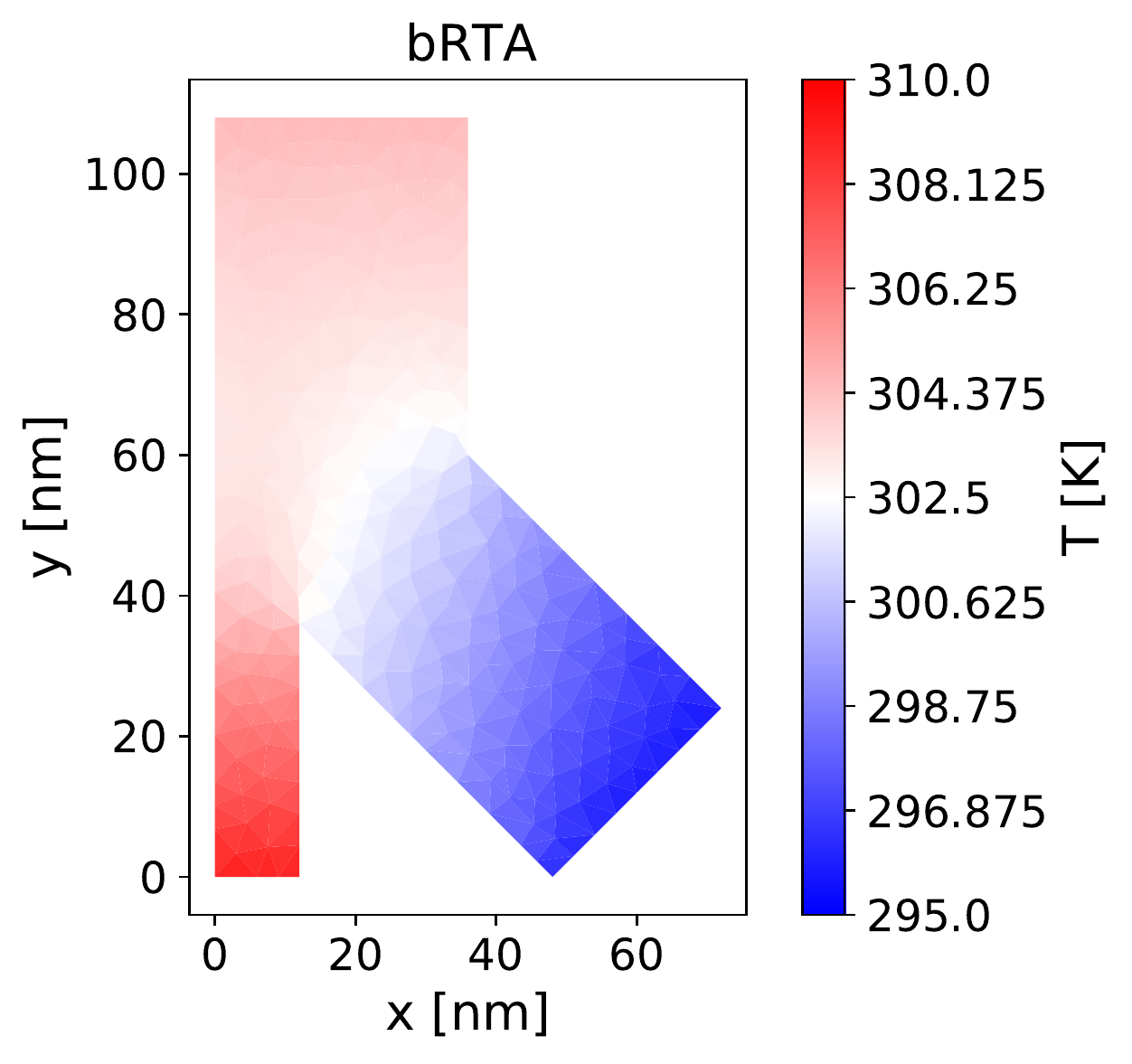}
		\label{pipe-T-b}
	\end{subfigure}
	\caption{Steady-state RTA and beyond-RTA temperature profiles for the configuration depicted in Fig.~\ref{pipe}.}
	\label{pipe-Ts}
\end{figure*}

\begin{figure*}
	\begin{subfigure}[b]{0.45\textwidth}
		\centering
		\includegraphics[width=\textwidth]{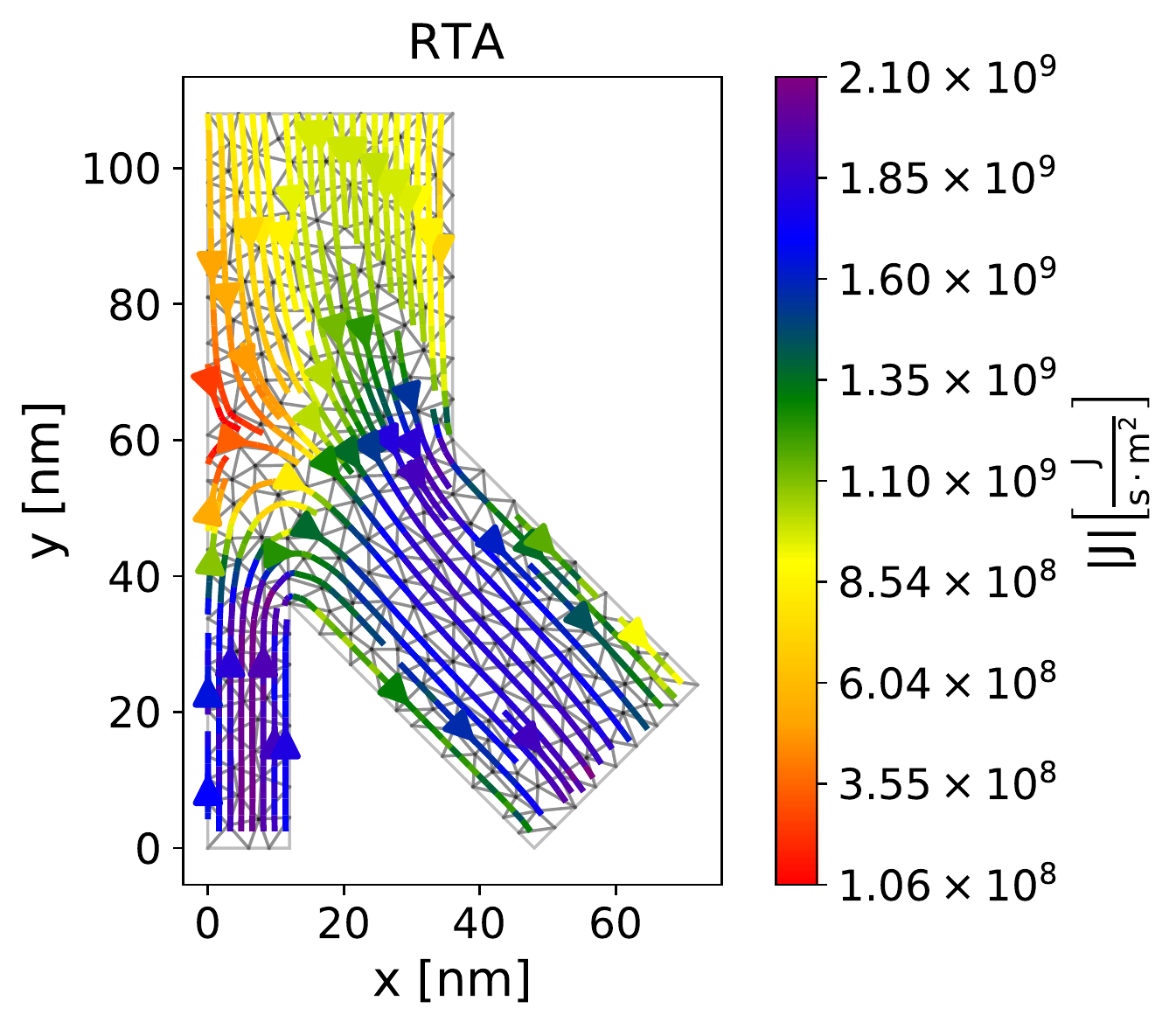}
		\label{pipe-J}
	\end{subfigure}
	\begin{subfigure}[b]{0.45\textwidth}
		\centering
		\includegraphics[width=\textwidth]{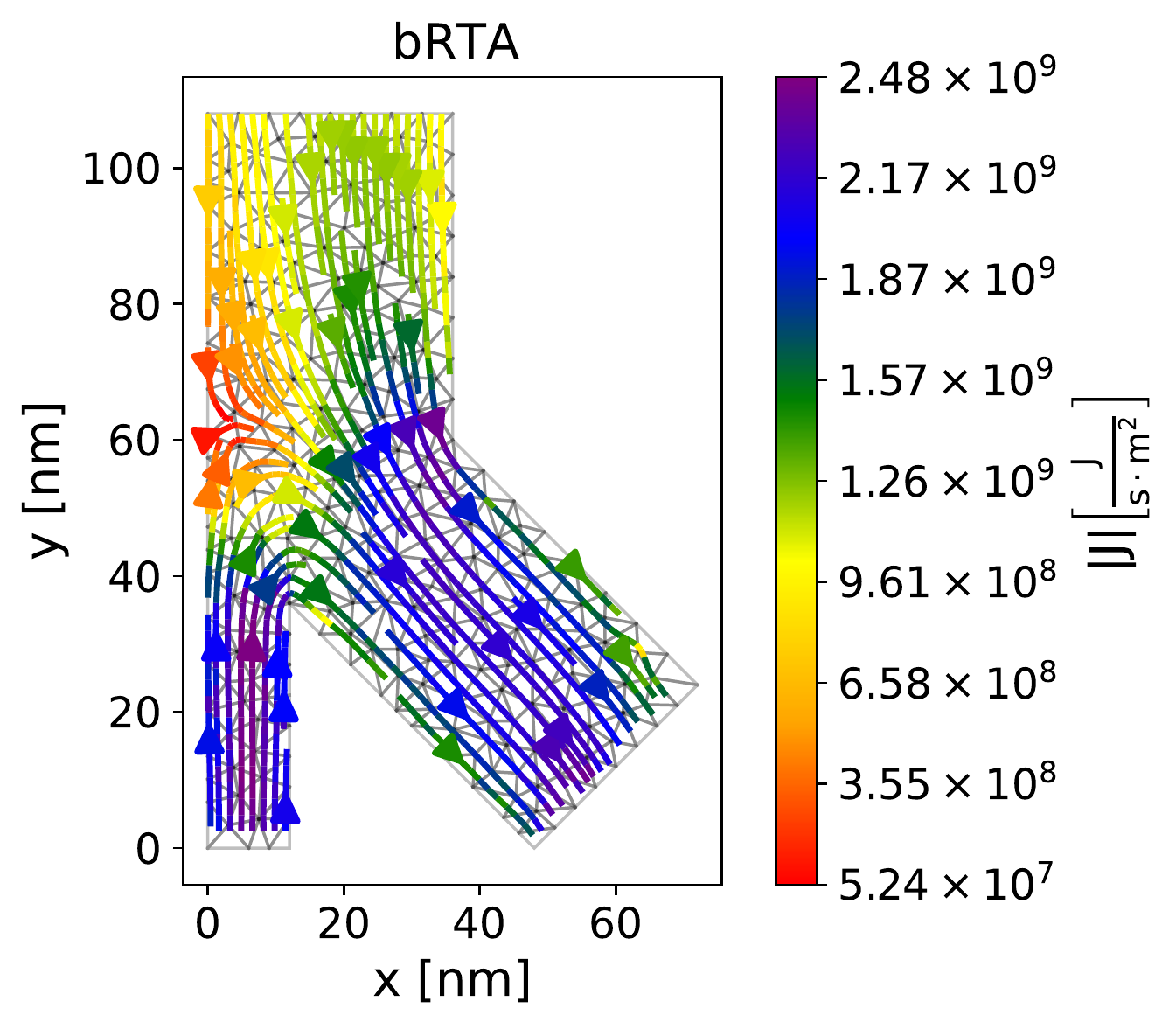}
		\label{pipe-J-b}
	\end{subfigure}
	\caption{Steady-state RTA and beyond-RTA heat fluxes for the configuration depicted in Fig.~\ref{pipe}.}
	\label{pipe-Js}
\end{figure*}

The heat fluxes and temperature profiles for the multiterminal structure (Fig.~\ref{pipe}) are presented in Figs.~\ref{pipe-Ts} and~\ref{pipe-Js}, showing the capability of our simulator to properly account for several sources/drains (isothermal reservoirs). Indeed, for the bRTA case, there is even an additional population of phonons due to the initial conditions, as the initial temperature profile was set to the RTA estimate in order to accelerate its convergence to the steady-state.

\subsection{\label{res:junction}Results: Example of material junction}

\begin{figure}
	\centering
	\includegraphics[width=\linewidth]{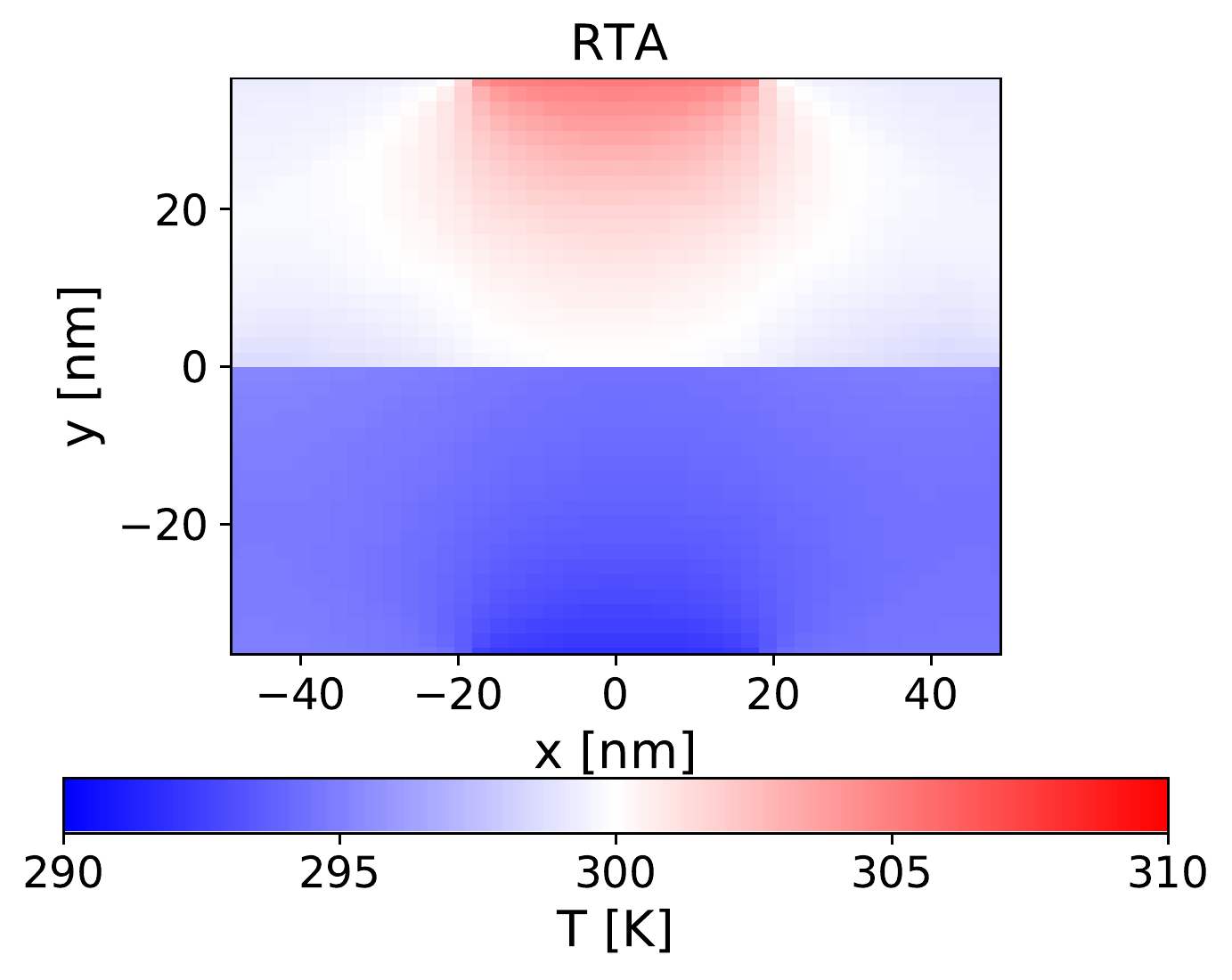}
	\caption{Steady-state RTA temperature profile for the configuration in Fig.~\ref{sandlevi}.}
	\label{sandleviT}
\end{figure}

\begin{figure}
	\centering
	\includegraphics[width=\linewidth]{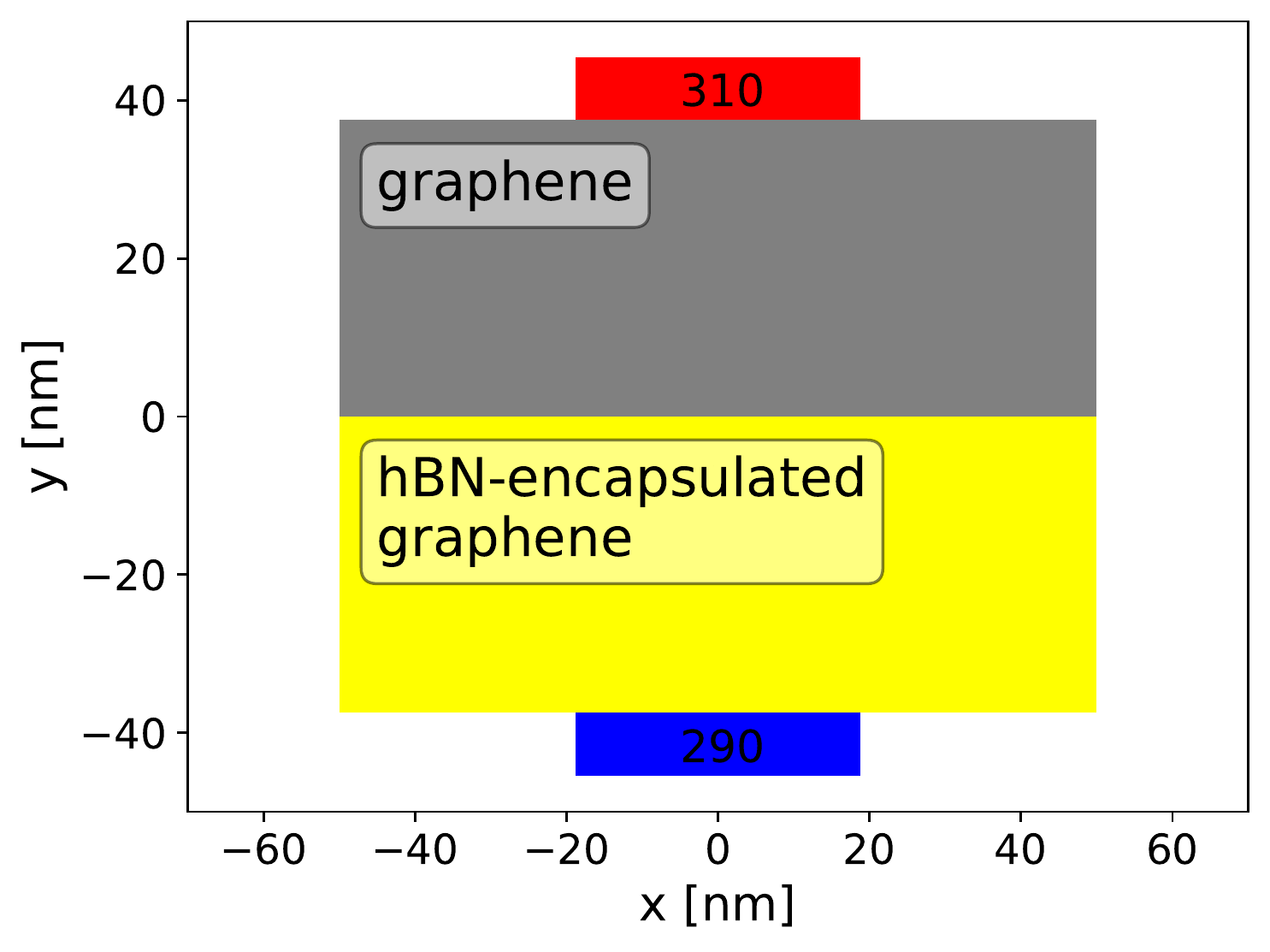}
	\caption{Structure containing a junction between graphene and hBN-encapsulated graphene and two isothermal boundaries at \SI{290}{\kelvin} (blue) and \SI{310}{\kelvin} (red).}
	\label{sandlevi}
\end{figure}

Finally, to show the capability of our improved RTA simulator to describe devices with different materials, we present the temperature profile (Fig.~\ref{sandleviT}) in a finite device structure composed by graphene on one half and h-BN encapsulated graphene on the other half (see Fig.~\ref{sandlevi}). The first-principles data needed for this calculation were obtained from Ref.~\citenum{almadatabase} in the case of graphene, where those properties had been calculated on a $80\times80\times1$ $q$-mesh with a broadening parameter of $1$ and a conventional thickness of \SI{0.345}{\nano\meter}~\cite{LandonJAP2014}. Regarding h-BN encapsulated graphene, the required first-principles data were obtained using density functional theory using Perdew-Burke-Ernzerhof functional~\cite{PBE} plus the D3~\cite{D3_1,D3_2} correction to energy due to van der Waals interactions between layers as implemented in VASP~\cite{VASP_1,VASP_2,VASP_3} with a $\Gamma$-centered $k$-mesh of $7\times7\times1$ points for minimization, and using Phonopy~\cite{phonopy} and thirdorder.py~\cite{ShengBTE} with a supercell of $7\times7\times1$ to obtain second and third order interatomic force constants, respectively. The phonon properties of h-BN-encapsulated graphene were calculated using a $40\times40\times1$  $\Gamma$-centered $q$-mesh with a broadening parameter of $1$ and setting the stack thickness to \SI{1.001}{\nano\meter}~\cite{LandonJAP2014,PeaseAC1952,GollaAPL2013}. 

\section{\label{sec:Conclusions}Conclusions}
In this work we have presented \texttt{BTE-Barna}, a software package that extends the \texttt{almaBTE} package to calculate the thermal properties of devices and systems based on 2D materials. We have showcased the new capabilities with an extensive set of tests and examples. For instance, the package was used to highlight the differences in the heat flux profile for the case of Poiseuille flow in a nanoribbon, for the case of RTA and beyond the RTA. Amid all new features the most relevant are:
\begin{enumerate}
	\item The iterative solver has been extended to provide the effective conductivity for nanoribbons (and nanowires in the case of 3D materials).
	\item The RTA Monte Carlo simulator was updated so that now it can address finite and/or periodic 2D systems under the effect of thermal gradients and isothermal reservoirs. Moreover, it now provides information for transient to steady-state for finite systems.
	\item A beyond-RTA Monte Carlo simulator for 2D systems was implemented to provide an accurate description for those cases in which RTA fails.
\end{enumerate}
This package is published as \texttt{almaBTE}'s fork and is freely available to download at \url{https://github.com/sousaw/BTE-Barna}. Overall, we expect \texttt{BTE-Barna} to provide a new set of tools for design and prediction of thermal transport/management in 2D based devices and systems.

\section*{Acknowledgements}
M.\,R.-M. and X.\,C. acknowledge financial support by Spain's Ministerio de Ciencia, Innovaci\'on y Universidades under Grant No.\,RTI2018-097876-B-C21 (MCIU/AEI/FEDER, UE), and the EU Horizon2020
research and innovation program under grant GrapheneCore3 881603.
M.R.-M. acknowledges financial support by the Ministerio de Educaci\'on, Cultura y Deporte programme of Formaci\'on de Profesorado Universitario under Grant No. FPU2016/02565 and Ministerio de Universidades under grant EST19/00655, as well as the kind hospitality of the Institute of Materials Chemistry, TU Wien.

\appendix

\section{\label{inputNANOS}Effective thermal conductivity in simple nanosystems: inputs, outputs and executable}
The calculation of the effective thermal conductivity $\kappa_{\mathrm{nano}}$ for nanowires and nanoribbons at both the RTA and beyond RTA levels is implemented in the \texttt{kappa\_Tsweep\_nanos} executable. The calculation is performed as in  \texttt{kappa\_Tsweep}~\cite{almaBTE}, but using $\tau_{\lambda}^{\text{nano}}$ in place of $\tau_{\lambda}^{0}$. Because of boundaries breaking crystal symmetry (i.e.: $S_\lambda^{\text{nano}}$ and consequently $\tau_{\lambda}^{\text{nano}}$ not possessing crystal symmetry), one needs to solve the linear system in the full Brillouin Zone; hence, \texttt{kappa\_Tsweep\_nanos} first recalculates and symmetrizes all bulk processes in the whole $q$-mesh, and if beyond RTA is required it solves the linear system using sparse matrices.
The executable parameters are the same as for \texttt{kappa\_Tsweep} but with an additional parameter to control the number of TBB threads in which the recalculation and symmetrization will take place.
\begin{alltt}
	kappa_Tsweep_nanos input.xml N\(\mathrm{\sb{threads}}\)
\end{alltt}
Regarding input files, it uses the same format as \texttt{kappa\_Tsweep} but with some additions:
\lstset{language=XML}
\begin{lstlisting}
<system name="nanoribbon" L="500.0"/>
<AnharmonicIFC name="FORCE_CONSTANTS_3RD" 
scalebroad="1.0"/>
\end{lstlisting}
\begin{enumerate}
	\item system/name [string]: Indicates the type of nanosystem. Accepted values: \verb|nanowires| and \verb|nanoribbons|.
	\item system/L [double,units(nm)]: limiting length of the system, radius for nanowires and width for nanoribbons.
	\item AnharmonicIFC/name [string]: path to file containing third-order force constants. Only needed for beyond calculations.
	\item AnharmonicIFC/scalebroad [double]: broadening parameter for adaptative smearing. Only needed for beyond-RTA calculations.
\end{enumerate}
Finally \texttt{kappa\_Tsweep\_nanos} will produce a csv output file (\verb|{h5filename}_{systemname}_|\\\verb|L_{L}_{u}_{Tmin}_{Tmax}.Tsweep|) with the following format
\begin{equation*}
\begin{matrix}
\begin{aligned}
T[K],   & ~~\kappa_\text{nano}^\mathrm{RTA}\left[\mathrm{\frac{W}{m\cdot K}}\right], & ~~\kappa_\text{nano}^\mathrm{beyond\;RTA}\left[\mathrm{\frac{W}{m\cdot K}}\right] \\
T_{0},  & ~~\kappa_\text{nano}^\mathrm{RTA}(T_{0}),                                  & ~~\kappa_\text{nano}^\mathrm{beyondRTA}(T_{0})                                    \\
T_{1},  & ~~\kappa_\text{nano}^\mathrm{RTA}(T_{1}),                                  & ~~\kappa_\text{nano}^\mathrm{beyondRTA}(T_{1})                                    \\
\vdots, & ~~\vdots,                                                           & ~~\vdots                                                                   \\
T_{n},  & ~~\kappa_\text{nano}^\mathrm{RTA}(T_{n}),                                  & ~~\kappa_\text{nano}^\mathrm{beyondRTA}(T_{n})                                    \\
\end{aligned}
\end{matrix}
\end{equation*}
where ${u}$ is the transport direction.

\section{\label{inputfiles}Monte Carlo simulators and analyzers: inputs, outputs and executables}
\subsection{Geometry files}
Files containing geometries are given in XML format. We provide a toy example to show geometry file input format:
\lstset{language=XML}
\begin{lstlisting}
<Geometry>
	<number_of_boxes Ngeom="4"/>
	<Box>
		<MaterialID name="black_P"/>
		<boxid id="0"/>
		<Vertices dim="2" npoints="4">
			6.0  6.0
			6.0  7.0
			7.0  6.0
			7.0  7.0
		</Vertices>
		<initCnd Teq="300." 
		  Tinit="301.0"/>
	</Box>
	<Box>
		<MaterialID name="black_P"/>
		<boxid id="1"/>
		<Vertices dim="2" npoints="4">
			6.0  6.0
			6.0  7.0
			5.0  6.0
			5.0  7.0
		</Vertices>
		<initCnd Teq="300."/>
		<Translate_to id="2">
			1.0 0.0 0.0
		</Translate_to>
	</Box>
	<Box>
		<MaterialID name="black_P"/>
		<boxid id="2"/>
		<Vertices dim="2" npoints="4">
			8.0  6.0
			8.0  7.0
			7.0  6.0
			7.0  7.0
		</Vertices>
		<initCnd Teq="300."/>
		<Translate_to id="1">
			-1.0 0.0 0.0
		</Translate_to>
	</Box>
	<Box>
		<MaterialID name="black_P"/>
		<boxid id="3"/>
		<Vertices dim="2" npoints="4">
			5.0  6.0
			5.0  5.0
			8.0  6.0
			8.0  5.0
		</Vertices>
		<initCnd Teq="303.0"/>
		<Reservoir/>
	</Box>
</Geometry>
\end{lstlisting}

We now list all possible parameters in geometry files:
\begin{enumerate}
	\item number\_of\_boxes/Ngeom [integer]: Indicates the number of boxes.
	\item Box/MaterialID/name [string]: Box material
	\item Box/Vertices/dim [integer]: system dimension (Accepted values: 2).
	\item Box/Vertices/npoints [integer]: number of vertices.
	\item Box/Vertices [double, array(dim,npoints),units(nm)]: Box vertices (NOTE: they need to form a convex hull).
	\item Box/initCnd/Teq [double,units(K)]: Reference temperature.
	\item Box/initCnd/Tinit [double,optional,units(K)]:  Temperature to initialize the box population out of reference (default: Teq). RTA simulator ignores this.
	\item Box/Reservoir : if present it indicates that the box is an isothermal reservoir.
	\item Box/Translate\_to/id [int]: It indicates that any particle entering that box is translated to ``id" box.
	\item Box/Translate\_to [double,array(1,3),units(nm)]: Translation vector applied to any particle entering that box.
\end{enumerate}

\subsection{\label{exec:MCRTA2D}RTAMC2D}
The \verb|RTAMC2D| executable implements the general RTA algorithm of Sec.~\ref{sec:Methods_RTA} and the specialization described in Sec.~\ref{RandrianalisoaAlgo} to obtain the steady-state for periodic structures under a thermal gradient. It has as command line inputs:
\begin{alltt}
	RTAMC2D input.xml N\(\mathrm{\sb{threads}}\) [N\(\mathrm{\sb{runs}}\)]
\end{alltt}
\begin{enumerate}
	\item input.xml [string]: xml file containing the input.
	\item $\mathrm{N_{threads}}$ [integer]: number of TBB threads for simulation.
	\item $\mathrm{N_{runs}}$ [integer,optional]: number of repetitions to be done, to prevent table creation and data loading (default: 1).
\end{enumerate}
\subsubsection{input.xml}
We provide now another toy example of input.xml files together with an explanation of its variables:
\lstset{language=XML}
\begin{lstlisting}
<RTA_MC2d>
	<geometry file="grta01.xml"/>
	<gradient x="0.2" y="0.0"/>
	<convergence energy="-1.0" flux="-1.0"/>
	<time dt="0.5" maxtime="100000"/>
	<material name="black_P" 
	  database="black_monolayer_50_50_1.h5" 
	  thickness="0.5" T0="300.0"/>
		<material name="biblack_P" 
		  database="biblack_50_50_1.h5" 
		  thickness="1.0" T0="300.0"/>
	<layer material="black_P" 
	  layer_name="black_bare" 
	  atoms="0 1 2 3"/>
	<layer material="biblack_P" 
	  layer_name="black_mid" 
	  atoms="0 1 2 3"/>
	<layer material="biblack_P" 
	  layer_name="black_alone" 
	  atoms="4 5 6 7"/>
	<layers_connection layer_A="black_mid" 
	  layer_B="black_bare"/>
	<particles N="10000000"/>
	<ballistic/>
	<spectral>
		<resolution ticks="500">
		<location   bin="0">
	<spectral/>
</RTA_MC2d>
\end{lstlisting}
\begin{enumerate}
	\item geometry\_file [string]: XML file containing geometry data.
	\item gradient/x [double,units(K/nm)]: x-component of homogeneous thermal gradient applied to all boxes (NOTE: requires y-component and activates the specialized algorithm to obtain steady-state for extended systems).
	\item gradient/y [double,units(K/nm)]: y-component of homogeneous thermal gradient applied to all boxes (NOTE: requires x-component and activates the specialized algorithm to obtain steady-state for extended systems).
	\item convergence/energy [double,optional]: if given, the loop of the specialized algorithm to obtain steady-state for extended systems with an applied thermal gradient is broken when relative difference in deviational energy density is lower than this threshold (default: -1.0).
	\item convergence/flux [double,optional]: if given, the loop of the specialized algorithm to obtain steady-state for extended systems with an applied thermal gradient is broken when relative differences of heat fluxes are lower than this threshold (default: -1.0).
	\item time/dt [double,units(ps)]: spacing for mesh in time.
	\item time/maxtime [double,units(ps)]: Maximum time for particles.
	\item material/name [string]: material id.
	\item material/database [string]: path to \texttt{almaBTE}'s database containing phonon properties.
	\item material/thickness [double,units(nm)]: real thickness in z-direction, used to correct DFT lattice vector in z-direction.
	\item material/T0 [double,units(K)]: reference temperature of whole system (NOTE: it must be the same for all materials).
	\item particles/N [integer]: Number of particles to be simulated (NOTE: be aware that even numbers are required for specialized algorithm to obtain steady-state for extended systems with an applied thermal gradient in order for sources not to add energy into the system).
	\item spectral/resolution/ticks [integer,optional]: Number of divisions for spectral decomposition of fluxes and deviational temperature.
	\item spectral/location/bin [integer,optional]: Box id in which spectral decomposition is calculated.
	\item layer/material [string,optional]: name of material in which layer is localized. This is for LDMM.
	\item layer/layer\_name [string,optional]: name of layer. This is for LDMM.
	\item layer/atoms [int,optional]: identity of atoms in the layer (follows the order of POSCAR). This is for LDMM.
	\item layers\_connection/layer\_A [string,optional]: name of layer connected with layer\_B (of same entry). There can be multiple entries of layer\_connection if more than one connection is present. This is for LDMM.
	\item layers\_connection/layer\_B [string,optional]: name of layer connected with layer\_A (of same entry). This is for LDMM.
	\item ballistic [optional]: if given, intrinsic scattering is deactivated.
\end{enumerate}
\subsubsection{Output files}
In the case of specialized algorithm to obtain steady-state for periodic structures under thermal gradient a file called \verb|steady_state_T0K_run_irun.csv|, where $T0$ is the reference temperature and $irun$ is the simulation id within the $N_{runs}$. This file contains in csv format the following:
\begin{equation*}
\begin{matrix}
\begin{aligned}
T_{0},   & ~~T_{1},   & \dots, & ~~T_{Nboxes}   \\
J_{x,0}, & ~~J_{x,1}, & \dots, & ~~J_{x,Nboxes} \\
J_{y,0}, & ~~J_{y,1}, & \dots, & ~~J_{y,Nboxes} \\
\end{aligned}
\end{matrix}
\end{equation*}
Additionally, if spectral decomposition is activated four files will be generated per selected box:
\begin{enumerate}
	\item \verb|steady_deltaT_omega_ibox_T0K_run_irun.csv|: contains the frequency grid in the first column and the deviational temperature per frequency in second one.
	\begin{equation*}
	\begin{matrix}
	\omega_{0},   & ~~\omega_{1},   & \dots, & ~~\omega_{Nticks}   \\
	\Delta T_{0}, & ~~\Delta T_{1}, & \dots, & ~~\Delta T_{Nticks} \\
	\end{matrix}
	\end{equation*}
	\item \verb|steady_jx_omega_ibox_T0K_run_irun.csv|: analogous to temperature file, but with heat flux (\SI{}{\J\per\square\meter\per\second}) in x-direction.
	\item \verb|steady_jy_omega_ibox_T0K_run_irun.csv|: analogous to temperature file, but with heat flux (\SI{}{\J\per\square\meter\per\second}) in y-direction.
	\item \verb|steady_fd_q_ibox_T0K_run_irun.csv|: contains the $q_x$ and $q_y$ grid in the first and second rows, respectively, and the deviational phonon population per $q$-point in the following one.
	\begin{equation*}
	\begin{matrix}
	\begin{aligned}
	\#q_{x,0},        & ~~q_{x,1},        & ~~\dots, & ~~q_{x,N_q-1}        \\
	\#q_{y,0},        & ~~q_{y,1},        & ~~\dots, & ~~q_{y,N_q-1}        \\ 
	f^d(q_0), &~~f^d(q_1), &~~\dots, & ~~f^d(q_{N_q-1}) \\
	\end{aligned}
	\end{matrix}
	\end{equation*}
\end{enumerate}
On the other hand, the general algorithm produces three output files per run:
\begin{enumerate}
	\item \verb|temperature_T0K_run_irun.csv|: contains the temperature in \SI{}{\kelvin} per boxes (column) at given times (rows). First column of each row gives the midle point of the time bin.
	\item \verb|jxT0K_run_irun.csv|: analogous to temperature file, but with heat flux (\SI{}{\J\per\square\meter\per\second}) in x-direction.
	\item \verb|jyT0K_run_irun.csv|: analogous to temperature file, but with heat flux (\SI{}{\J\per\square\meter\per\second}) in y-direction.
\end{enumerate}
plus four additional files per selected box if spectral decomposition is conducted:
\begin{enumerate}
	\item \verb|deltaT_omega_ibox_T0K_run_irun.csv|: contains the frequency grid in the first column and the deviational temperature per frequency in at each time step in the following ones.
	\begin{equation*}
	\begin{matrix}
	-1,    & ~~\omega_{0},        & ~~\omega_{1},        & ~~\dots, & ~~\omega_{Nticks}        \\
	t_0,   & ~~\Delta T_{0}(t_0), & ~~\Delta T_{1}(t_0), & ~~\dots, & ~~\Delta T_{Nticks}(t_0) \\
	t_1,   & ~~\Delta T_{0}(t_1), & ~~\Delta T_{1}(t_1), & ~~\dots, & ~~\Delta T_{Nticks}(t_1) \\
	\vdots & ~~\vdots             & ~~\vdots             & ~~\ddots & ~~\vdots                 \\
	t_f,   & ~~\Delta T_{0}(t_f), & ~~\Delta T_{1}(t_f), & ~~\dots, & ~~\Delta T_{Nticks}(t_f) \\
	\end{matrix}
	\end{equation*}
	where $t_i$ is the middle point of each time grid bin.
	\item \verb|jx_omega_ibox_T0K_run_irun.csv|: analogous to temperature file, but with heat flux (\SI{}{\J\per\square\meter\per\second}) in x-direction.
	\item \verb|jy_omega_ibox_T0K_run_irun.csv|: analogous to temperature file, but with heat flux (\SI{}{\J\per\square\meter\per\second}) in y-direction.
	\item \verb|fd_q_ibox_T0K_run_irun.csv|: contains the $q_x$ and $q_y$ grid in the first and second rows, respectively, and the deviational phonon population per $q$-point at each time step in the following ones.
	\begin{equation*}
	\begin{matrix}
	\#q_{x,0},        & ~~q_{x,1},        & ~~\dots, & ~~q_{x,N_q-1}        \\
	\#q_{y,0},        & ~~q_{y,1},        & ~~\dots, & ~~q_{y,N_q-1}        \\ \\
	t_1,   &f^d(q_0,t_1), &\dots, & f^d(q_{N_q-1},t_1) \\
	\vdots  &\vdots & \ddots & \vdots  \\
	t_f,   & f^d(q_0,t_f), &\dots, & f^d(q_{N_q-1},t_f) \\
	\end{matrix}
	\end{equation*}
\end{enumerate}

\subsection{PropagatorBuilder}
The \verb|PropagatorBuilder| executable implements the building of the $B$-matrix and subsequent $P(\Delta t)$ calculation using the Krylov subspace method. It has as command line inputs:
\begin{alltt}
	[mpi] PropagatorBuilder database.h5 IFC3 T \\ \(\verb|     |\) \(\Delta\)t N\(\mathrm{\sb{threads}}\)
\end{alltt}
\begin{enumerate}
	\item database.h5 [string]: \texttt{almaBTE}'s database for that material.
	\item IFC3 [string]: third-order force constants file in sparse format~\cite{ShengBTE}.
	\item T [double,units(K)]: reference temperature.
	\item $\mathrm{\Delta t}$ [double, units(ps)]: time step for propagator calculation.
	\item $\mathrm{N_{threads}}$ [integer]: number of TBB threads to be used in the simulation.
\end{enumerate}
\subsubsection{Output files}
PropagatorBuilder generate two binaries: \textalltt{materialname_\(\mathrm{N\sb{a}}\)_\(\mathrm{N\sb{b}}\)_1_TK_{\(\mathrm{\Delta}\)t}ps.B.eigen.bin} and \textalltt{materialname_\(\mathrm{N\sb{a}}\)_\(\mathrm{N\sb{b}}\)_1_TK_{\(\mathrm{\Delta}\)t}ps.P.eigen.bin}, which contain the $B$ and $P(\Delta t)$ matrices. 

\subsection{\label{app:progbRTA} beRTAMC2D}
The \verb|beRTAMC2D| executable implements the bRTA algorithm. It has the following command line inputs:
\begin{alltt}
	beRTAMC2D input.xml N\(\mathrm{\sb{threads}}\)
\end{alltt}
\begin{enumerate}
	\item input.xml [string]: xml file containing the input.
	\item $\mathrm{N_{threads}}$ [integer]: number of TBB threads to be used in the simulation.
\end{enumerate}
\subsubsection{input.xml}
We provide now another toy example of input.xml files together with an explanation of its variables:
\lstset{language=XML}
\begin{lstlisting}
<beRTAMC2D>
	<geometry file="grta01.xml"/>
	<gradient x="0.2" y="0.0"/>
	<time dt="0.5" maxtime="100000"/>
	<material name="black_P" 
	  database="black_monolayer_50_50_1.h5" 
	  thickness="0.5">
		<propagator T="300" 
		 file="bP.P.eigen.bin"/>
	</material>
	<Eeff Ed="4.0e-26" particles="3600" 
	  Tmax="301.0" Tmin="299.0"/>
</beRTAMC2D>
\end{lstlisting}
\begin{enumerate}
	\item geometry\_file [string]: XML file containing geometry data.
	\item gradient/x [double,units(K/nm)]: x-component of homogeneous thermal gradient applied to all boxes (NOTE: requires y-component).
	\item gradient/y [double,units(K/nm)]: y-component of homogeneous thermal gradient applied to all boxes (NOTE: requires x-component).
	\item time/dt [double,units(ps)]: time step.
	\item time/maxtime [double,units(ps)]: Maximum time for simulation.
	\item material/name [string]: material id.
	\item material/database [string]: path to \texttt{almaBTE}'s database containing phonon properties.
	\item material/thickness [double,units(nm)]: real thickness in z-direction, used to correct DFT lattice vector in z-direction.
	\item material/propagator/T [double,units(K)]: reference temperature used to calculate the propagator.
	\item material/propagator/file [string]: path to propagator matrix binary.
	\item Eeff/particles [integer]: division of energy guess, the calculation of the guess is done individually per box and the minimum value is selected. Not needed if Eeff/Ed is provided.
	\item Eeff/Tmin [double,units(K)]: minimum temperature used to calculate the deviational energy per particle. Not needed if Eeff/Ed is provided.
	\item Eeff/Tmax [double,units(K)]: maximum temperature used to calculate the deviational energy per particle. Not needed if Eeff/Ed is provided.
	\item Eeff/Ed [double,units(J)]: deviational energy per particle, it is not needed if Eeff/particles, Eeff/Tmin and Eeff/Tmax are provided. (NOTE: the \verb|RTAMC2D| calculated value for same system and initial conditions is a proper value).
	\item RTA [optional]: if given, the RTA is used for scattering. 
\end{enumerate}

\subsubsection{Outputs: standard output}
Simulation data is dumped to standard output, with lines without physical information starting with \#. Data for lines with physical information is structured as follows:
\begin{align*}
istep ~ t ~ N_{particles} ~ \rho^d_{0} ~ J_{x,0} ~ J_{y,0} ~ \ldots ~ \rho^d_{N-1} ~ J_{x,N-1} ~ J_{y,N-1},
\end{align*}
with $istep$ being the MC loop steep, and the energy density and fluxes are given for the $N$ boxes forming the system. All quantities are given in SI except for time which is given in ps.
\subsubsection{Outputs: properties.msgpack.bin}
The particle distribution per box and time step is dumped in sparse binary format in \verb|properties.msgpack.bin| using using MessagePack library~\cite{msgpack}. Each time step information in the file starts by a std::size\_t value ($size_{block}$) followed by \# indicating the number of characters comprising the block. Blocks are comprised of arrays of $3 + 2N_{boxes}$ elements, with the three first ones being: time (double), deviational energy per particle (double) and one array with the $T_\text{ref}$ for all boxes. This is followed by a structure for each box composed of an std::size\_t indicating the number of non-zero elements in the distribution function vector and one array containing non-zero pairs (except in the case of empty boxes in which a dummy one is created) of an std::size\_t indicating the phonon mode and one integer containing the number of net particles in that mode with its sign indicating if they are positive or negative particles. 
This file can be processed with \verb|dist_reader| executable to obtain the spectral decomposition of the deviational temperature and heat flux at selected boxes and times.

\subsection{dist\_reader}
The \verb|dist_reader| executable allows the extraction of spectral resolved quantities. It has as command line inputs:
\begin{alltt}
	dist_reader input.xml properties.msgpack.bin
\end{alltt}
\begin{enumerate}
	\item input.xml [string]: xml file containing the input.
	\item properties.msgpack.bin [string]: file containing distribution function at given times as produced by \verb|beRTAMC2D|.
\end{enumerate}
it will provide the spectral decomposition of the deviational temperature as well as those of fluxes, with the same format as \verb|RTAMC2D| spectral decomposition files, but with time column referring to instantaneous time rather than the middle of time bin.
\subsubsection{input.xml}
\verb|input.xml| is the same as in the case of \verb|beRTAMC2D|, with the following extra terms:
\begin{enumerate}
	\item spectral/resolution/ticks [integer]: Number of divisions for spectral decomposition of fluxes and deviational temperature.
	\item spectral/time/t [double]: Times at which the spectral decomposition will be calculated. It can be used more than one time
	\item spectral/location/bin [integer]: Box id in which spectral decomposition is calculated. It can be used more than one time.
\end{enumerate}
\subsubsection{Output files}
Spectral decomposition of temperature is printed in \verb|deltaT_omega_BoxID.csv| and fluxes in \verb|jx_omega_BoxID.csv| and \verb|jy_omega_BoxID.csv|. The $q$-resolved deviational phonon distribution function is printed in \verb|fd_BoxID.csv|. Data format is the same of the RTA spectral decomposed and $q$-resolved quantities, but this time with $t$ indicating the instantaneous time and not the middle of time bin.

\section{\label{APPnanowires}Suppression factors for nanowires}

Suppression factors for nanowires ($S^\mathrm{nw}_{\lambda}$) can be calculated, analogously to the case nanoribbons, by evaluating the integrals in Eq.~8 of Ref.~\citenum{LiPRB2012} in cylindrical coordinates, yielding: 
\begin{gather}
	S^\mathrm{nw}_{\lambda} = 1 - \frac{2 M^\mathrm{nw}_{\lambda} }{R^2}\left[M^\mathrm{nw}_{\lambda} \left(e^{-\frac{R}{M^\mathrm{nw}_{\lambda}}} -1\right)+R\right] \\
	M^\mathrm{nw}_{\lambda} = \lVert {v}_\lambda - ({v}_\lambda\cdot {u}){u} \rVert \tau_{\lambda}
\end{gather}
where $R$ is the nanowire radius.

\section{\label{rtaalgo}RTA algorithm}
We now describe all possible events that a particle may take part of during the RTA simulation. 
\subsection{Particle generation}
The deviational power introduced in the system is calculated as $\left|\dot{E}^\mathrm{tot}\right|=\left|\dot{E}^\mathrm{{\nabla_r}T}\right|+\left|\dot{E}^\mathrm{iso}\right|$ where $\left|\dot{E}^\mathrm{{\nabla_r}T}\right|$ and $\left|\dot{E}^\mathrm{iso}\right|$ are the deviational power due to applied thermal gradients and isothermal walls, and are given by:
\begin{multline}
	\left|\dot{E}^\mathrm{iso}\right| = \sum_{j,i} |\dot{E}_{i,j}^\mathrm{iso}| = \sum_{j,i}\frac{\delta z_{j} L_{j}}{N_{q}V_\mathrm{uc}}\left|(n^0_i(T_{j}) -  n^0_i(T_\text{ref}))\right| \times \\ {v}_i \cdot {\hat{e}}^\mathrm{out}_{j} \Theta({v}_i \cdot {\hat{e}}^\mathrm{out}_{j}),
\end{multline}
where $j$ is the isothermal wall index, $i$ runs over modes at the wall, $\delta z_{j}$ is the material's thickness, $L_j$ the wall length, $N_q$ the number of $q$-points at which phonon properties are calculated, $\hbar$ is the reduced Planck constant, $V_\mathrm{uc}$ is the unit cell volume and ${\hat{e}}^\mathrm{out}_{j}$ is a unit vector pointing outwards from the wall, and
\begin{multline}
	\label{gradGen}
	\left|\dot{E}^\mathrm{{\nabla_r}T}\right| = \sum_{j,i}|\dot{E}_{i,j}^\mathrm{{\nabla_r}T}| =\sum_{j,i}\frac{V_{j,\mathrm{box}}}{N_{q}V_\mathrm{uc}} \times \\ \left|{v_i} \cdot {\nabla_r}T\right| \frac{\partial n^0_i(T_\text{ref})}{\partial T}. 
\end{multline}
Then particles with mode $i$ are generated at $j$ from source $s$ with a probability $|\dot{E}_{i,j}^\mathrm{s}|/|\dot{E}^\mathrm{tot}|$ and $\sigma = \mathrm{sgn}(E_{i,j}^s)$.

\subsection{Free flight}
Particles evolve ballistically from ${\mathrm{r_0}}$ (${\mathrm{r_f}} = {\mathrm{r_0}} + {\mathrm{v_i}} t_\mathrm{flight}$) where $t_\mathrm{flight}$ is defined as $\mathrm{min}\{\tau_i ln(R),t_{b}\}$, where $t_b$ is the time the particle needs to encounter a boundary, interface or an isothermal wall and $R$ is random number in range $(0,1]$. 

\subsection{Interface scattering}
If a particle arrives to an interface between two different materials we use the diffuse mismatch model \cite{SwartzRevModPhys1989,ReddyAPL2005} to determine the output state and side in a $A-B$ interface as:
\begin{multline}
	P_{k,A\rightarrow k',C} = \frac{8\pi^3}{V_{\mathrm{uc},C}N_{q,C}}|{v}_k\cdot {n}| \delta(\omega_k'-\omega_k)\times \\ \left(\Theta({v}_k\cdot {n})\delta_{BC} + \Theta(-{v}_k\cdot {n})\delta_{AC}\right) \Bigg/ \\ \Bigg[ \sum^{A,B}_{D}\frac{8\pi^3}{V_{\mathrm{uc},D}N_{q,D}}\sum_{j}|{v}_j\cdot {n}|\times\\ \delta(\omega_j-\omega_k)\left(\Theta({v}_j\cdot {n})\delta_{BD} + \Theta(-{v}_j\cdot {n})\delta_{AD}\right)\Bigg]
\end{multline}
where ${n}$ is the vector normal to the interface, pointing out of $A$ into $B$, and $C\in\{A,B\}$. For systems with layered stacks the LDMM of Section \ref{sandwichDMM} can be used, thus multiplying the probability $P_{k,A\rightarrow k',C}$ by the coupling constant $\mathscr{C}_{k,k'}^{A,C}$

\subsection{Boundary scattering}
A particle encountering a boundary undergoes a full diffusive boundary scattering process after which the outgoing mode is determined following Lambert's cosine law [see Eq.~(\ref{lambert})].

\subsection{Absorption by reservoirs}
Whenever a particle reaches a reservoir, it is absorbed by it, thus terminating its trajectory. 

\subsection{Intrinsic scattering}
If particles end their free flight segment without encountering a boundary, interface or isothermal reservoir, they undergo intrinsic scattering, in which the phonon mode ($k'$) is resampled from the distribution $\frac{C_{k'}(j)/\tau_{k'}(j)}{\sum_i C_i(j)/\tau_k(j)}$ of the $j$-th computational box.

\subsection{Particle termination}
Besides being terminated when contacting reservoirs, particles are also terminated if their simulation time exceeds a maximum threshold, in order to alleviate computational burden. This is justified as trajectories contributing to heat flux reach values in the order of the statistical deviation after a relatively small number of scattering events~\cite{PeraudAPL2012,PeraudARHT2014}.

\section{\label{app:bRTA} Beyond RTA}
Overall the implemented bRTA algorithm is quite similar to that of the RTA described in \ref{rtaalgo} except for scattering and sampling (see Sec.~\ref{sec:Methods_beyondRTA}), as well as the following aspects:
\begin{enumerate}
	\item Initial temperature: It allows for initial temperatures different to the reference one.
	\item Free flight: $t_{\mathrm{flight}}$ is defined as $\mathrm{min}\{\Delta t,t_b\}$ being $\Delta t$ the time step used for $P(\Delta t)$ calculation.
	\item Interface scattering: It is not implemented.
	\item Boundary scattering: Particles reaching the boundaries are used to calculate the boundary temperature ($T_{\mathrm{wall}}$) via cubic spline interpolation of $E_{out}\Delta t / (L_{border}\delta z ) = \frac{1}{V_{uc}}\sum_{i} {v}_i\cdot {\hat{e}}_{\perp}^\mathrm{in} (n^0_i(T_{wall})-n^0_i(T_\text{ref})) [{v}_i\cdot{\hat{e}}_{\perp}^\mathrm{in} > 0]$, where ${\hat{e}}_{\perp}$ is a vector normal to the boundary, pointing inside the material. Then, $T_{\mathrm{wall}}$ is used to create the distribution from which particles are randomly drawn during the time step.
\end{enumerate}

\subsection{Initial temperature}

An initial number of particles is generated due to an initial temperature profile ($t=$\,\SI{0}{\pico\second}). The number of particles ($N^j$) introduced in the $j$-th box is calculated as:
\begin{multline}
N^j = \sum_{i} |N_i^j| = \sum_{i} \Big| \frac{V_{j,\mathrm{box}}}{N_{q}V_\mathrm{uc}\varepsilon_d}  \times  \\ (n^0_i(T_{j}) -  n^0_i(T_\text{ref}))\Big|
\end{multline}
where $i$ is the phonon mode. Then particles with mode $i$ are generated at $j$ with a probability $|N_i^j\mathrm|/N^j$ and $\sigma = \mathrm{sgn}(N_{i})$.

\subsection{Particle cancellation}
As the scattering step creates new particles, computational cost will scale to infinity as simulation progresses. That is prevented by a cancellation scheme in which particles in the same box and identical mode but with opposite sign are removed~\cite{LandonThesis}. It should be noted that cancellation is not performed at each time step, but periodically, so as to achieve a balance between the increased number of particles and the cancellation cost. 

\subsection{Storage of particle distributions}
A histogram of particles is saved at each time step, enabling the calculation of frequency- and mode-resolved properties. Writing to disk is handled by a dedicated thread and performed in a sparse format using the MessagePack library~\cite{msgpack}.

\section{\label{RTALVDSMC}RTA-bRTA}
The RTA version of the bRTA algorithm is exactly equal to the full version, except that the scattering algorithm for each particle in a given state $k$ is simplified to generate a random number $R$ in range $[0,1)$ and if $R < 1 - e^{B_{kk}\Delta t} $ the particle is scattered and resampled from the distribution $\frac{C_{k'}(j)/\tau_{k'}(j)}{\sum_i C_i(j)/\tau_k(j)}$ of the $j$-th computational box. Contrary to the \verb|RTAMC2D| implementation, this RTA implementation allows for multiple reference temperatures and initial temperature profiles. Despite those advantages, we notice that \verb|RTAMC2D| implementation is much more efficient for steady-state calculations, with time-independent sources.

\section{\label{nanoribbons-ZZ}Thermal properties of phosphorene ZZ-nanoribbons}
\begin{figure}
	\centering
	\includegraphics[width=\linewidth]{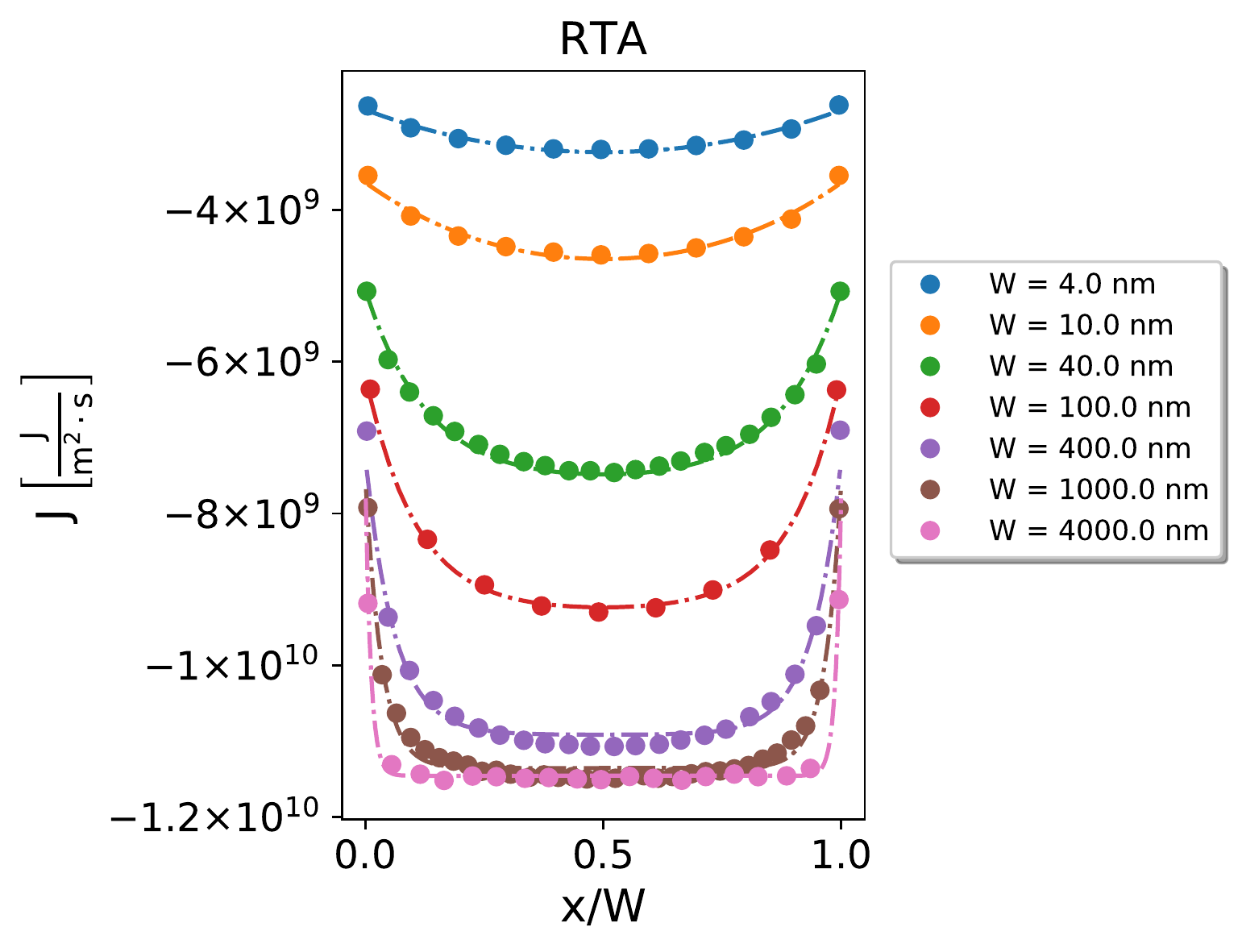}
	\caption{Fitting to Eq.~\ref{Seillitto} (lines) of RTA-MC calculated heat flux (points) as a function of normalized position for phosphorene ZZ nanoribbons of different widths under the effect of $\nabla_\mathrm{ZZ}T = \SI{0.2}{\kelvin\per\nano\meter}$.}
	\label{ZZ-nrhydroRTA}
\end{figure}

\begin{figure}
	\centering
	\includegraphics[width=\linewidth]{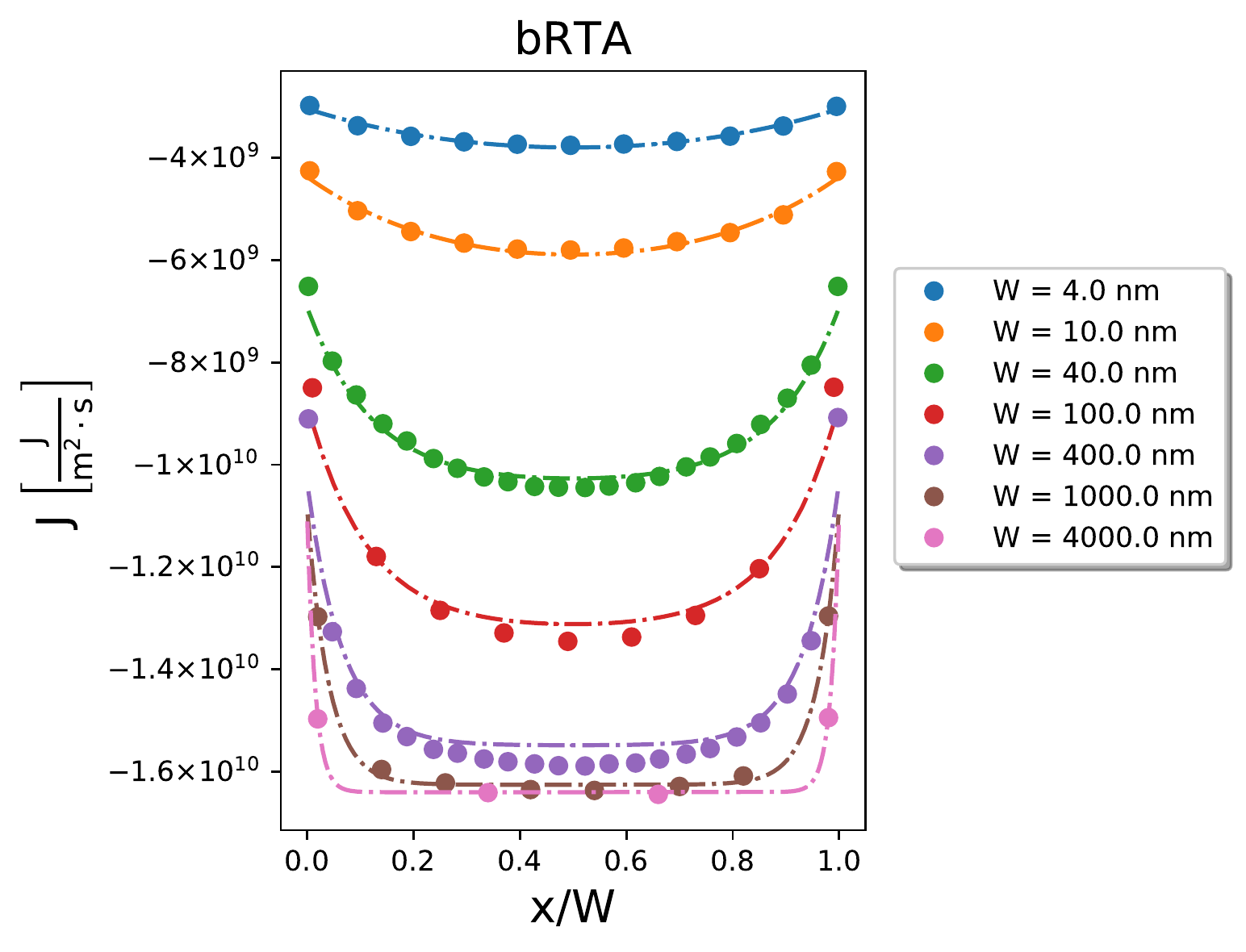}
	\caption{Fitting to Eq.~\ref{Seillitto} (lines) of bRTA calculated heat flux (points) as a function of normalized position for phosphorene ZZ nanoribbons of different widths under the effect of $\nabla_\mathrm{ZZ}T = \SI{0.2}{\kelvin\per\nano\meter}$.}
	\label{ZZ-nrhydroBERTA}
\end{figure}

\begin{figure}
	\centering
	\includegraphics[width=\linewidth]{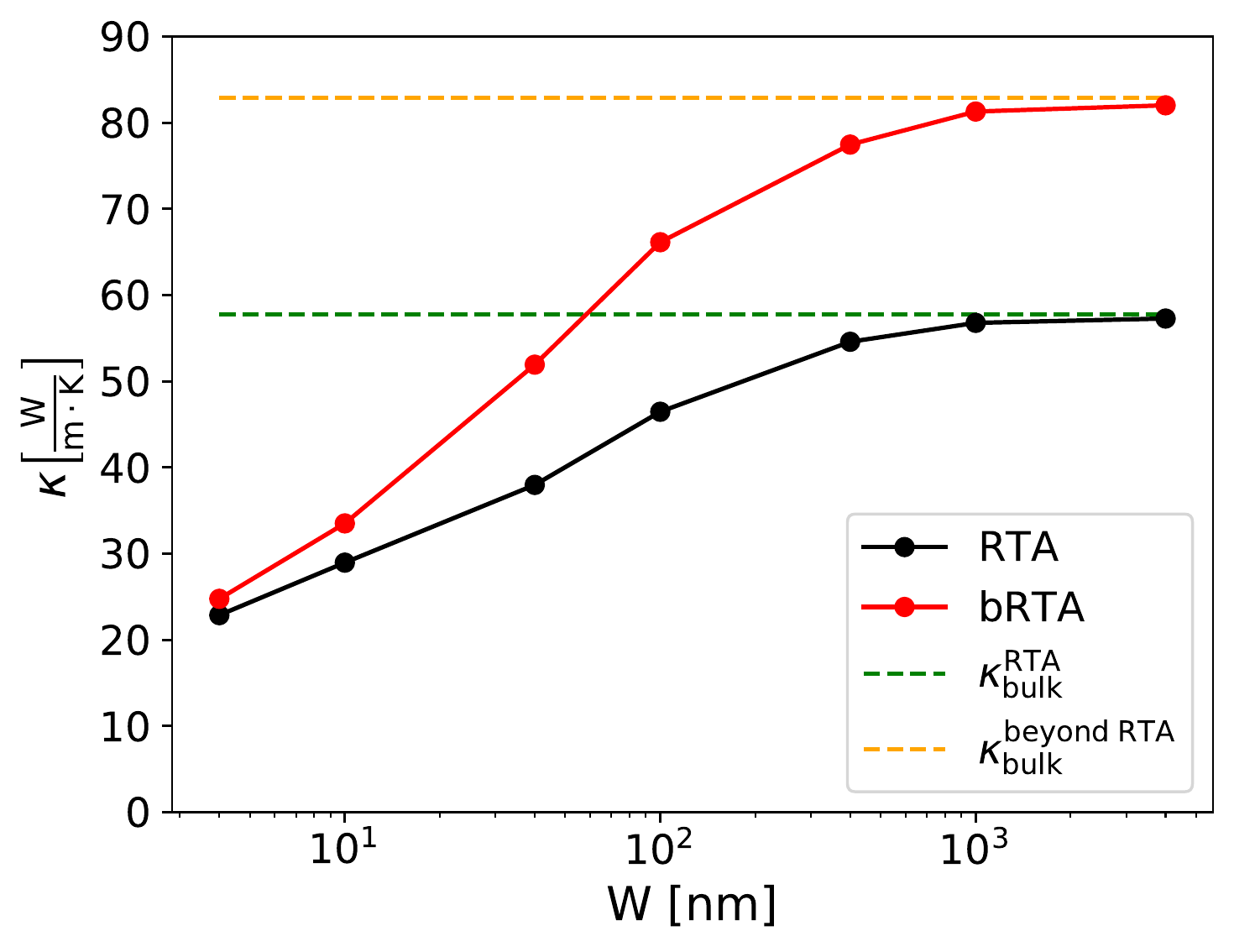}
	\caption{Fitted thermal conductivity as a function of ZZ nanoribbon width for RTA (black) and bRTA (red) MC calculations. RTA (green) and beyond RTA (orange) bulk values are given for reference.}
	\label{ZZ-kapparibbons}
\end{figure}

\begin{figure}
	\centering
	\includegraphics[width=\linewidth]{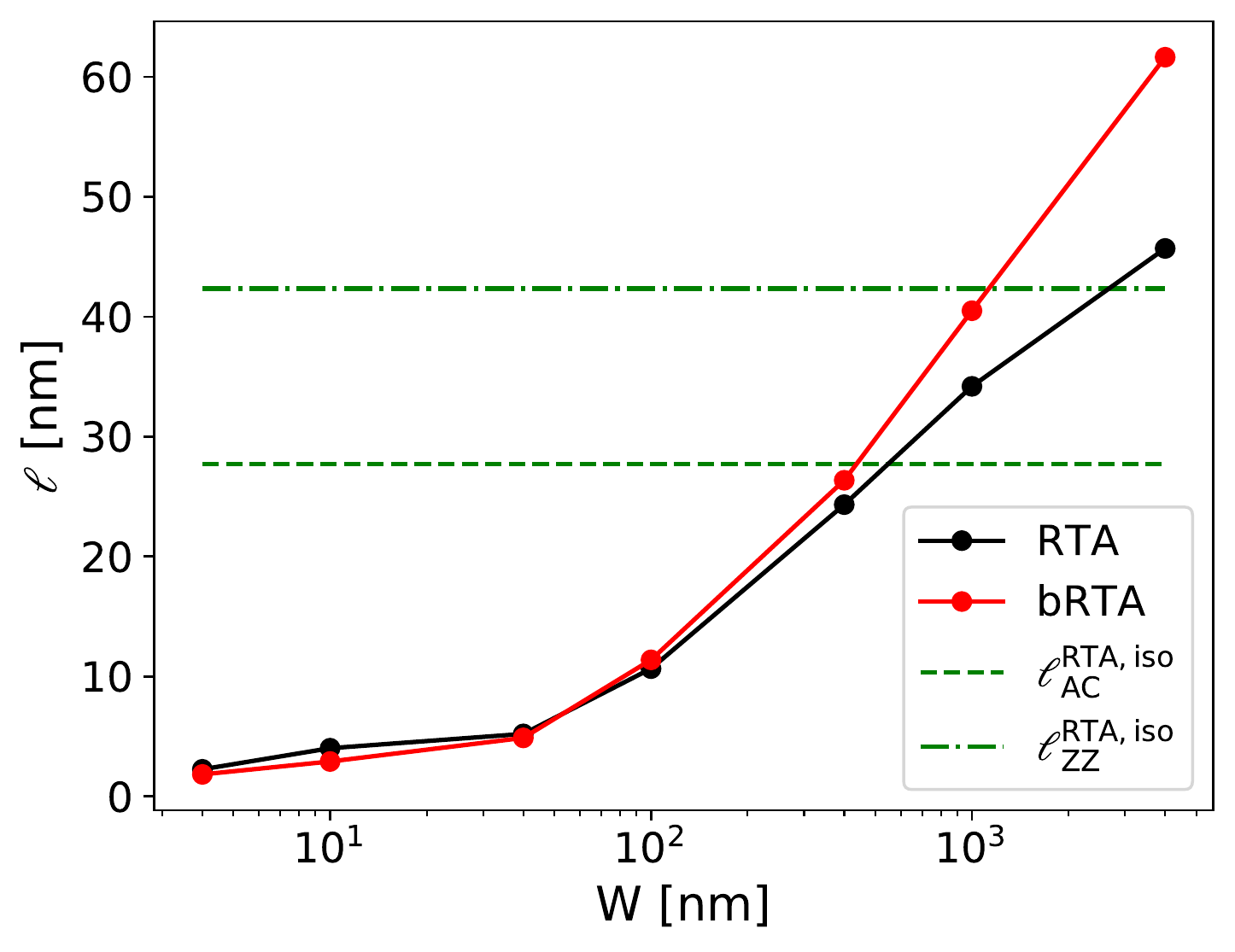}
	\caption{Fitted non-local distance $\ell$ as a function of ZZ nanoribbon width for RTA (black) and bRTA (red) MC calculations. RTA bulk values of $\ell$ calculated using Sendra et. al.'s formula~\cite{SendraPRB2021} are given for reference (green).}
	\label{ZZ-lgkribbons}
\end{figure}

Here we provide, for reference, the results for the heat flux profiles and fits to the mesoscopic Eq.~\ref{Seillitto} of ZZ-nanoribbons having several widths under an applied thermal gradient. Those results (see Figs.~\ref{ZZ-nrhydroRTA}-\ref{ZZ-lgkribbons}) do not provide any additional information regarding the analysis done in Sec.~\ref{Sec:nanoribbons}, with the AC findings and conclusions also valid for ZZ nanoribbons.


\clearpage
\section*{References}
\bibliographystyle{./myelsarticle-num.bst}
\bibliographystyle{elsarticle-num}
\bibliography{ref.bib}

\begin{thebibliography}{10}
\bibitem{broadening}
Conservation energy deltas are approximated by Gaussian functions (
  $\delta(\omega_i\pm\omega_j-\omega_k) \approx
  \frac{1}{\sqrt{2\pi}\sigma_{ijk}} \exp \left(
  \frac{-(\omega_i\pm\omega_j-\omega_k)^2}{2\sigma_{ijk}^2} \right)$ with $
  \sigma_{ijk} = \frac{a}{\sqrt{12}} \lVert \left( G^{T}_{\mu\alpha}
  N_{\mu\mu}^{-1} \right)^T \cdot (\boldsymbol{v}_j -
  \boldsymbol{v}_k)_{\alpha} \rVert$, where $i$, $j$ and $k$ are the phonon
  modes taking part in the three-phonon process, $\mu$ indicates a
  reciprocal-space lattice vector, $\alpha$ indicates a cartesian axis,
  $G_{\mu\alpha}$ is the reciprocal lattice, $N_{\mu\mu}$ is a diagonal matrix
  with the q-grid size and $a$ is the scalebroad factor) which, because of how
  $\sigma_{ijk}$ is calculated, breaks the microscopic reversibility as one
  process can be registered but not the reciprocal or it can be registered, but
  with different value. Moreover, the way smearing is constructed from products
  of velocities over reciprocal lattice vectors can cause symmetry equivalent
  pairs/triplets to give different values.
\end{thebibliography}

\end{document}